\documentclass[%
 reprint,
 amsmath,amssymb,
 aps,
prb,
]{revtex4-2}
\usepackage{braket}
\usepackage{comment}

\usepackage{graphicx}
\usepackage{tipa}
\usepackage{amsfonts}
\usepackage{bbding}
\usepackage{bbold}
\usepackage{bbm}
\usepackage[T1]{fontenc}
\usepackage{amsmath}
\usepackage{color}
\usepackage{soul}
\usepackage{ulem}
\usepackage{bbm} 
\usepackage{tensor}
\usepackage{dsfont}
\usepackage{slashed}
\usepackage{hyperref}
\usepackage[all]{hypcap}
\hypersetup{colorlinks=true,
   linkcolor=blue,
    citecolor=blue,
    filecolor=magenta,
    urlcolor=cyan}

\PassOptionsToPackage{unicode}{hyperref}
\PassOptionsToPackage{naturalnames}{hyperref}
  
\begin{document}

\title{\textcolor{black}{Is curved graphene a possible answer to the problem of graphene's diverging magnetic susceptibility?}}
\author{Abdiel de Jesús Espinosa-Champo\texorpdfstring{${}^{a,c}$}{}}
\author{Gerardo G. Naumis\texorpdfstring{${}^{b}$}{}}
\affiliation{\texorpdfstring{${}^{a}$}{} Posgrado de Ciencias Físicas, Universidad Nacional Autónoma de México, Apartado Postal 20-364 0100,Ciudad de México, México}

\affiliation{\texorpdfstring{$^{b}$}{} Departamento de Sistemas Complejos, Instituto de Física, Universidad Nacional Autónoma de México, Apartado Postal 20-364 01000, Ciudad de México, México.}
\author{ Pavel Castro-Villarreal\texorpdfstring{$^{c}$}}{}
\email[e-mail: ]{pcastrov@unach.mx }
\thanks{author to whom correspondence should be addressed.}
\affiliation{\texorpdfstring{$^{c}$}{}Facultad de Ciencias en Física y Matemáticas, Universidad Autónoma de Chiapas, Carretera Emiliano Zapata, Km. 8, Rancho San Francisco, C. P. 29050, Tuxtla Gutiérrez, Chiapas, México.}

\begin{abstract}
\textcolor{black}{A study of curved graphene in the presence and absence of a real magnetic field is conducted to determine the magnetization and magnetic susceptibility. Utilizing a Dirac model, the Landau level energy corrections are found. These results are compared with those obtained from a tight-binding model analysis, showing good agreement with the Dirac model. The obtained spectra are then used to calculate the free energy, magnetization, and magnetic susceptibility as functions of the external magnetic field and curvature. The resulting de Haas van Alphen (dHvA) effect exhibits distinctive signatures due to the curvature of graphene, including a resonance effect when the pseudomagnetic and the real magnetic fields are equal. Considering that curvature induces effective pseudomagnetic fields, a mechanical effect stemming from an electronic contribution is found, resulting in a pseudo-de Haas van Alphen (pdHvA) effect without needing an external magnetic field. This effect is associated with oscillating (electronic) forces opposing deformations. These forces, divergent in flat graphene, suggest that graphene (without a substrate) attains mechanical equilibrium through local corrugations. These mechanical deformations prevent the theoretically calculated pristine graphene's diamagnetic divergence at low temperatures, indicating that corrugations produce a finite, experimentally measurable magnetic susceptibility. The divergent susceptibility becomes apparent only when such corrugations are removed using various strategies. }
\end{abstract}

\maketitle

\section{Introduction \label{Sec:Introduction}}

The de Haas van Alphen effect (dHvA) is a fascinating phenomenon that has attracted the attention of many physicists over the years \cite{Haas1930, NingMa2019, Zhang_2010, Wilde2006, Vagner2006, pluzhnikov2007, Pratama2021, Shoenberg1959, Shoenberg1984Book, lukyanchuk2011, kryuchkov2016, Dong2020, Champel2001, Manninen2022, Shoenberg1952, Shoenberg1957, Shoenberg1960, Holstein1973, Shoenberg1969, Shoenberg1988}. It refers to periodic oscillations in metal magnetization when subjected to a magnetic field at low temperatures \cite{Haas1930, Vagner2006, Shoenberg1984Book, Fu2011, Sharapov2004, Slizovskiy2012, Li2015}. These oscillations are caused by quantization of the electron energy levels in the magnetic field \cite{Landau_1930, Shoenberg1959, Shoenberg1984Book}, and their frequency is proportional to the cross-sectional area of the Fermi surface of the sample \cite{Onsager1952}.

The de Haas van Alphen effect was first discovered by Wander Johannes de Haas and Pieter M. van Alphen in 1930 while studying bismuth magnetization in a high magnetic field \cite{Haas1930}. Later, it was realized that the effect is a general phenomenon that occurs in any metal or semiconductor with a Fermi surface \cite{Shoenberg1960, Shoenberg1959, pluzhnikov2007, Shoenberg1952}. In two-dimensional electron gases (2DEGs), the dHvA effect is fascinating because the quantization of electron energy levels is more pronounced as a result of the dimensionality reduction of the system \cite{Holstein1973, Shoenberg1984Book, Peierls_1933, Champel2001}. This makes the effect a powerful tool for studying the electronic properties of quasi- and two-dimensional materials \cite{Manninen2022, Fu2011, kryuchkov2016, lukyanchuk2011, pluzhnikov2007,  Liu2019}. This effect has also been extensively studied in 2DEGs formed at the interface between two semiconductors \cite{Wilde2006, Wilde2009, Wang2009, Vagner2006}. 

One of the most exciting aspects of the de Haas van Alphen effect in 2DEGs is the observation of sawtooth oscillations \cite{Peierls_1933, Champel2001}, which are caused by the fractional filling of the Landau levels (LLs) under a magnetic field. These oscillations provide a powerful tool for studying the fractional quantum Hall effect \cite{Meinel1999}, which is a striking manifestation of strong electron-electron interactions in 2DEGs. 

A natural setting to study the dHvA effect is on two-dimensional materials beyond the 2DEGs. In recent years, there has been a growing interest in exploring this effect in two-dimensional materials beyond 2DEGs. For example, this effect has been experimentally observed in graphene \cite{Manninen2022, Li2015}. 

Graphene is a two-dimensional material comprising a single layer of carbon atoms arranged in a honeycomb lattice with remarkable properties \cite{Novoselov2004, Wallace1947}. For example, when graphene is deformed, the Dirac points are separated, and the Fermi surface becomes a series of deformed circles \cite{Maurice2013, Maurice2015, Naumis2017}. { \color{black}These deformations in graphene produce a pseudomagnetic field \cite{VozmedianoCortijo2007, CastroNeto2009, Vozmediano2012, Naumis2017, Pavel2017, Abdiel2018, Abdiel2021, Li2020, Bitan2011, Bitan2013, Pablo_Morales2023}, which can lead to pseudo-Landau levels (P-LL) \cite{VozmedianoCortijo2007, CastroNeto2009, Vozmediano2012, Wagner2022, Li2020, Naumis_2024,Yin2017,Sandler20201,Sandler2018,Wakker2011,Zhou2023,HSU2020,Gentile2022,Liu2023,NancySandler2019}}. Also induce corrections to the Quantum Hall effect (QHE) \cite{Wagner2022}, pseudomagnetic QHE \cite{bhagat2019}, anomalous dHvA effect in strained graphene \cite{NingMa2019, Dong2020},  Klein tunneling on \textcolor{black}{negatively} curved \textcolor{black}{graphene sheets} 
\cite{Victor2023}
, flat bands \cite{RomanTaboada2017,RomanTaboada2017b,RomanTaboada_2017JPC,Mao2020, Manesco_2021, Manesco_2021_2, Milanovic2020, Sandler2023}
and many other geometrical effects on the electronic properties  \cite{RamonCarrillo2014,   RamonCarrillo2016, Naumis2017, Abdiel2018, Abdiel2021, Monteiro2023}, which allow posing the concept of {\it curvatronics}, that is, consider the curvature as a tunable parameter to control the electronic properties of the material \cite{Cariglia2017}. Specifically, this enables curved graphene to have several potential applications. One possible application is in the field of valleytronics, which aims to use the valley degree freedom of electrons in graphene rather than their charge to transmit and process information, and the curvature could be used to polarize the valley of electrons, allowing the creation of new types of valleytronic devices \cite{Stegmann_2019, NingMa2019, RamonCarrillo2014}. Understanding the magnetic, electronic, and thermodynamic properties of curved graphene is crucial for a comprehensive understanding of this material's behavior.

 Because curvature induces effects similar to those found for magnetic fields, we present a study of curved graphene under magnetic fields in this work. In particular, we will show that curvature induces discrete spectra of the allowed energies (LLs) and a discontinuity in the magnetization with periodicity $1/B$ leading to an effective dHvA effect. {\color{black}  For free-standing graphene, this allows us to propose a potential curvature-induced mechanism to address the issue of the theoretically calculated divergence of magnetic susceptibility at low temperatures. It is important to note that the suggested mechanism is rooted in corrugations. This does not preclude the experimental observation of such divergence, as curvature can permanently be eliminated through strain or encapsulation using other materials \cite{Vallejo_2021}.} 

We have organized this paper as follows. In Sec. \ref{sec2}, we present the simplest model to describe curved graphene under magnetic fields, both in its continuum model using the Dirac equation formalism and in the tight-binding model. In Sec. \ref{sec3}, we calculate the LLs for two curvature regimes using the continuum model and show their effects on the Local Density of States (LDOS) obtained with the tight-binding model. In Sec. \ref{sec4}, we briefly review the dHvA effect in flat graphene with a constant charge carrier density and near-zero temperature. This review establishes the techniques to be used in the following sections. In Sec. \ref{Sec:DHVA-CURVED-GRAPHENE}, we study the dHvA effect in strongly curved graphene and under a real magnetic field. {\color{black} In Sec. \ref{Sec: Physical differences between magnetization and pseudomagnetization}, we briefly discuss the physical differences between magnetization and pseudo-magnetization.}
In Sec. \ref{sec:pseudo-dhva}, we present the mechanical effect produced by the electronic contribution in graphene and how this gives rise to a pseudo-de Haas van Alphen effect (pdHvA) due only to curvature. Finally, in Section \ref{sec5}, we present our conclusions and future perspectives.

\section{Graphene electronic models }\label{sec2}

This section presents the details of the models we will consider and how they relate. {\color{black} This section is divided into two parts. In the first, we provide the Dirac Hamiltonian for curved graphene, and in the second, we study the same system from the perspective of a tight-binding Hamiltonian. It is worthwhile to mention that such a system can also be studied by using perturbation theory \cite{Wakker2011, Sandler20201}.   However, the curved-space method offers several distinctive advantages. Specifically, this approach allows for the natural handling of non-planar geometries and local deformations, such as bumps, which are common in corrugated graphene structures.
In comparison to perturbation theory, often used to address slightly perturbed systems around a known base state, the curved-space method provides a description where the effects of curvature dominate the behavior of electrons in graphene. While perturbation theory has proven effective in specific contexts \cite{Wakker2011, Sandler20201},  the curved-space method has the potential to offer a unique perspective, allowing to analytically study more complex phenomena such as dislocations \cite{Victor2023}. Furthermore, it may facilitate establishing analogies with systems from other fields of physics in a natural manner \cite{Gallerati_2022, Morresi_2020, Pablo_Morales2023}.}

\subsection{Dirac model for curved graphene in an external  magnetic field}
\begin{figure*}
    \centering
    \includegraphics[scale=0.55]{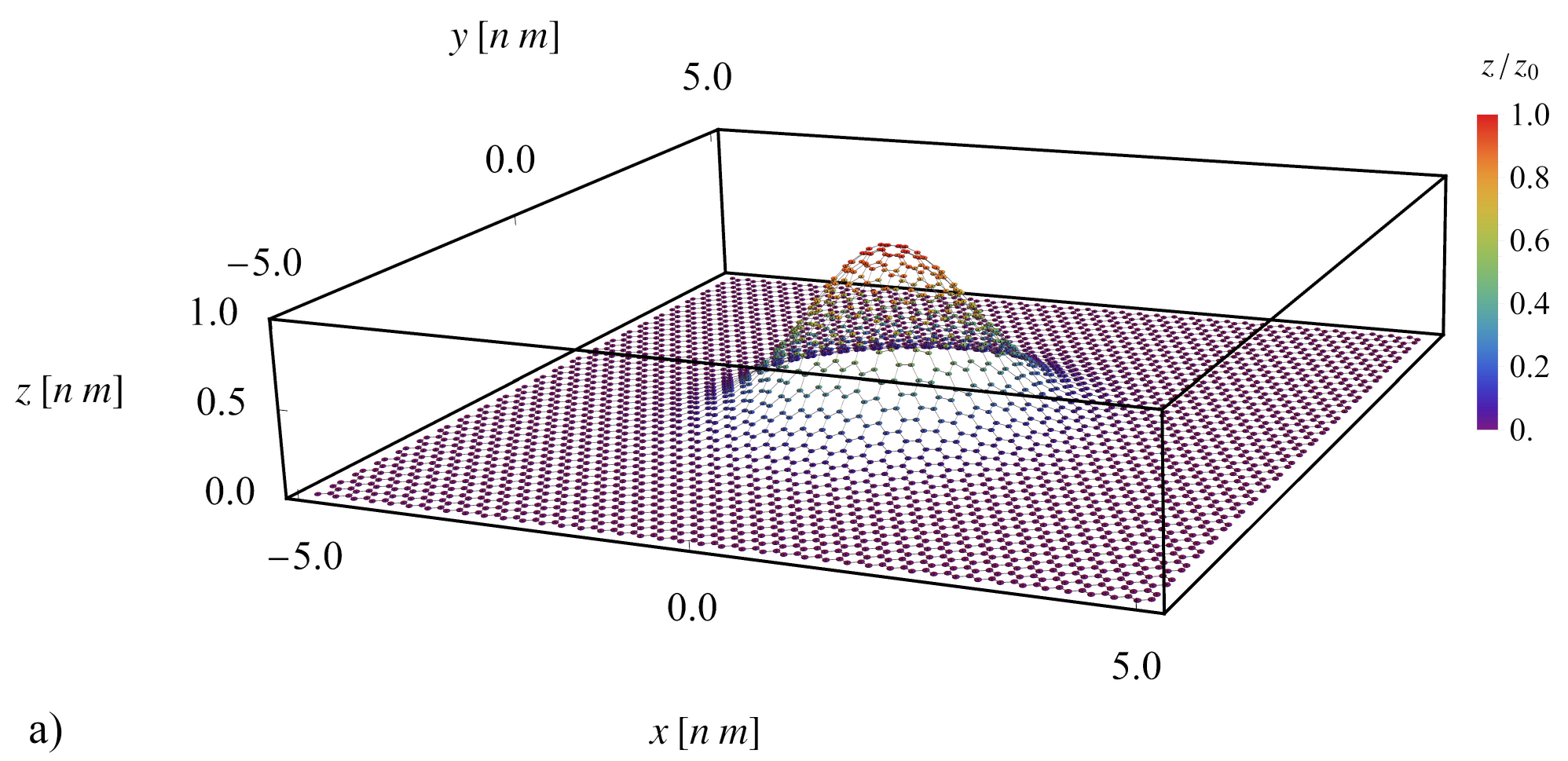}
     \includegraphics[scale=0.55]{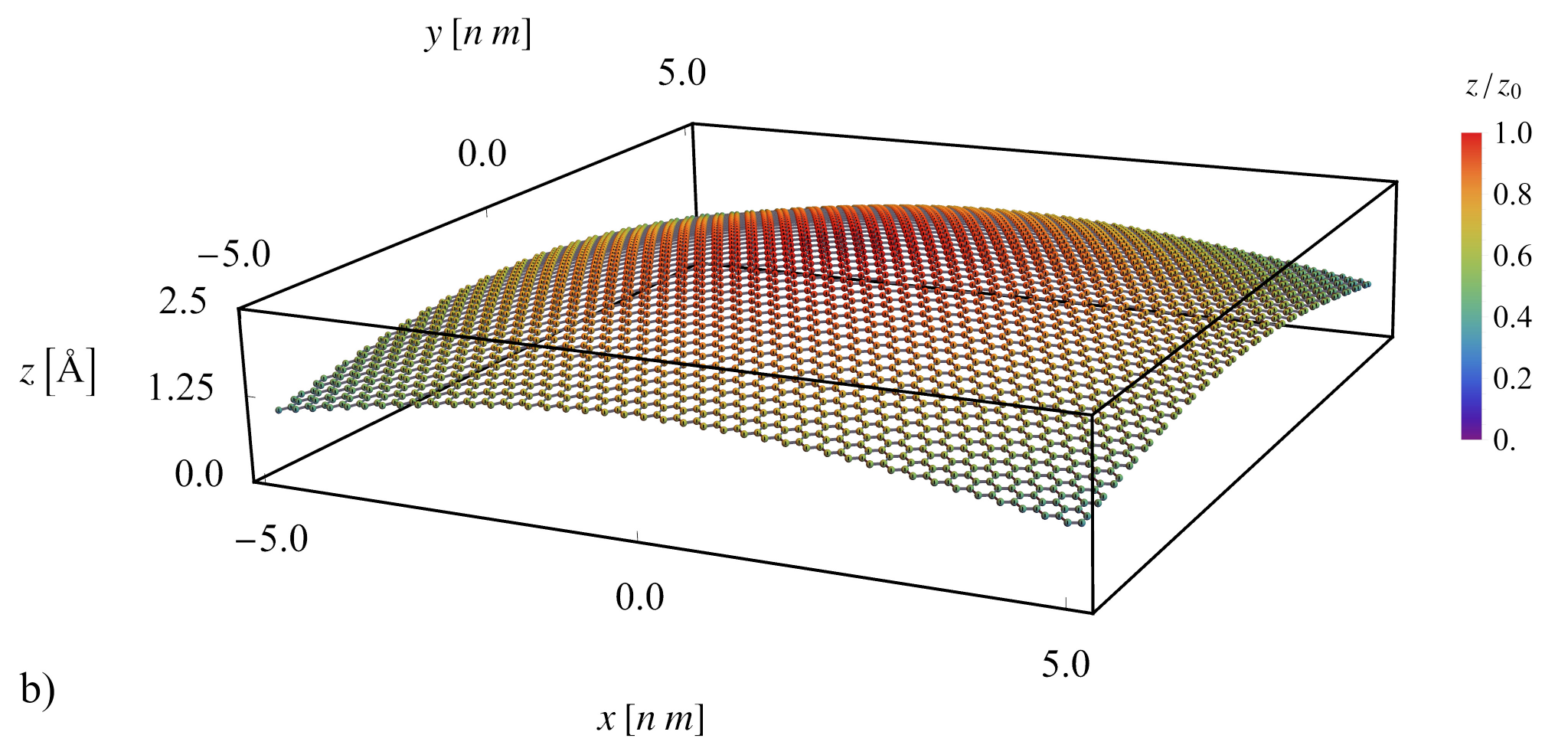}
    \caption{Graphene sheet with dimensions $L_{x}=L_{y}= 10$ nm deformed in such a way as to generate a Gaussian bump whose profile is given by Eq. \eqref{ap.a.11}. As mentioned in Appendix \ref{Sec: Ap.A.Geometric}, the parameter that controls the type of regime established is the ratio between the height of the bump $z_{0}$ and the standard deviation $\varpi^{2}$, that is, $(z_{0}/ \varpi^{2})^{2}=R(r=0 \text{ nm})$. a) For this case, we set $z_{0}= 1$ nm and $\varpi=1$ nm such that $(z_{0}/ \varpi^{2})^{2}= 1 \text{ nm}^{-2} $ establishes a strong curvature regime, while for b) we set $z_{0}= 0.25$ nm and $\varpi=5$ nm such that $(z_{0}/ \varpi^{2})^{2}= 1 \times 10^{-4} \text{ nm}^{-2} $ establishes a weak curvature regime. {\color{black} The color scale indicates the vertical position $z$ of each site in relation to the maximum height $z_0$.}} 
    \label{fig:Fig4}
\end{figure*}

This section introduces the simplest model to describe the electronic degrees of freedom of a curved graphene sheet under a uniform magnetic field. In Fig. \ref{fig:Fig4}, we present two limiting examples of such deformations. To introduce a model for describing such systems, we begin with the Dirac equation in the presence of electromagnetic fields for massless Dirac fermions \cite{Pollock:2010zz,frankel2004geometry, Le2019, Wagner2022, Boada2011, pnueli_spinors_1994, le_electric_2019, olpak_dirac_2012, burgess_fermions_1993, brandt_dirac_2016, le_relativistic_2019, ferrari_schrodinger_2008, Pavel2017}
\begin{equation}\label{eq:II.1}
i \hbar \underline{\gamma}^{\alpha} \left( \nabla_{\alpha}-i \frac{q}{\hbar} A_{\alpha}\right)\psi=0,
\end{equation}
where $q$ is the charge's particle, and the electromagnetic vector potential, denoted as $A_{\alpha}$, is defined on a $2+1$ dimensional curved space-time $\mathbb{M}$. The Dirac matrices $\gamma^{\mathcal{A}}$ satisfy the Clifford algebra $\{ \gamma^{\mathcal{A}}, \gamma^{\mathcal{B}} \}= 2 \eta^{\mathcal{AB}} \mathbb{I}_{2 \times 2}$, with $\eta_{\mathcal{AB}}$  the Minkowski metric tensor and $\underline{\gamma}^{\alpha}(x)= \gamma^{\mathcal{A}} e^{\alpha}_{\mathcal{A}}(x)$.
The set $\{ e^{\alpha}_{\mathcal{A}}(x) \}$ consists of dreibeins associated with each coordinate patch of $\mathbb{M}$. Here, the capital and italic Latin indices $\mathcal{A}$ represent the Minkowski flat coordinates, while the Greek indices $\alpha$ indicate the local curved coordinates.

The covariant derivative for the spinor representation of the Lorentz group $SO(2,1)$ is given by $\nabla_{\alpha}= \partial_{\alpha}+ \Omega_{\alpha}$, where $\Omega_{\alpha}= \frac{1}{4} \omega_{\alpha}^{\mathcal{AB}} s_{\mathcal{AB}}$ serves as the spin connection. The components $\omega_{\mathcal{AB}}^{\alpha}$ form elements of the $1-$form satisfying the Maurer-Cartan equations \cite{nakahara2018geometry}. Meanwhile, $s_{\mathcal{AB}}=\frac{1}{2}[\gamma_{\mathcal{A}}, \gamma_{\mathcal{B}} ]$ represents the pseudo-spin operator. Thus, both $\omega_{\alpha}^{\mathcal{AB}}$ and $e_{\mathcal{A}}^{\alpha}(x)$ carry the geometric essence of the Dirac equation. The metric tensor of the space-time $\mathbb{M}$ can be expressed using dreibeins, denoted by $g_{\alpha \beta}=e^{\mathcal{A}}_{\alpha}(x) e^{\mathcal{B}}_{\beta}(x) \eta_{\mathcal{AB}}$.

 Here, we consider a stationary spacetime with global structure $\mathbb{M}=\mathbb{R}\times \Sigma$, whose spatial sector $\Sigma$ is a two-dimensional curved surface, {\it i.e.} with a metric given by $ds^{2}=-v_{F}^{2}dt^{2}+g_{ij} dx^{i}dx^{j}$, being $g_{ij}$ the metric tensor of $\Sigma$ with $i,j=1,2$, and $v_{F}$ is the Fermi velocity. In addition, we consider that the electromagnetic potential $A_{\alpha}$ has only spatial components, that is,  $A_{0}=0$. In the following, we proceed to separate the indices $\mathcal{A}$ and $\alpha$ into time and spatial components, $\{0, a\}$ and $\{0,j\}$, respectively. Thus, the dreibeins for this metric are $e^{0}_{0}=1, e^{i}_{0}=e^{0}_{a}=0$; and $e^{i}_{a} \neq 0$ such that $g_{ij}=e^{a}_{i}e^{b}_{j} \delta_{ab}$. Also, it can be shown that $\omega^{0}_{jb}=\omega^{a}_{0b}=\omega^{a}_{j0}=0$ and $\omega^{a}_{jb} \neq 0$, with spatial indices $a,b,i,j$ and $k$. This implies that the covariant derivative is reduced to $\nabla_{0}= \partial_{0}$ and $\nabla_{j}= \partial_{j}+\frac{1}{4} \omega^{ab}_{j}s_{ab}$.
This implementation of index decomposition on the Dirac equation (\ref{eq:II.1}) results in a Schrödinger-like equation
$i \hbar \partial_{0} \psi = \mathcal{H}
\psi$, 
where $ \mathcal{H}=- i \hbar v_{F} \gamma_{0} \underline{\gamma}^{j}(x) \boldsymbol{\nabla}_{j}$ with $ \boldsymbol{\nabla}_{j} \equiv  \nabla_{j}- i \frac{q}{\hbar} A_{j}$. 
This decomposition has been studied in previous works \cite{Boada2011,KERNER2012,Pavel2017,Wagner2022}. Additionally, the effective tensorial and space-dependent Fermi velocity $v_{a}^{\textit{eff},j}(x)=v_{F} e^{j}_{a}(x)$ can be obtained using $\underline{\gamma}^{j}(x)=\gamma^{a} e^{j}_{a}(x)$ \cite{Vozmediano2012, Pavel2017}. 
We now have all the components to write a field-theoretic Hamiltonian for the curved sheet of graphene as 
\textcolor{black}{\begin{equation} \label{eq:II.4} 
\hat{H}= \sum_{\xi=\pm} \sum_{\sigma=\uparrow, \downarrow}\int d^{2} x\,\, \sqrt{g}\,\, \psi^{\dag}_{\sigma,\xi}\mathcal{H}_{\xi}\psi_{\sigma, \xi},
\end{equation} }
\textcolor{black}{where $\sigma$ and $\xi$ label  the spin and valley indexes. The operators  $\mathcal{H}_{+}$ and $\mathcal{H}_{-}$ represents two Dirac operators corresponding to each valley $K$ and $K^{\prime}$, respectively. The difference between these operators resides in the Dirac matrix representations in each valley. 
We choose the 
particular representations of the Dirac matrices 
$\gamma^{0}_{\xi}=\gamma^{0}=-i \sigma_{3}, \gamma^{1}_{\xi}\gamma^{0}_{\xi}= \sigma^{1}$, and $\gamma^{2}_{\xi}\gamma^{0}_{\xi}= \xi\sigma^{2}$ with $\sigma_{3}$ and $\sigma^{a}$, $a=1,2$, being the standard Pauli matrices}\textcolor{black}{; this implies, particularly, that in the $K^{\prime}$ valley pseudo-spin operator has opposite sign with respect to $K$ valley. For compact notation, we introduce the Dirac operators in each valley explicitly using the valley index $\xi$}

\begin{eqnarray}
    \mathcal{H}_{\xi}=- i \hbar v_{F} \gamma_{0} \underline{\gamma}^{j}_{\textcolor{black}{\xi}}(x) \boldsymbol{\nabla}_{j}^{{\color{black} \xi}}, ~~{\rm with}~~
    \boldsymbol{\nabla}_{j}^{{\color{black} \xi}} \equiv  \nabla_{j}^{{\color{black}\xi}}- i \frac{q}{\hbar} A_{j}.
\end{eqnarray}
In this representation, for flat graphene, it is easy to show that $\mathcal{H}_{+}=- i \hbar v_{F} \sigma_{a} (\partial_{a} - i \frac{q}{\hbar} A_{a})$, which is the low energy limit of the Wallace tight-binding model \cite{CastroNeto2009, Wallace1947} in the presence of electromagnetic fields once the movement of the Dirac cone tip is taken into account \cite{Maurice2013,Maurice2015,Naumis2017,Naumis_2024}. This Hamiltonian $\hat{H}$ is also similar to the cosmological model proposed by Vozmediano \textit{et al.} \cite{VozmedianoCortijo2007}.

We remark that the curved sheet of graphene is considered here also under a real, external magnetic induction field ${\bf B}$, which is defined in the Euclidean space $\mathbb{R}^{3}$. The field ${\bf B}$ can be expressed in terms of the $U(1)$ gauge field ${\bf A}$ as usual ${\bf B}={\rm rot} {\bf A}$, where ${\rm rot}$ is the rotational operator on vector fields in $\mathbb{R}^{3}$. It is clear that the field ${\bf B}$ determines the vector field $A_{j}$ on the curved surface $\Sigma$. For this purpose, it is imperative to introduce a few extrinsic elements of the curved sheet of graphene geometry. So, let us call ${\bf X}:\mathbb{D}\subset \mathbb{R}^{2}\to \Sigma\subset\mathbb{R}^{3}$ a parameterization of the surface, where $\mathbb{D}$ is a domain and ${\bf X}(x^{i})$ is vector position in $\mathbb{R}^{3}$ on a certain point $p$  of the surface $\Sigma$. Since the charge carriers move intrinsically on the surface, their electromagnetic moment $\frac{q}{\hbar} A_{j}$ must be tangent to the surface, therefore the magnetic potential can be written as $A_{j}(x^{i})={\bf e}_{j}(x^{i})\cdot {\bf A}({\bf X}(x^{i}))$, where  ${\bf A}({\bf X})$ is the $U(1)$ gauge field at the point $p$ on the surface, and $\{{\bf e}_{j}(x^{i})=\partial_{j}{\bf X}(x^{i})\}$ is a set of tangent vectors at $p$.  

{\color{black} We add that due to the introduction of a complex phase in Eq. \eqref{eq:II.1}, the application of a real magnetic field in graphene breaks the time-reversal symmetry. The pseudomagnetic field that arises from the graphene's curvature is different in the sense that it does not break such symmetry \cite{Naumis2017}. This can be seen from Eq. \eqref{eq:II.4} as the graphene valleys $K$ and $K'$ are related by time-reversal symmetry once the real magnetic field is removed. This matter will be studied in more detail in the context of the resulting spectrum.}

\subsection{Tight-binding (TB) model for a curved graphene in a magnetic field}

To better understand and compare some of the results from the previously discussed effective model, we also explore a numerical implementation of a tight-binding (TB) model for curved graphene under a real, external magnetic field. 

The TB Hamiltonian of the model is defined as follows,
\begin{equation} \label{eq:II.5}
    \mathcal{H}= \sum_{\braket{nm}} t_{nm}(\boldsymbol{r}) \hat{c}^{\dag}_{\boldsymbol{r}_{n}} \hat{c}_{\boldsymbol{r}_{m}}+h.c.
\end{equation}
Here, $\braket{nm}$ represents the sum over the neighbors with positions $\boldsymbol{r}_{n}$ and $\boldsymbol{r}_{m}$ that satisfy $\left| \boldsymbol{r}_{n}- \boldsymbol{r}_{m}\right|^{2} - \left(\boldsymbol{r}_{n}- \boldsymbol{r}_{m}\right)_{z}^{2} \leq a_{c}^{2} $ with $a_{c}=1.42$ {\AA} the interatomic distance in flat graphene; $\hat{c}_{\boldsymbol{r}_{n}}^{\dag} (\hat{c}_{\boldsymbol{r}_{n}})$ is the creation (annihilation) operator and $t_{mn}(\boldsymbol{r})$ is the hopping integral between the $n$-th and $m$-th sites, given by

\begin{equation} \label{eq:II.6}
    t_{nm}(\boldsymbol{r})= t_{0} e^{ i \frac{2 \pi}{\Phi_{0}} \int_{\boldsymbol{r}_{n}}^{\boldsymbol{r}_{m}} \boldsymbol{A}_{nm} \cdot d \boldsymbol{l}} e^{- \overline{\beta} \left(\left| \boldsymbol{r}_{n}- \boldsymbol{r}_{m}\right|^{2}-a_{c} \right)/a_{c}},
\end{equation}
where $t_{0}=-2.8$ eV. $\boldsymbol{A}_{nm}$ is the vectorial potential along the path that joins sites $n-th$ and $m-the$, $\Phi_{0}=h/e$ is the magnetic flux quantum, $\overline{\beta}=3.37$ is the Grüneisen parameter  \cite{Naumis2017,Stegmann_2019} . The Fermi velocity in flat graphene can be computed as $v_{F}= \frac{3 t_{0} a_{c}}{2 \hbar} \approx 9.06104 \times 10^{5} \text{ m/s}$. In the present study, the eigenvalues and eigenfunctions resulting from the TB model Hamiltonian, given by Eq. \eqref{eq:II.5}, were obtained using the Pybinding package \cite{pybinding}. 

{\color{black}
Note that, as a consequence of curvature, the $p_{z}$ orbitals become misaligned \cite{Sandler2018}. To account for this effect, it is necessary to introduce an additional correction term to the hopping parameters \cite{RomanTaboada_2017JPC}, such that the new hopping parameters are given by
\begin{eqnarray} \label{eq:correction by misalignement pz orbitals}
\tilde{t}_{nm}(\boldsymbol{r})&= [1+\kappa (1- \boldsymbol{\hat{N}}_{n} \cdot \boldsymbol{\hat{N}}_{m})]t_{nm}(\boldsymbol{r})
\end{eqnarray}
where $\kappa \approx 0.4$, $\boldsymbol{\hat{N}}_{i}$ is the unit normal vector to the curved graphene, given by
\begin{equation} \label{eq:normal vector at the i-th site}
    \boldsymbol{\hat{N}}_{i}= \frac{\hat{e}_{z}- \nabla z_{i}}{\sqrt{1+|\nabla z_{i}|^{2}}},
\end{equation}
with $z_{i}$ representing the height of the $i$-th site, $\nabla=(\partial_{x}, \partial_{y})$ is the two-dimensional gradient operator, and $\hat{e}_{z}$ is the unit vector perpendicular to the flat graphene. Here, the term $\kappa (1- \boldsymbol{\hat{N}}_{i} \cdot \boldsymbol{\hat{N}}_{j})$ accounts for the change in relative orientation between the $\pi$ orbitals  \cite{RomanTaboada_2017JPC}.  The numerically obtained maximum value of this term in the strongly curved systems considered in this work is approximately,
\begin{equation} \label{eq: numerical value of correction}
    \kappa (1- \boldsymbol{\hat{N}}_{i} \cdot \boldsymbol{\hat{N}}_{j}) \approx 3.9 \times 10^{-3} \ll 1.
\end{equation}
In fact,  Eq. \eqref{eq:II.6} can be used for systems with curvature $R \leq 1 \,\, \text{nm}^{-2}$ and involving more atoms to optimize numerical calculations. From these previous considerations, we will neglect $\pi$ orbital misalignment effects in what follows.}

\section{Spectra in flat and curved graphene and density of states (DOS)}\label{sec3}

{\color{black} This section is divided into two subsections. In the first subsection, we calculate the LLs for flat graphene under a real external magnetic field, while in the second one, we obtain the LLs for curved graphene under such an external magnetic field.}

\subsection{Landau Levels in flat graphene under a magnetic field revisited.}
We first consider the Dirac equation for a 2+1 Minkowskian spacetime in a transverse induction magnetic field, $\boldsymbol{{B}}=(0,0, B)$, and $\boldsymbol{{A}}=(0, Bx,0)$ its corresponding vector potential in the Landau gauge. The one-particle Hamiltonian is then given by \cite{Pratama2021, Zhang_2010, Sharapov2004, Stegmann_2019}

\begin{equation} \label{eq:III.1}
\mathcal{H}=  v_{F} \hat{\boldsymbol{\sigma}} \boldsymbol{\cdot} \hat{\boldsymbol{\pi}},
\end{equation}
in which $\hat{\boldsymbol{\sigma}}=(\xi \sigma_{1}, \sigma_{2})$, where $\sigma_{j}$ is the $j$-th Pauli matrix, $\xi= +1 (-1)$ for the K (K') valley,  and $\hat{\boldsymbol{\pi}}$ is the canonical momentum with the Peierls substitution ($\hat{\boldsymbol{\pi}}=\hat{p}+e \boldsymbol{A}$). 
From the Schrödinger-Dirac time-independent  equation $\mathcal{H} \boldsymbol{\Psi}= \varepsilon \boldsymbol{\Psi}$, we obtain the known energy spectra \cite{Zhang_2010,McClure1956}

\begin{figure}
    \centering
     \includegraphics[scale=0.43]{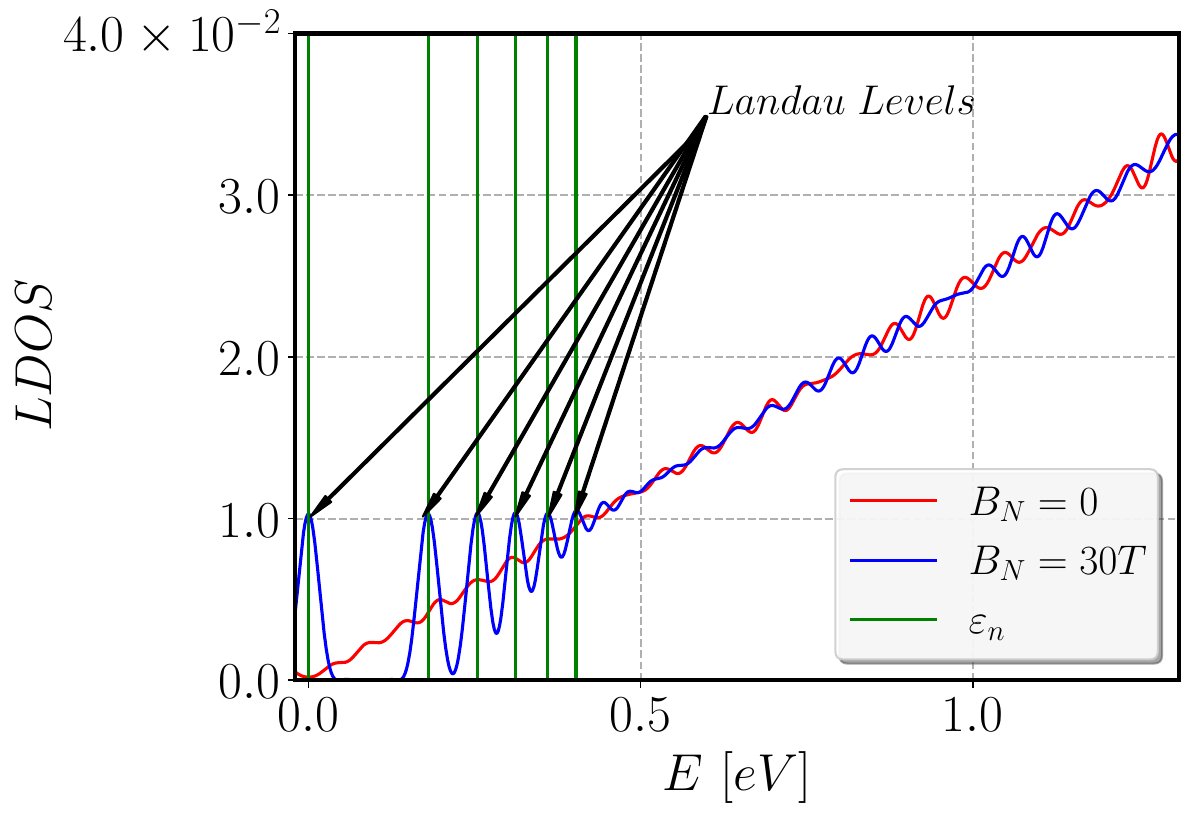}
    \caption{
    Numerical calculation of the Local Density of States (LDOS) obtained from the tight-binding Hamiltonian Eq. (\ref{eq:II.5}) for a flat graphene sheet with dimensions $L_{x}=L_{y}=70$ nm without a magnetic field (red lines) and with a magnetic field $B_{z}=30$ T (blue lines) perpendicular to the $xy$ plane. The field strength has been exaggerated for pedagogical reasons and to visualize the LLs (green lines) obtained from the Dirac model Eq. \eqref{eq:III.2}. It should be noted that for $E>0.5$ eV, the cases for flat graphene with and without a magnetic field are very similar. This is due to the limitation of the Dirac approximation, where $2 \hbar \omega_{c} \sim 0.4 $ eV. In other words, as the magnetic fields become stronger, the continuum model becomes more effective at capturing information at higher energies compared to the flat graphene without magnetic fields. Furthermore, the oscillations observed in the case without a magnetic field are due to the finite size effects of the sample. {\color{black} As the LDOS is symmetric with respect to the zero energy, the plot is exclusively presented for $E>0$.}}
    \label{fig:Fig1}
\end{figure}

\begin{equation} \label{eq:III.2}
\varepsilon_{n}= \text{sign}(n) v_{F} \sqrt{2|n|\hbar e B}, \,\, n \in \mathbb{Z},
\end{equation}
 and which does not depend on the valley index, so there is a valley degeneration. These energies (\ref{eq:III.2}) are related to the emergence of LLs. In addition, in Fig. \ref{fig:Fig1} it is shown the Local Density of States (LDOS) for a finite sample of graphene with area $10^2~{\rm nm}^2$, which was numerically calculated via diagonalization of the Hamiltonian Eq. (\ref{eq:II.5}). In particular, in this figure, we compare the smallest energy eigenvalues from Eq. \eqref{eq:III.2} with the results obtained from the numerical results of LDOS. 
 

\subsection{Landau Levels in curved graphene under a real magnetic field.}

This section is devoted to adapting the methods used in the previous section to find the spectrum of curved graphene under a magnetic field. Here, we shall use a local frame defined by the so-called Riemann normal coordinates (RNCs) $y=x-x'$, where $x'$ is a fiducial point that can be chosen as the origin \cite{Eisenhart1997}. We carry out the transformation \textcolor{black}{$\tilde{\psi}_{\sigma, \xi}=g^{\frac{1}{4}}\psi_{\sigma, \xi}$} in the Hamiltonian (\ref{eq:II.4}) to capture the geometrical data coming from the area element, thus,  (\ref{eq:II.4}) can be cast in the form
\textcolor{black}{\begin{equation}  \label{eq:IV.1}
\hat{H}= \sum_{\xi=\pm} \sum_{\sigma=\uparrow, \downarrow}\int d^{2} y\,\,  \tilde{\psi}^{\dag}_{\sigma,\xi}\tilde{\mathcal{H}}_{\xi}\tilde{\psi}_{\sigma, \xi},
\end{equation} }
where \textcolor{black}{$\tilde{\mathcal{H}}_{\xi}=g^{\frac{1}{4}}\mathcal{H}_{\xi}g^{-\frac{1}{4}}$}.  

Now, let be \textcolor{black}{$\tilde{\mathcal{H}}_{F}= \tilde{\mathcal{H}}_{\xi}/ \hbar v_{F}$}. Further simplification can be achieved by taking the square of $\tilde{\mathcal{H}}_{F}$, that is, $\tilde{\mathcal{H}}_{F}^{2}=g^{\frac{1}{4}}\mathcal{H}_{F}^{2}g^{-\frac{1}{4}}$, where now using the Clifford algebra implies  \textcolor{black}{$\mathcal{H}_{F}^{2}=- \gamma_{0} \slashed{D}_{\xi} \gamma_{0} \slashed{D}_{\xi}=- \slashed{D}_{\xi}^{2}$}, and using the Schrödinger-Lichnerowicz formula for $\slashed{D}^{2}_{\xi}$(see Appendix \ref{appendix B}) we obtain
\textcolor{black}{\begin{eqnarray}\label{eq:IV.2}
\tilde{\mathcal{H}}_{F}^{2}&=& - g^{\frac{1}{4}}\boldsymbol{\nabla}^{\xi}_{i} g^{ij} \boldsymbol{\nabla}^{\xi}_{j}g^{-\frac{1}{4}} + \frac{1}{4}R-\frac{\xi q}{2 \hbar }\sigma_{3}\epsilon_{ij}F^{ij},
\end{eqnarray}}
where $R$ is the Ricci scalar curvature, $\epsilon_{ij}$ is the 2nd order Levi-Civita tensor and $F_{ab}=\nabla_{a}A_{b}-\nabla_{b}A_{a}$ is the covariant magnetic strength tensor. This tensor is related to the external magnetic induction field ${\bf B}$ applied to the curved graphene sheet. The expression of $\tilde{\mathcal{H}}^{2}_{F}$ has the same structure obtained in the context of quantum field theory in curved space \cite{Parker2009}. The first term of (\ref{eq:IV.2}) can be simplified in such a manner that \cite{Pavel2017} 
\textcolor{black}{\begin{eqnarray}\label{eq:IV.3}
\tilde{\mathcal{H}}_{F}^{2}&=&-\boldsymbol{\nabla}^{\xi}_{i}g^{ij}\boldsymbol{\nabla}^{\xi}_{j}- g^{-\frac{1}{4}}\partial_{i}\left(g^{\frac{1}{2}}g^{ij}\partial_{j}(g^{-\frac{1}{4}})\right)+ \frac{1}{4}R\nonumber\\
&-&\frac{q\xi}{2 \hbar }\sigma_{3}\epsilon_{ij}F^{ij}.
\end{eqnarray}}
This operator is the starting point to analyze the weak and strong curvature approximations, depicted in Fig. \ref{fig:Fig4}. We will consider a slight curvature perturbation from the flat Hamiltonian in the weak approximation. In contrast, for the strong curvature limit, the pseudo-magnetic field associated with the curvature will be comparable to the magnetic field's value.

Next, for the aforementioned approximations, we use the fact that the metric tensor and spin connection (using RNCs) can be written as a Taylor series with coefficients given in terms of the covariant derivative of the Riemann tensor \cite{Parker2009, Muller1999}. The first terms of these series expansions are
\textcolor{black}{\begin{equation} \label{eq:IV.4}
\begin{split}
g^{ij}=&\delta^{ij}-\frac{1}{3}R\indices{^{i}_{kl}^{j}}y^{k}y^{l}+ \ldots, \\
\Omega^{\xi}_{j}=& \frac{\xi}{4}y^{k}R\indices{_{kj}^{ab} }s_{ab}+ \ldots,
\end{split}
\end{equation}}
where the dots indicate higher order terms $\mathcal{O}(y^{n})$ with $n \geq 3$, and the coefficients $R_{ijkl}$ are the components of the Riemann curvature tensor evaluated at the fiducial point $x^{\prime}$. \textcolor{black}{Here $s_{ab}$ is the pseudo-spin operator in $K$ valley.}

Additionally, we approximate the magnetic potential $A_{j}$. According to the previous section, the gauge field can be written as $A_{j}={\bf e}_{j}\cdot {\bf A}$. Now, for a uniform magnetic induction field ${\bf B}$ let us choose the symmetric gauge thus the $U(1)$ vector potential ${\bf A}$ is given by ${\bf A}=\frac{1}{2}{\bf X}\times {\bf B}$, where recall that ${\bf X}$ are the embedding functions of the curved surface $\Sigma$.
Although the magnetic field ${\bf B}$ is uniform, the magnetic potential $A_{a}$ may have a nonlinear dependence on the local coordinates of the surface. Expressing the embedding functions ${\bf X}(y)$ in terms of the RNC is not difficult to show that ${\bf X}(y)\approx {\bf e}_{b}y^{b}+O\left(y^2\right)$, thus $A_{a}=\frac{1}{2}\left({\bf e}_{a}\times {\bf e}_{b}\right)\cdot {\bf B}$. After using identity ${\bf e}_{a}\times {\bf e}_{b}=\sqrt{g}\epsilon_{ab}{\bf N}$ and $\sqrt{g}=1$ at the fiducial point, we can find the first approximation of the curved magnetic potential $A_{a}=\frac{1}{2}B_{N}\epsilon_{ab}y^{b}+O(y^{2})$, where it can be shown that quadratic terms involve tangent components of ${\bf B}$ and the extrinsic curvature tensor. The extrinsic curvature corrections are out of the present analysis's scope and will be analyzed elsewhere.   
In the expression for $A_{a}$, $B_{N}={\bf B}\cdot{\bf N}$ is the external magnetic field ${\bf B}$ along the normal direction  ${\bf N}$ to the tangent plane at the fiducial point $x^{\prime}$ belonging to the surface patch.

\begin{figure*}[t]
\includegraphics[scale=0.4]{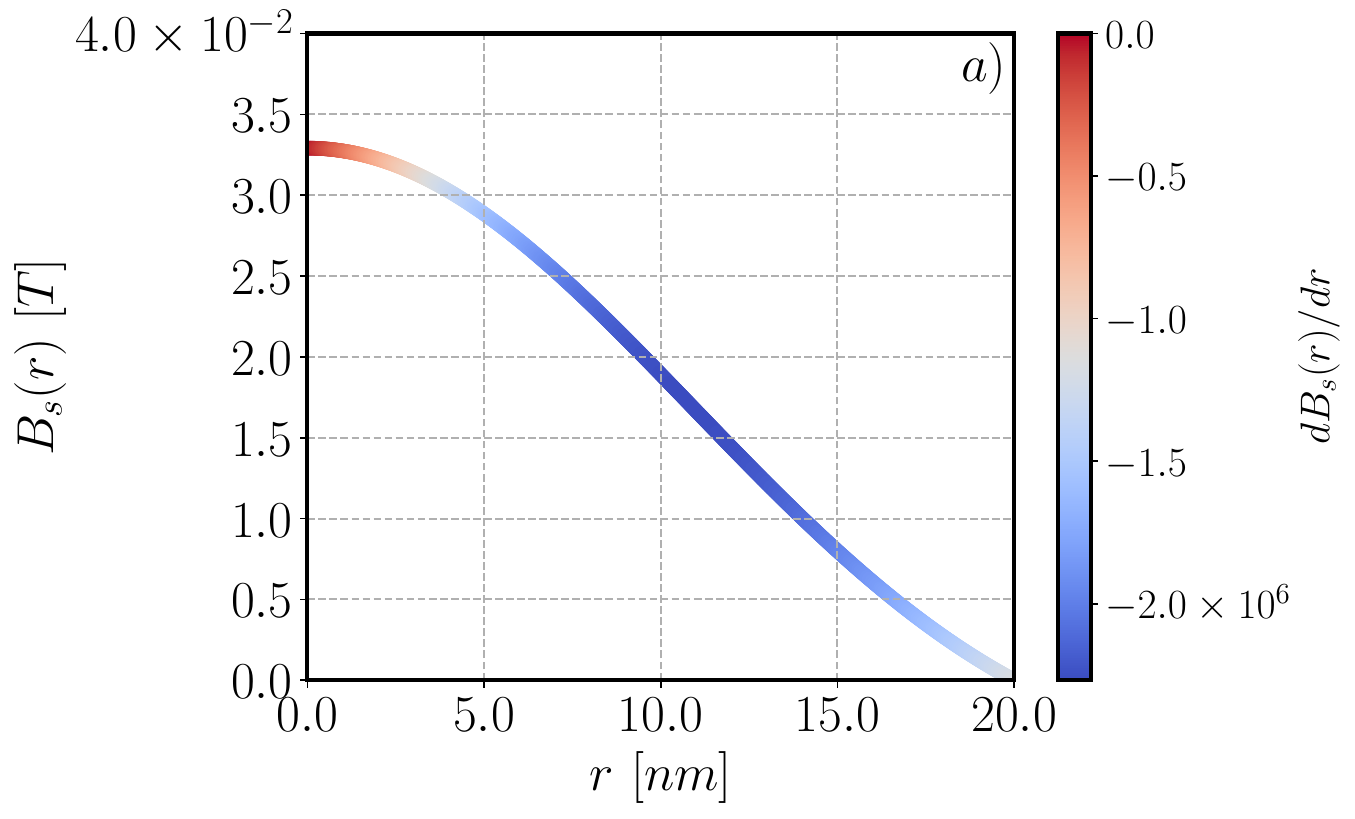}
\includegraphics[scale=0.4]{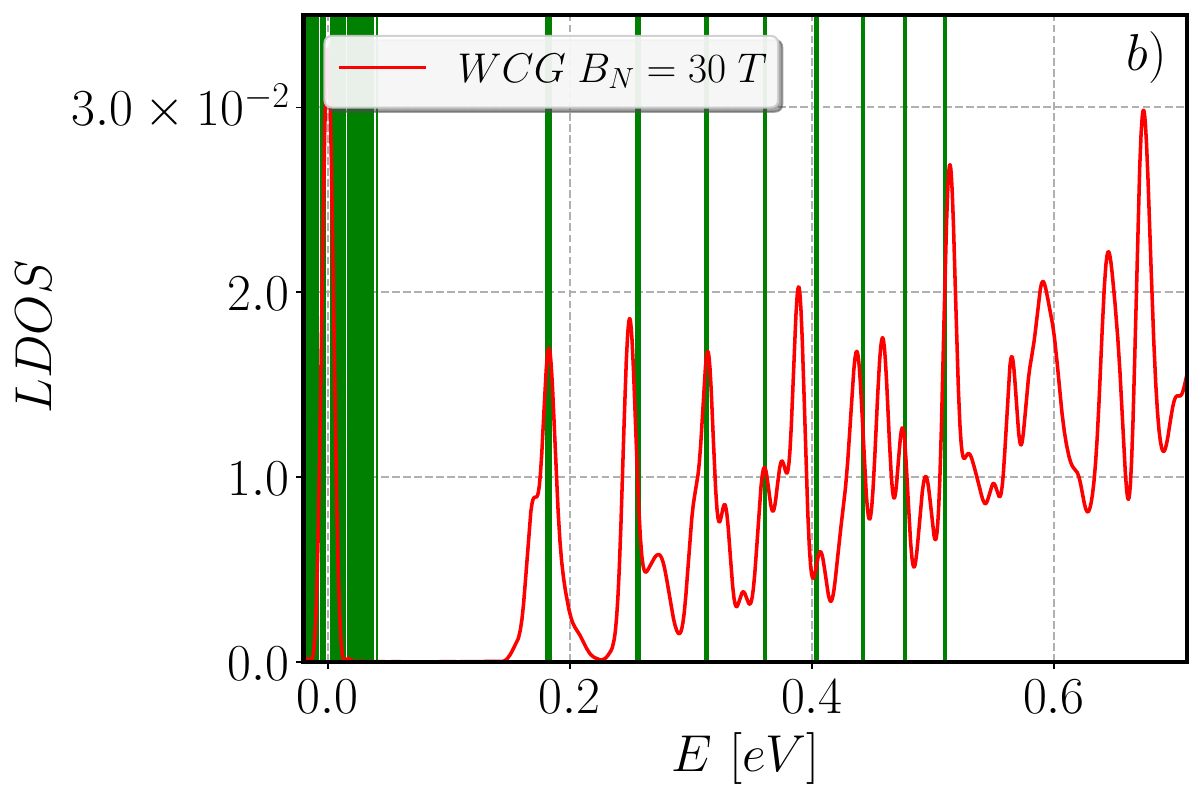}\\
\includegraphics[scale=0.4]{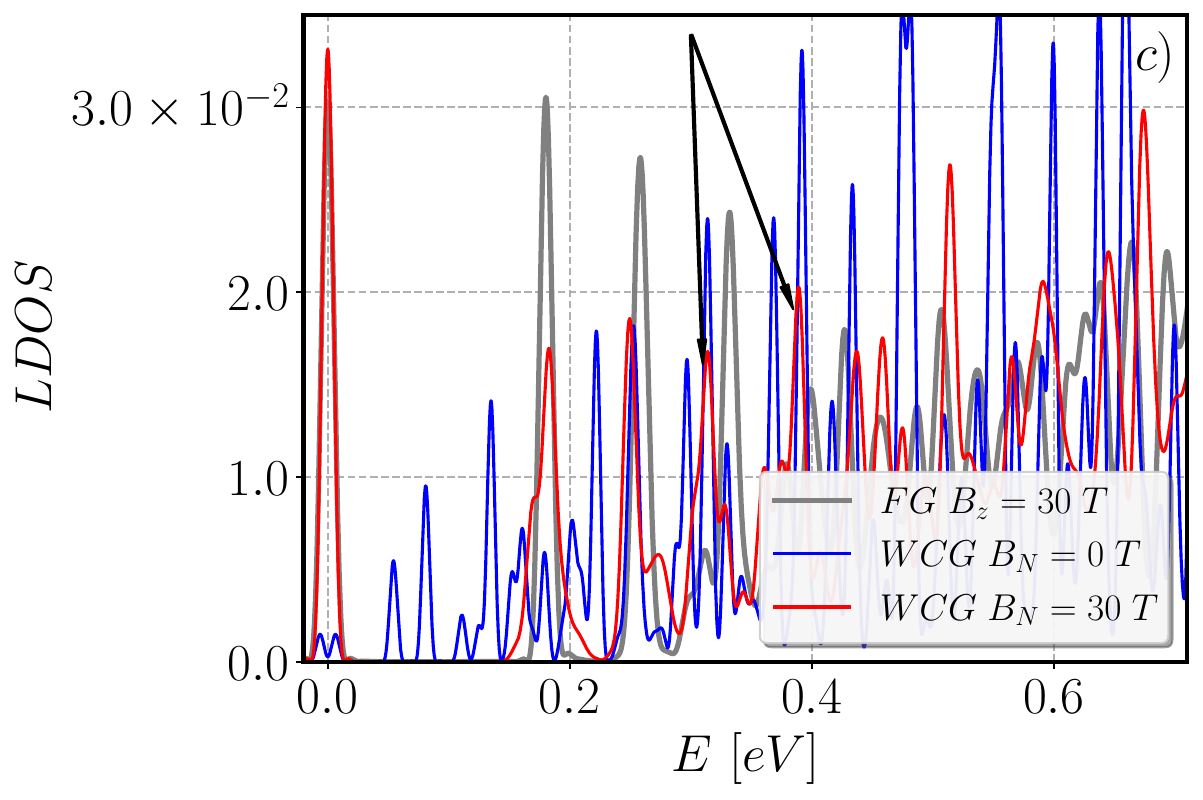}~~~~~~~~~~
\includegraphics[scale=0.4]{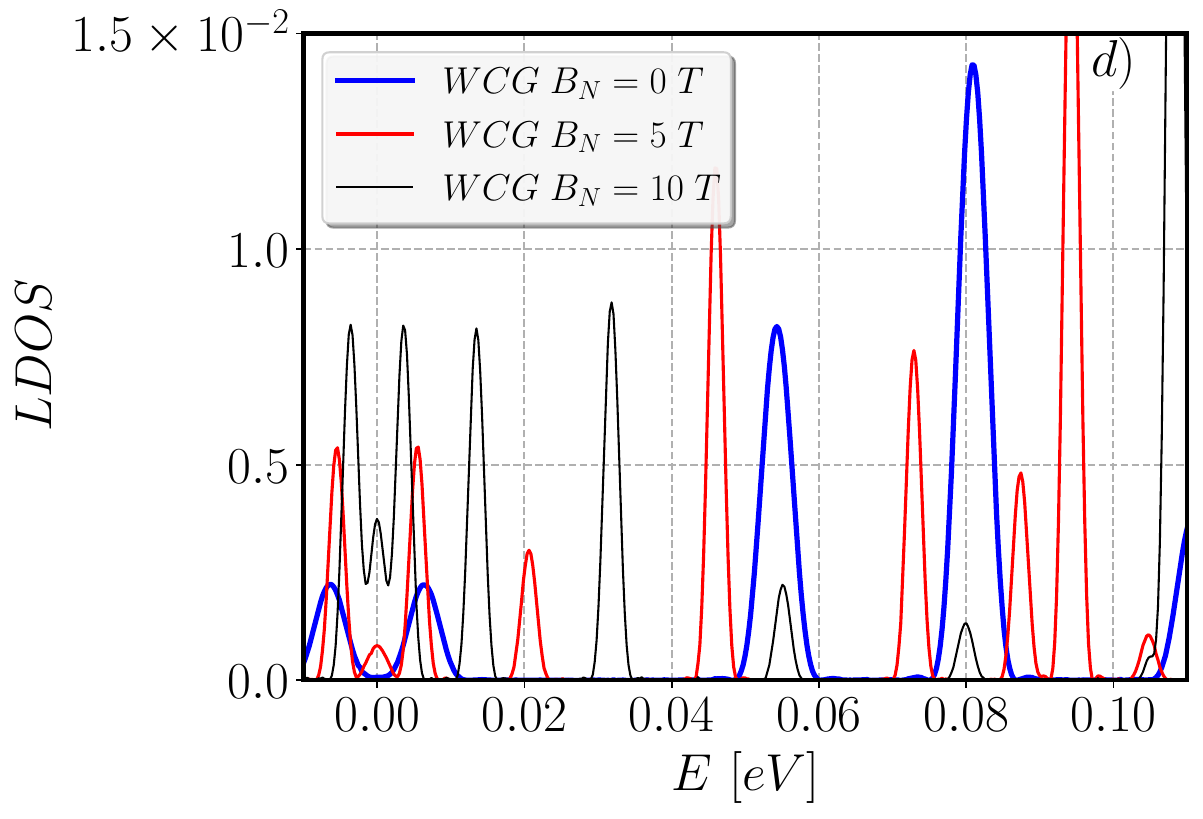}
    \caption{\textcolor{black}{Collection of plots for the} weak pseudomagnetic field case for a graphene sheet with dimensions $L_{x}=L_{y}=40$ nm; numerical comparison between the results obtained from the TB model and the Dirac model. 
    a) Pseudomagnetic field profile obtained from a Gaussian deformation \textcolor{black}{with parameters $z_{0}=4$ nm and $\varpi=20$ nm}. The most significant energy changes concerning flat graphene are produced by states close to the origin where $B_{s} \approx 0.033 $ T. The color code of the curve indicates the field derivative, showing two separate regions as explained in the text. b) LDOS obtained from the LLs $\left|n\right|=0,\ldots,7$ and $m_{max}=11$ using the Dirac equation model in curved space with $B_{s,max} \approx 0.033$ T (see Eq. \eqref{eq:IV.11}), compared to the LDOS obtained from the TB model. Observe how the continuous model predicts almost all of the prominent peaks for low energies. c) {\color{black} LDOS obtained from the TB model for three representative cases, flat graphene under a very strong magnetic field ($B_N=30 T$) (gray lines), weak curvature regime under the same external magnetic field (red lines) and without the external magnetic field (blue lines). }The contribution associated with the curvature is visible in the peaks indicated by arrows, which are shifted with respect to the flat graphene peaks. {\color{black} d) Zoom near $E=0$, showing the evolution of the LDOS as the inversion symmetry is broken by a gradual increase of the external magnetic field. Notice how  Landau levels with $n=0$ arise as the temporal inversion is broken. In all cases, the plots are exclusively presented for $E>0$ as the LDOS is symmetric with respect to the zero energy.} }
    \label{fig:Fig5}
\end{figure*}

\subsubsection{The weak curvature regime}

Our starting point is to consider that, for a surface with metric $g_{ij}$, the Riemann curvature tensor can be written as $R_{mklj}=\frac{R}{2}\left(g_{ml}g_{kj}-g_{mj}g_{kl}\right)$, thus the Riemann curvature tensor and the spin connection can be written at the fiducial point $x^{\prime}$ as 
\begin{equation} \label{eq:IV.7}
\begin{split}
    R_{mklj}&= \frac{R}{2} \left( \delta_{ml} \delta_{kj}- \delta_{mj} \delta_{kl} \right)= \frac{R}{2} \epsilon_{mk} \epsilon_{lj}\\
    \Omega_{j}& = i {\color{black} \xi} \frac{R}{8} y^{l} \epsilon_{lj} \sigma_{3}, \text{ here } i= \sqrt{-1}
\end{split}
\end{equation}
and we can rewrite the $\boldsymbol{\nabla}$ operator as 
\begin{equation} \label{eq:IV.8}
\boldsymbol{\nabla}_{l}= i \left( \frac{1}{\hbar}\boldsymbol{\pi}_{l} -i \Omega_{l} \right),   
\end{equation}
where $\boldsymbol{\pi}_{l}$ is the canonical momentum with the Pierls substitution using $B_{N}$ instead of $B$. Thus, from Eqs. \eqref{eq:IV.3}, \eqref{eq:IV.4}, \eqref{eq:IV.7},  and \eqref{eq:IV.8},  we obtain that
\begin{equation} \label{eq:IV.9}
    \begin{split}
        \tilde{\mathcal{H}}_{F}^{2}&= \frac{1}{\hbar^{2}}\boldsymbol{\pi}_{l} g^{lj} \boldsymbol{\pi}_{j} - g^{lj} \Omega_{l} \Omega_{j} -\frac{i}{\hbar} \left( \boldsymbol{\pi}_{l} g^{lj} \Omega_{j}+ \Omega_{l} g^{lj} \boldsymbol{\pi}_{j} \right) \\
& + \textcolor{black}{\frac{1}{12} R} - \frac{{\color{black}\xi}q}{2 \hbar} \sigma_{3} \epsilon_{ij} F^{ij}.
    \end{split}
\end{equation}

Considering only the first-order expansion in $R$, we obtain a ``{}Hamiltonian"{} $\tilde{\mathcal{H}}_{{\color{black}\xi}}^{2}=\hbar ^{2} v_{F}^{2} \tilde{\mathcal{H}}_{F}^{2}=\hat{H}_{0}+\hat{H}_{I}$ such that $\hat{H}_{0}=v_{F}^{2} \boldsymbol{\pi}^{2}$ is the square flat graphene Hamiltonian, which corresponds to the square of (\ref{eq:III.1}) and the \textcolor{black}{perturbative} term
{\small\begin{eqnarray} \label{eq:IV.10}
\hat{H}_{I} \approx \left(v_{F}^{2} \frac{R}{6}L^{2}+v_{F}^{2} \hbar^{2} \frac{R}{12}\right)\mathbb{1}+ {{\color{black}\xi}}\left(v_{F}^2 \hbar \frac{R}{4}L+e \hbar v_{F}^{2} B_{N}\right)\sigma_{3},\nonumber\\
\end{eqnarray}}
where $L=\epsilon_{ij}y^{i}{p}^{j}$ is an angular momentum operator like in two dimensions. It can be seen that the first term within the second parenthesis is a pseudo-Rashba effect, and the second term is similar to a type of Zeeman effect. \textcolor{black}{In the weak approximation, we also have neglected terms such as $RB_{N}y^{i}y_{i}$, and $RB_{N}^{2} \left(y^{i}y_{i}\right)^{2}$ coming from the first and third term of Eq. (\ref{eq:IV.9}). }

{\color{black}Note that for the valleys $K(K')$, the valley index must be $\xi=1$ ($\xi=-1$) in Eq. \eqref{eq:IV.10}, which turns on a change in the sign of the last two terms due to a change in the third term of Eq. \eqref{eq:IV.9}. Taken into account this observation}, the eigenvalues of $\tilde{H}_{K(K')}$ are given by (see Appendix \ref{Ap.Weak})
\begin{equation} \label{eq:IV.11}
    E_{n,m,\tau, \xi, \pm }= \pm  \hbar \omega_{c}\sqrt{  n+\frac{1}{2}+{\color{black}\overline{\eta}}  \frac{1+\lambda m}{2} + \frac{\lambda}{3} \left(m^{2}+ \frac{1}{2} \right)},
\end{equation}
 where we defined the cyclotron frequency, $\omega_{c}$, using the equation
\begin{equation*}
    \hbar \omega_{c } \equiv \sqrt{2 e \hbar v_{F}^{2} B_{N}},
\end{equation*}
the pseudomagnetic field,
\begin{equation}\label{eq:Bsdef}
    B_{s} \equiv \frac{\hbar |R|}{4e},
\end{equation}
and the ratio of the pseudo and real magnetic fields,
\begin{equation*}
    \lambda \equiv \text{sign}(R) \frac{B_{s}}{B_{N}},
\end{equation*}
where $n =0, 1, 2, \ldots$ is the radial quantum number.
\textcolor{black}{  The pseudo-spin coupling index $\overline{\eta}$  is defined by}

{\color{black}

\begin{equation} \label{eq: pseudo-spin coupling index}
    \overline{\eta} \equiv  \xi \tau = \pm 1
\end{equation}
that contains spatial and reciprocal information through the valley index $\xi$ and the pseudo-spin index $\tau= \pm 1$, which labels the eigenvalues of $\sigma_{3}$.} 

The quantum number $m$ gives the splitting of each LL according to $m=-m_{max}, \ldots, m_{max}$.  As follows from the work by Ruiz et. al. \cite{Pavel2017}, $m_{max}= \lfloor e (B_{N}+B_{s})S/2 \pi \hbar \rfloor $. If $l$ is the angular momentum quantum number, then $m=l-n$.

In Fig. \ref{fig:Fig5} a), we show the pseudomagnetic field profile for a Gaussian bump defined by the height function (\ref{ap.a.11}) with parameters $z_{0}=4~{\rm nm}$, $\varpi=20~{\rm nm}$.  We observe that the states near the origin make the most significant contributions to the energy change compared to flat graphene. Therefore, in Fig.  \ref{fig:Fig5} b), we compare the LLs obtained from the TB model (\ref{eq:II.5}) with those from the Curved Dirac model (cf. Eq. \eqref{eq:IV.11}). For this case, we have considered the first $|n|=0, \ldots,6$ levels, and the degeneracy number is $m_{max}=2$. This degeneration is visible for energy $E \approx 0$. It is important to note that the contribution associated with the curvature is evident in the peaks indicated by the arrows in Figure \ref{fig:Fig5} c). These peaks are observed to be shifted compared to the peaks of flat graphene. 
\begin{figure*}
    \centering
\includegraphics[scale=0.43]{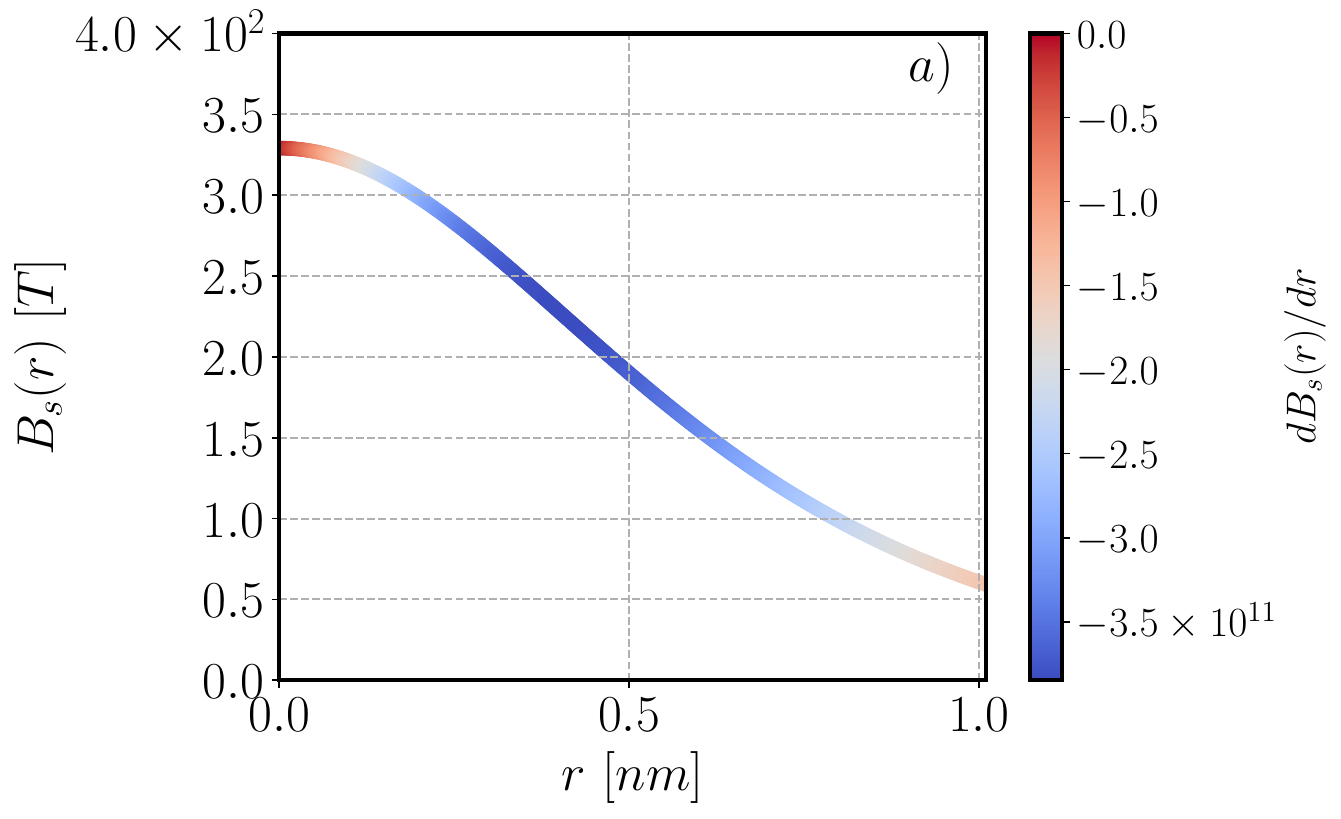}
\includegraphics[scale=0.43]{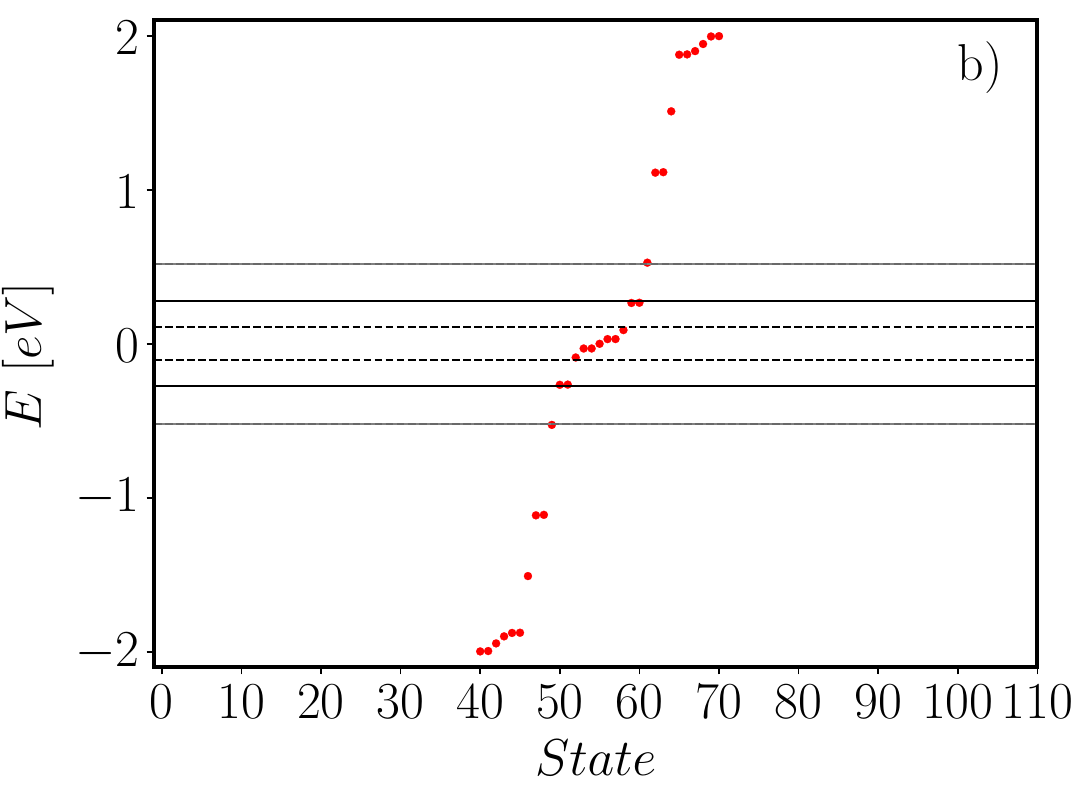}\\
~~~~~~~~~~\includegraphics[scale=0.45]{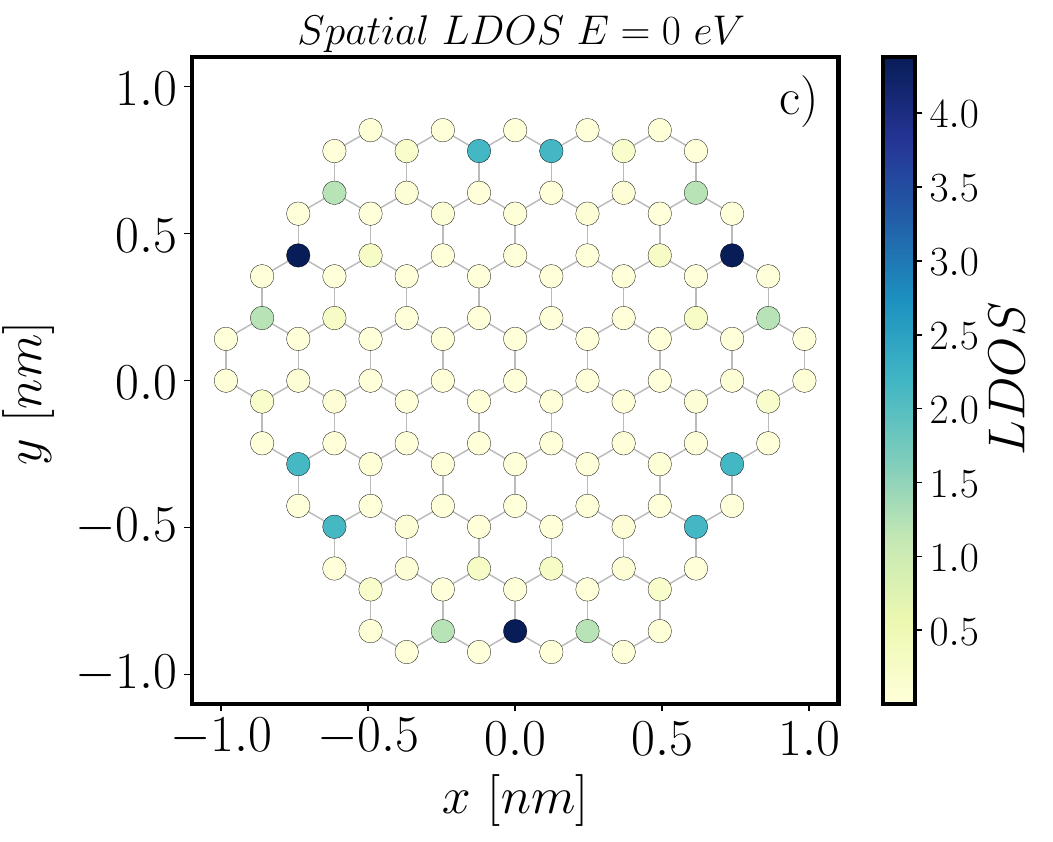}~~~~~~~~~~~~~~~~~~~~~
\includegraphics[scale=0.45]{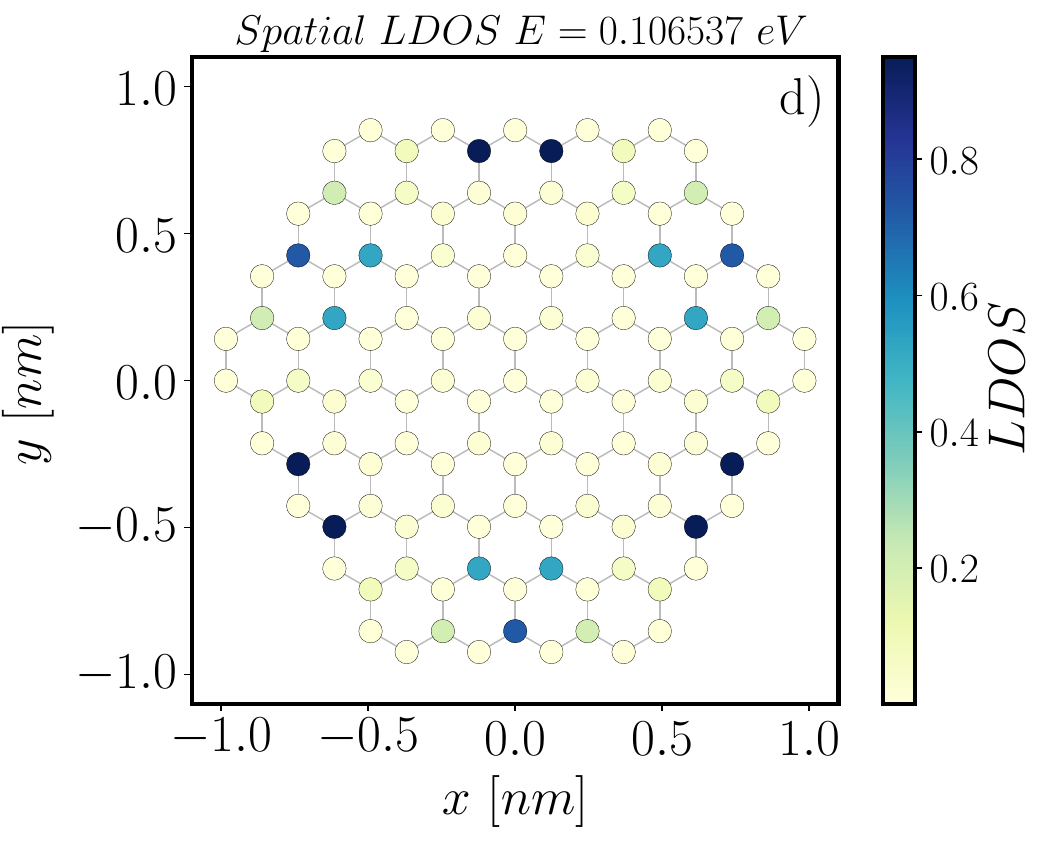}\\
~~~~~~~~~~\includegraphics[scale=0.45]{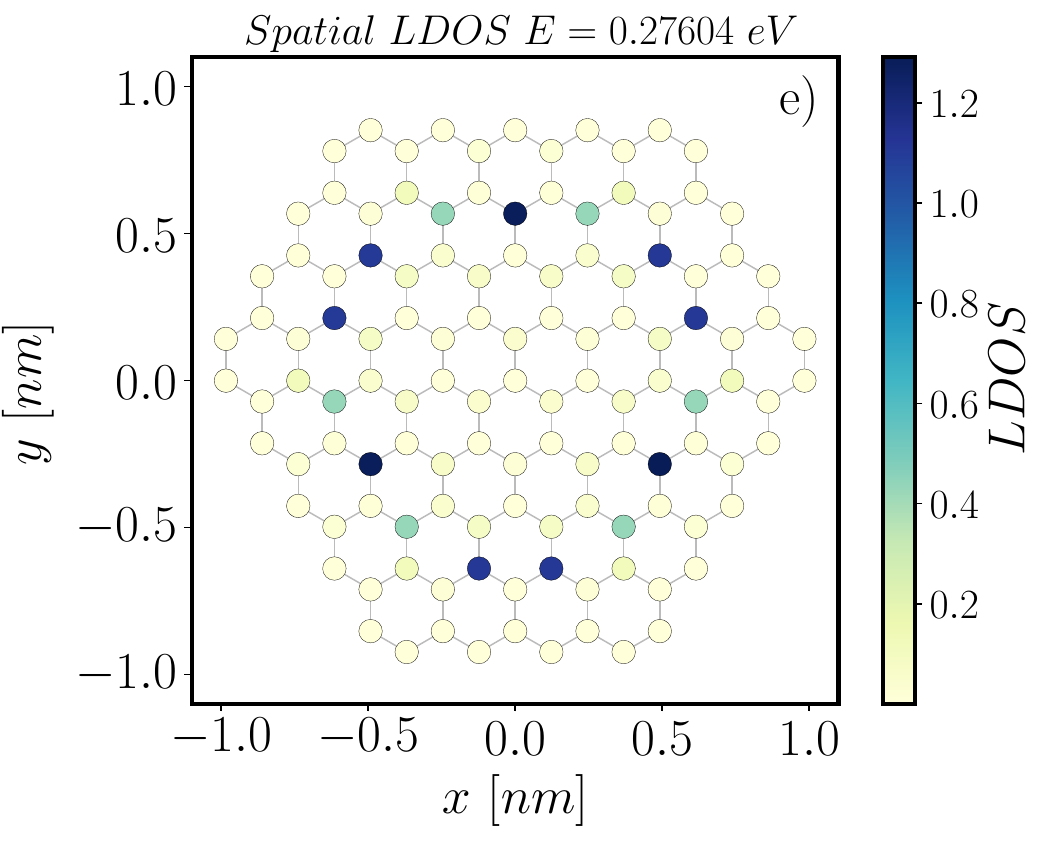}~~~~~~~~~~~~~~~~~~~~~
\includegraphics[scale=0.45]{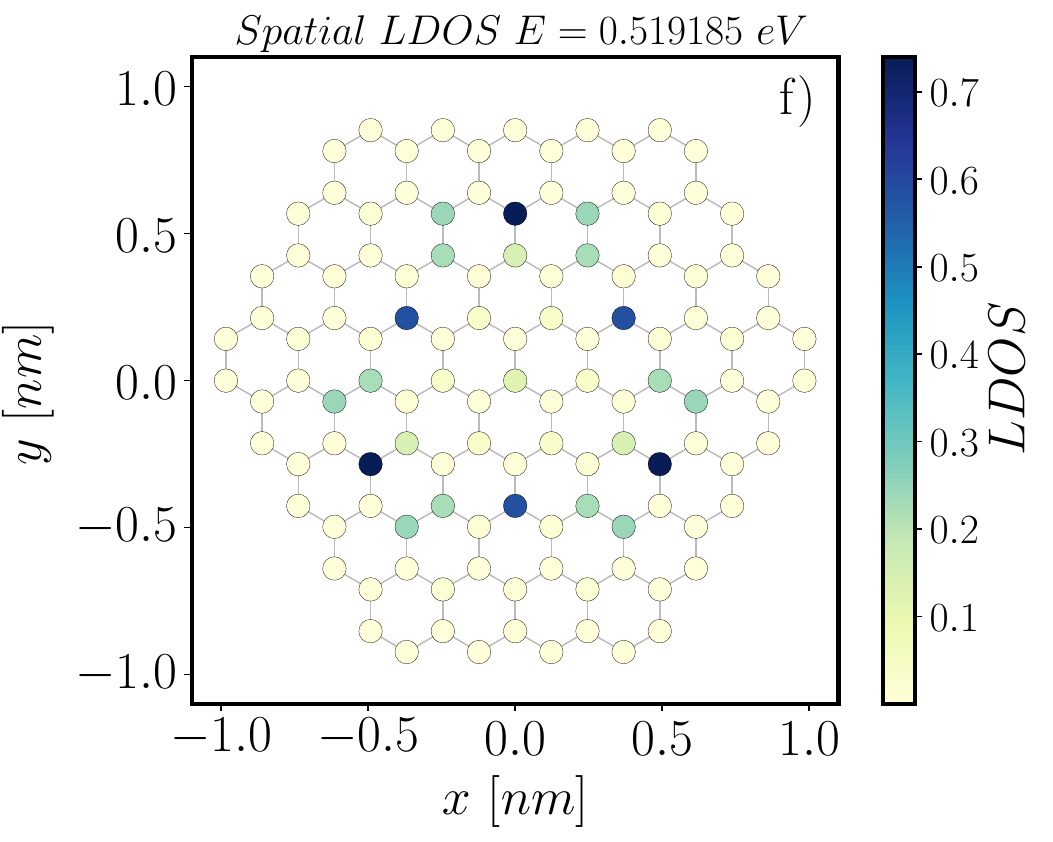}\\
    \caption{\textcolor{black}{Collection of plots for the} strong pseudomagnetic field case for a graphene sheet with dimensions $L_{x}=L_{y}=2$ nm; numerical comparison between the results obtained from the TB model and the Dirac model. a) The pseudomagnetic field profile $B_{s}$ is shown as a function of radial distance $r$. b) The red dots are the energy of the states for the graphene nanodisk deformed by a Gaussian bump with parameters $z_0 = 4~{\rm nm}$ and $\varpi = 2~{\rm nm}$ obtained using the TB model under magnetic field $B_{N}=30$ T. Additionally, the horizontal lines represent the eigenvalues obtained using the Dirac model {Eq. (\ref{eq:IV.16})}. The dashed lines correspond to $\tau=-1$ while the solid lines represent $\tau=1$. c), d), e), and f) are panels of the graphene nanodisk corresponding to energy values of the LDOS $E=0 ~{\rm eV}$, $E=0.106537~{\rm eV}$, $E=0.27604~{\rm eV}$ and $E=0.51985~{\rm eV}$, respectively. 
    } 
    \label{fig:Fig6}
\end{figure*}

\subsubsection{The strong curvature regime}
In the strong curvature corrugation approach, a quadratic shape of the local geometry is still maintained so that the Hamiltonian operator \eqref{eq:IV.3}, with an external magnetic field, is reduced to 
\begin{equation} \label{eq:IV.12}
\begin{split}
\tilde{\mathcal{H}}_{F}^{2}= \delta^{ij} \left( \frac{{\pi}_{i}}{\hbar}+{{\color{black} \xi}}\frac{R}{8}\epsilon_{il}y^{l}\sigma_{3} \right) \left( \frac{\pi_{j}}{\hbar}+{{\color{black} \xi}}\frac{R}{8}\epsilon_{jk}y^{k}\sigma_{3} \right) \\
- \frac{R}{6 \hbar^{2}} \epsilon_{ik}\epsilon_{lj} {p}^{i} y^{k}y^{l} {p}^{j}+ \frac{R}{12}+ {{\color{black} \xi}} \frac{e}{\hbar}B_{N}\sigma_{3},
\end{split}
\end{equation}
where the second term  can be simplified after using the commutation relation $[y^{l},{p}^{j}]= i \hbar \delta^{ij}$, that is, 
\begin{equation} \label{eq:IV.13}
    \epsilon_{ik}\epsilon_{lj} {p}^{i} y^{k}y^{l} {p}^{j}= - {L}^{2},
\end{equation}
with ${L}= \epsilon_{ij} y^{i} {p}^{j}$ a two-dimensional angular momentum like operator. Clearly, $\tilde{\mathcal{H}}_{\xi}^{2}$ can be written as 
\begin{equation}\label{eq:IV.14}
    \tilde{\mathcal{H}}_{\xi}^{2}= v_{F}^{2} \left( \begin{array}{cc}
        \hat{h}^{2}_{\tau_{R}} &0  \\
         0& \hat{h}^{2}_{-\tau_{R}} 
    \end{array}\right),
\end{equation}
with
\begin{equation} \label{eq:IV.15}
    \hat{h}^{2}_{\tau_{R}}= \delta^{ij} \hat{\Pi}_{i}^{(\tau_{R})} \hat{\Pi}_{j}^{(\tau_{R})}+ \frac{R}{6} \left( \hat{L}^{2}+ \frac{\hbar^{2}}{2} \right)+ \tau_{R} \rm{sign}(R) e \hbar B_{N},
\end{equation}
where 
\begin{equation}
\label{eq: Pi tau}
\hat{\Pi}_{i}^{(\tau_{R})}=\hat{p}_{i}+ \frac{e}{2} B_{T}^{(\tau_{R})}\epsilon_{li} y^{l}.
\end{equation}
The total effective magnetic field is given by the sum of the external magnetic field $B_{N}$ and its pseudomagnetic counterpart $B_{s}$,
\begin{equation}
\label{eq: BT tau}
B_{T}^{(\tau_R)} \equiv B_{N}+ \tau_{R} B_{s},
\end{equation}
where we defined the factor $\tau_{R} \equiv  {\color{black} \overline{\eta}}\,\,\text{sign}(R)= \pm 1$. {\color{black}Here recall $\overline{\eta}$ is the pseudo-spin coupling index (Eq. \eqref{eq: pseudo-spin coupling index}), which is the product of the pseudo-spin index, $\tau$, and the valley index, $\xi$}. {\color{black} Notice that such effective fields and the valley-dependent Landau levels have been recently measured in graphene with a nanoscale ripple under an external magnetic field \cite{ExperimentalValley}.}
 
 Therefore, the eigenvalues of $\tilde{\mathcal{H}}_{K(K')}$ are given by (see Appendix \ref{Ap.Strong}), 
\begin{equation} \label{eq:IV.16}
\begin{split}
    E_{n,m,\tau, \xi, \pm }&= \pm  \hbar \omega_{c,\tau_R}\\
    & \times \left[  n_{\tau_{R}}+ \frac{1}{2}+ \frac{\lambda_{\tau_{R}}}{3} \left(m_{\tau_{R}}^{2}+ \frac{1}{2} \right)+ \overline{\eta} \Theta_{\tau_{R}} \right]^{1/2},
\end{split}
\end{equation}
where now we have two possible cyclotron frequencies depending on the value of $\tau_{R}$,
\begin{equation} \label{eq:ciclotron frequency curvature}
    \hbar \omega_{c ,\tau_R} \equiv  \sqrt{2 e \hbar v_{F}^{2} |B_{T}^{(\tau_R)}|},\hspace{2mm} m_{\tau_{R}}=l_{\tau_{R}}-n_{\tau_{R}}.
\end{equation}
We also defined,
\begin{eqnarray}\label{eq:SCGlamba}
    \lambda_{\tau_R}&=& \text{sign}(R) \frac{B_{s}}{|B_{T}^{(\tau_R)}|}\nonumber\\
    \Theta_{\tau_{R}}&=&\frac{B_{N}}{2|B_{T}^{(\tau_R)}|}
\end{eqnarray}
and $m_{\tau_{R}}=- m_{max, \tau_{R}}, \ldots, m_{max, \tau_{R}}$.

As follows from the work of Ruiz et. al. \cite{Pavel2017},
\begin{equation}\label{defmax}
 m_{max, \tau_{R}}= \frac{e |B_{T}^{(\tau_{R})}|S}{2 \pi \hbar}.   
\end{equation}
 The pseudomagnetic field $B_s$ definition is the same as in Eq. \eqref{eq:Bsdef}.
 Note that as stated below Eq. \eqref{eq:V.4}, the occupation number is inversely proportional to $|B_{T}^{(\tau_{R})}| S$ while $m_{\rm max, \tau_{R}}$ is proportional. Thus, by changing the curvature values $R$ and/or the area $S$, we will have a different eigenvalue behavior. 

 Specifically, employing a Gaussian bump deformation results in regions with curvatures of different signs. Consequently, the eigenvalues of these distinct regions become mixed. To effectively compare our Dirac model with the TB model, it is advisable to focus on a section of the material corresponding to a domain exhibiting the highest curvature.

In Figure \ref{fig:Fig6}, we consider a Gaussian deformation of the graphene sheet with parameters $z_0 = 4~{\rm nm}$ and $\varpi = 2~{\rm nm}$. The condition for the biggest curvature is a region of radius $1 ~{\rm nm}$ around the bump center. In subfigure \ref{fig:Fig6} a), two spatial regions are established for the pseudomagnetic field $B_s$ as a function of the radial distance $r$ from the bump center. The region near the origin exhibits $B_s \approx 330$ T while the region near the edge has $B_s \approx 60$ T.

Once the biggest curvature region is identified, we proceed to compare the pseudomagnetic model with the TB calculation. In subfigure \ref{fig:Fig6} b), the red dots are the energies obtained by using the TB model (\ref{eq:II.5}) in a graphene nanodisk of radius $1 ~{\rm nm}$. States with zero energy are attributed to edge states due to the zigzag configuration of the boundary, as shown in \ref{fig:Fig6} c) and are not obtained from the Dirac model. States obtained from the TB with energies $E \approx 0.1065$ eV and $E \approx 0.2760$ eV correspond to the Dirac model LLs $n_{\tau_{R}= \mp 1}=0$ with $B_{s} = 60.75$ T (represented by horizontal black lines), respectively (see Eq. \eqref{eq:IV.16}). These states are localized near the region edges, as shown in d) and e). Finally, states with  $E \approx 0.5191$ eV correspond to the LLs $n_{\tau_{R}= \pm}=0$ with $B_{s} = 329.10$ T (gray lines, see Eq. \eqref{eq:IV.16}). These states are localized near the origin, as demonstrated in panel f).

{\color{black}  
Unlike the real magnetic field, the pseudomagnetic field has an opposite sign at each Dirac valley due to the time-reversal symmetry. Consequently, in the presence of both uniform real and pseudo fields, it is expected that the LLs will lose their valley degeneracy \cite{Bitan2011, Bitan2013}. This is supported by Eq. \eqref{eq:IV.16} where the {{pseudospin}} coupling index is given by $\overline{\eta}$, showing that for $B_{N} \neq 0$, different sequences of energies are obtained. In the case of $B_{N}=0$, Eq. \eqref{eq:IV.16} indicates the restoration of valley degeneracy as expected. Interestingly, for the case of flat graphene under a real magnetic field $B_{N} \neq 0$, the field shifts the LL sequences on each valley in a different fashion for each sublattice, i.e., zero energy solutions on different valleys have different pseudospin polarizations.
However, the shift of the squared energy is precisely given by the difference between LL squared energies, resulting only in a relabeling of the whole sequence, which turns out to be independent of the valley. Another way to understand this fact is to observe that in flat graphene under a real magnetic field, there is a Zeeman effect produced by the electron orbital motion \cite{katsnelson_2012}.  This results in the widely known and experimentally observed zero mode LLs.}

Before leaving this section, we emphasize that the existence of LLs induces changes in electronic properties and transport, leading to an electronic confinement effect, as has been observed in previous works \cite{RamonCarrillo2014, RamonCarrillo2016, Sandler2023}.


\section{The dHvA effect on flat  graphene}\label{sec4}

{\color{black} Here, we review how the dHvA effect is obtained in flat graphene at the low-temperature limit $T \to 0 $ K. In the first subsection, we establish the techniques needed, especially to compare them later with the curved situation. In the second subsection, we calculate the free energy, magnetization, and magnetic susceptibility of the flat graphene.}


\subsection{The Helmholtz free energy for a relativistic gas \label{sec4:Anew}}

To establish the dHvA effect in flat and curved situations, we compute the magnetization $M$. This is done by first obtaining the Helmholtz free energy per unit area, $\mathcal{F}$, which is given by 

\begin{equation}\label{eq:III.3}
\mathcal{F}(\mu, \rho)= \Omega(\mu,T)+\mu \rho,
\end{equation}
where $\mu$ refers to the chemical potential, $\rho$ is the electron concentration in the graphene sheet, and $\Omega(\mu, T)$ is the thermodynamic potential. We should be careful at this point because the expression for  $\Omega(\mu, T)$ needs to include considerations of the relativistic invariance of the Dirac equation. Indeed the adequate expression must read \cite{katsnelson_2012, CANGEMI1996582, Sharapov2004}
\begin{equation}\label{eq:III.4}
\Omega(\mu, T)= -k_{B} T \int_{-\infty}^{\infty} d \varepsilon \mathcal{D}(\varepsilon) \ln\left[ 2 \cosh \left(  \frac{\varepsilon-\mu}{2k_BT} \right)\right],
\end{equation}
where $\mathcal{D}(\varepsilon)$ {\color{black} is the total density of states (DOS) of graphene at finite temperature $T$, being $k_B$ the Boltzmann constant. It encompasses information related to impurity scattering, electron-electron interactions, and electron-phonon interactions. Due to these interactions, the LLs undergo broadening, and the delta function must be substituted with a Lorentzian function to account for the broadening induced by such interactions.} Then, the magnetization per unit area can be calculated by using one of the following expressions, \cite{reis2013fundamentals,beale2011statistical}
\begin{equation} \label{eq:III.5}
\begin{split}
M= - \left(\frac{\partial \mathcal{F}}{\partial B}\right)_{\rho,T},\\
M= - \left(\frac{\partial \Omega}{\partial B}\right)_{\mu,T}.
\end{split}
\end{equation}
In the first case, when $\rho$ is constant, $\mu$ oscillates as a function of $B$; while in the other case \cite{Sharapov2004}, $\partial \Omega(\mu, T)/ \partial \mu= - \rho$.
 
Here, we discuss only the dHvA effect in the fixed electron density case $\rho$. Thus,  we consider a system of $N$ electrons within a sample area $S$ moving in the magnetic induction field $\boldsymbol{B}$. Let the system remain at $T= 0~{\rm K}$, and accordingly, the full occupation of LLs obeys  \cite{Zhang_2010, Champel2001, Vagner2006}
\begin{equation} \label{eq:III.6}
    g(B) \sum_{n=0}^{n_{f}} f_{n}= g(B) (n_{f}+1)= \rho,
\end{equation}
where $n_{f}$ is the highest occupied LL,  $f_{n}=\left[1+ \exp(\beta(\varepsilon_{n}- \mu))\right]^{-1}$ is the Fermi-Dirac distribution, $g(B)=g_{s} g_{v} B /\Phi_{0}$ is the degeneracy of the LLs, $g_{s} (g_{v})=2$ is the spin (valley) degeneracy. We use this relation to obtain
\begin{equation} \label{eq:III.7}
g(B)= \frac{2eB}{\hbar \pi}.
\end{equation}

In the following, we provide an approximation to the thermodynamic potential $\Omega(\mu, T)$ in the limit when $T\to 0~{\rm K}$. Let us consider the integration domain in Eq. (\ref{eq:III.4}) as the union of the intervals $I_{-}=(-\infty, \mu)$ and $I_{+}=(\mu, \infty)$. For the sake of simplicity, $\mu>0$; thus, in the zero temperature limit, one has $2\cosh\left(\frac{\epsilon-\mu}{2 k_{B}T}\right)\simeq \frac{\mu-\epsilon}{2 k_{B}T}$ in $I_{-}$ whereas  $2\cosh\left(\frac{\epsilon-\mu}{2 k_{B}T}\right)\simeq \frac{\epsilon-\mu}{2 k_{B}T}$ in $I_{+}$. Therefore, one has the following expression for the thermodynamical potential,
\begin{eqnarray}
    \Omega\left(\mu, T=0~{\rm K}\right)&=&\frac{1}{2}\int_{-\infty}^{\mu}d \varepsilon \mathcal{D}_{0}(\varepsilon)\left(\epsilon-\mu\right)\nonumber\\&-&\frac{1}{2}\int_{\mu}^{\infty}d \varepsilon \mathcal{D}_{0}(\varepsilon) \left(\epsilon-\mu\right)\label{OmegaDed}
    \end{eqnarray}
{\color{black}where $\mathcal{D}_{0}(\varepsilon)$ is the DOS in the absence of scattering.}
We perform a further separation of the integration domain as $(-\infty, 0)\cup (0, \mu)$ in the first integral of last Eq. (\ref{OmegaDed}), and we add and subtract the integral $\frac{1}{2}\int_{0}^{\mu}d \varepsilon \mathcal{D}_{0}(\varepsilon) \left(\epsilon-\mu\right)$. Thus 
\begin{eqnarray}
    \Omega\left(\mu, T=0~{\rm K}\right)&=&\frac{1}{2}\int_{-\infty}^{0}d \varepsilon \mathcal{D}_{0}(\varepsilon)\left(\epsilon-\mu\right)\nonumber\\&-&\frac{1}{2}\int_{0}^{\infty}d \varepsilon \mathcal{D}_{0}(\varepsilon) \left(\epsilon-\mu\right)\nonumber\\&+&\int_{0}^{\mu}d \mathcal{D}_{0}(\varepsilon) \left(\epsilon-\mu\right).
\end{eqnarray}
Now we take advantage of the evenness of the DOS, $\mathcal{D}_{0}(-\varepsilon)=\mathcal{D}_{0}(\varepsilon)$ to make a change of variable $\epsilon\to-\epsilon$. Thus the first integral turns out as $-\frac{1}{2}\int_{0}^{\infty}d \varepsilon \mathcal{D}_{0}(\varepsilon)\left(\epsilon+\mu\right)$,  
implying the cancellation of the $\mu$ term from the first and second integrals. This procedure can also be implemented for the $\mu<0$ case. The result is given by,
\begin{equation} \label{eq:III.8}
    \Omega(\mu,T=\text{0 K})= - \int_{0}^{\infty} d \varepsilon \mathcal{D}_{0}(\varepsilon) \varepsilon + \int_{0^{+}}^{|\mu|} d \varepsilon \mathcal{D}_{0}(\varepsilon) (\varepsilon-|\mu|)
\end{equation}
This result is consistent with the treatment performed in other works \cite{Pratama2021,Sharapov2004}. As the DOS is given by 
\begin{equation} \label{eq:III.9}
\mathcal{D}_{0}(\varepsilon)= g(B) \sum_{n \in \mathbb{Z}} \delta(\varepsilon- \varepsilon_{n})
\end{equation}
the thermodynamical potential can be written as,
\begin{eqnarray}
    \Omega(\mu,T=\text{0 K})= - \sum_{n=0}^{\infty} g(B) \varepsilon_{n}+ \sum_{n=0}^{n_{f}} g(B)(\varepsilon_{n}-|\mu|)  \nonumber\\
\end{eqnarray}
This equality follows because, for a given $B$ at zero temperature, the chemical potential is a constant equal to the highest LL energy $\varepsilon_{n_{f}}$.  Now, from Eqs. \eqref{eq:III.2},\eqref{eq:III.3}, \eqref{eq:III.6} and \eqref{eq:III.8},  we obtain that $\mathcal{F}$ is simply the total energy of the system per unit area up to the highest LL
\begin{equation} \label{eq:III.11}
    \mathcal{F}=  \sum_{n=-\infty}^{n_{f}} g(B) \varepsilon_{n}.
\end{equation}

\subsection{ Revisiting the dHvA effect on flat graphene at  \texorpdfstring{$T=0~\rm{ K}$}{}\label{sec4:A}}

According to the preceding subsection for a given $B$, the chemical potential is equal to the highest LL energy $\varepsilon_{n_{f}}$, thus for the flat graphene one has $\mu= \varepsilon_{n_{f}}= \sqrt{2 e \hbar v_{F}^{2} B n_{f}}$, and 
the Helmholtz free energy per unit area $\mathcal{F}$, due to a magnetic field, is, from Eqs. \eqref{eq:III.6}, \eqref{eq:III.7}, 
\begin{equation} \label{eq:III.12}
\begin{split}
\mathcal{F}&= \sum_{n=- \infty}^{\left\lfloor\frac{ \hbar \pi \rho}{2eB} -1\right \rfloor} \left( \frac{2eB}{\hbar \pi} \right) \left( \text{sign}(n) \sqrt{2e \hbar v_{F}^{2} B |n|} \right)
\end{split}
\end{equation}
with $\lfloor\cdot \rfloor$ denotes the floor function.

We define $B_{0} \equiv  \hbar \pi \rho/ 2e $, $\hbar \omega_{c0} \equiv \sqrt{2 e \hbar v_{F}^{2} B_{0}}$,  $ \hbar \omega_{c} \equiv \sqrt{2 e \hbar v_{F}^{2} B}$, and $\Lambda= B/B_{0}$, where $B_{0}$ is the value of $B$ above which $n_{f}=0$, i.e. only the valence levels are occupied, and $\omega_{c} \hspace{1mm}(\omega_{c0})$ is the frequency of the cyclotron associated with the induction magnetic field $B \hspace{1mm}(B_{0})$. From Eq. \eqref{eq:III.12} we obtain a simplified free energy,
\begin{equation} \label{eq:III.13}
\begin{split}
\mathcal{F}&= \rho \hbar \omega_{c0}  \Lambda ^{3/2} \sum_{n=- \infty}^{\left\lfloor \frac{1}{\Lambda}-1\right\rfloor} \text{sign}(n) \sqrt{|n|}
\end{split}
\end{equation}
\begin{figure}[!ht]
    \centering
\includegraphics[scale=0.65]{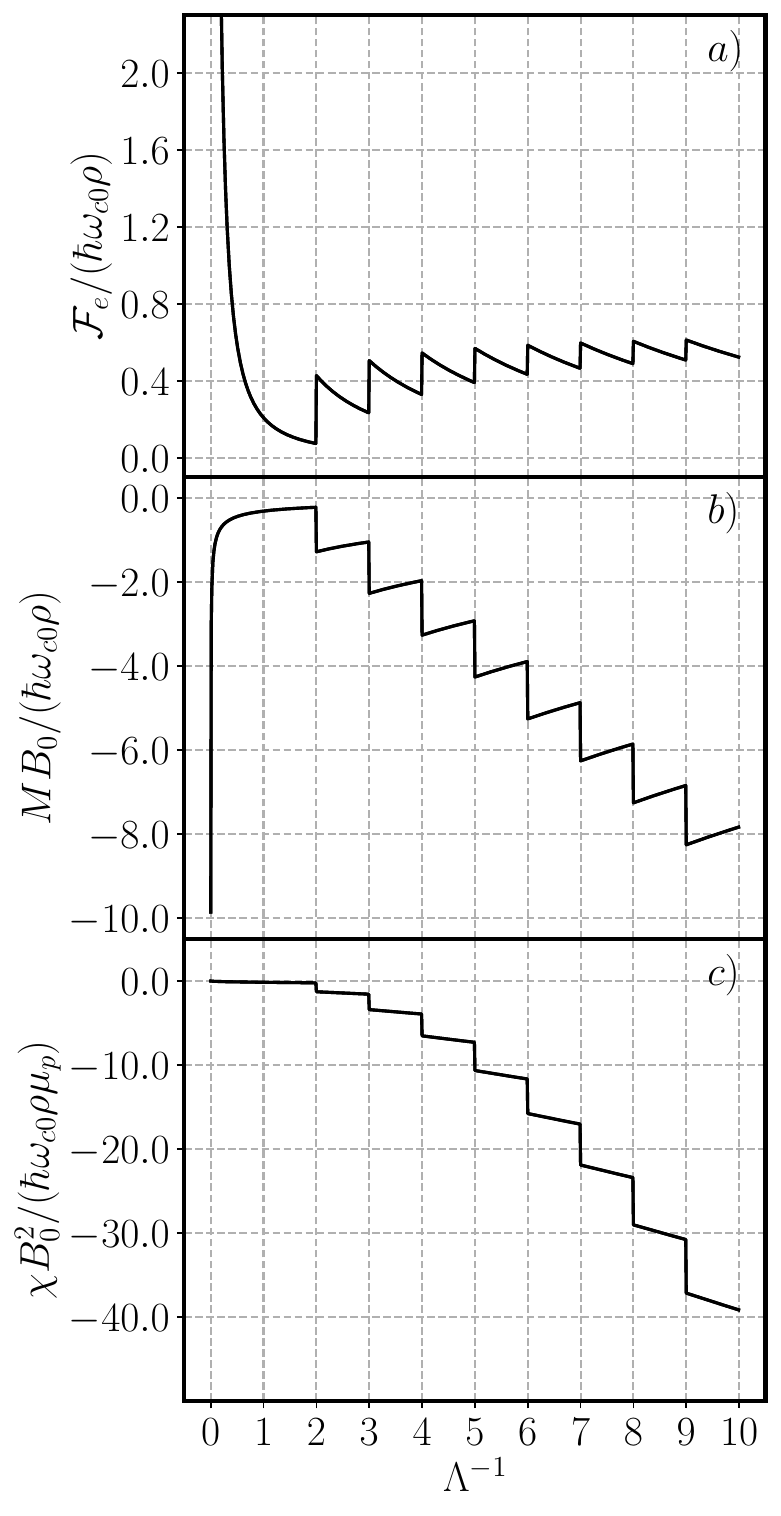}
    \caption{ Dimensionless plots of a) the Helmholtz free energy, obtained using Equation \eqref{eq:III.15}, b) the magnetization, obtained from Equation \eqref{eq:III.16}, and c) the magnetic susceptibility, obtained from Equation \eqref{eq:III.17}, as a function of the dimensionless parameter $\Lambda^{-1} = \frac{B_0}{B}$. These plots reveal the periodicity in $1/B$, as established by Onsager \cite{Onsager1952}, demonstrating the dHvA effect. Additionally, they illustrate the susceptibility's divergence as $\Lambda^{-1} \to \infty$, corresponding to $B \to 0$ \cite{Sharapov2004}. }
    
    \label{fig:Fig2}
\end{figure}
The previous series can be expressed using the Riemann zeta function, $\zeta_{R}$, and the Hurwitz zeta function, $\zeta_{H}$. 
Hence, according to the procedure detailed in Appendix \ref{firstAPP}, 
\begin{equation} \label{eq:III.15}
\begin{split}
\frac{\mathcal{F}}{\hbar \omega_{c0} \rho}&= \mathfrak{F}_{0}\left(\Lambda,0,1\right), 
\end{split}
\end{equation}
where the function $\mathfrak{F}_{0}\left(\lambda,\Delta, \overline{f}\right)$ is given by Eq. \eqref{eq:Ap.Function.General.Free.Energy} of Appendix \ref{firstAPP}. In Fig. \ref{fig:Fig2} a), we present a plot of $\mathcal{F}$ calculated from Eq. (\ref{eq:III.15}). We obtain the magnetization per unit area using the above equation (\ref{eq:III.15}), $M \equiv -\left.\left( \partial \mathcal{F} / \partial B \right)\right|_{\rho, T}$, thus following discussion in Appendix \ref{firstAPP}  
we obtain for the magnetization 
\begin{equation}\label{eq:III.16}
\frac{B_{0}}{\hbar \omega_{c0} \rho }M = \mathfrak{M}_{0}\left( \Lambda,0,1\right),  
\end{equation}
where the function $\mathfrak{M}_{0}(\lambda,\Delta,  \overline{f})$ is given in Appendix \ref{firstAPP} given by Eq. \eqref{eq:Ap.Function.General.Magnetization}.
Fig. \ref{fig:Fig2} b) presents a plot of Eq. (\ref{eq:III.16}) 
showing the typical oscillations of the magnetization $M$ as a function of $1/B$.  Increased temperature, impurity scattering, electron-electron interactions, and electron-phonon interactions cause the Landau Levels (LLs) to broaden. Consequently, the magnetization oscillation becomes less sharp \cite{Pratama2021, Sharapov2004}.

Let us now symbolize $\boldsymbol{B}= \mu_{p} \boldsymbol{H}$ and $ B_{0}= \mu_{p} H_{0}$, $\mu_{p}$ as being the magnetic permeability of free space, $\boldsymbol{H}$ stands for the magnetic field intensity. Thus, we obtain that the magnetic susceptibility per unit area, $\chi \equiv (\partial M/ \partial H)$, is
\begin{equation} \label{eq:III.17}
\frac{B_{0}^{2}}{\hbar \omega_{c0} \rho \mu_{p} }\chi = \mathfrak{S}_{0}\left( \Lambda,0,1 \right), 
\end{equation}
where the function $\mathfrak{S}_{0}(\lambda,\Delta, \overline{f})$ is given in Appendix \ref{firstAPP} by Eq. \eqref{eq:Ap.Function.General.Susceptibility}.
 As seen in Fig. \ref{fig:Fig2} c), the dependence $\sim \sqrt{B}$ of the magnetization implies that the susceptibility $\chi \propto B^{-1/2}$ diverges at zero field, a result similar to that obtained by Sergei G. Sharapov and his collaborators two decades ago \cite{Sharapov2004}. 
 {\color{black}  Experimentally, measuring such a striking result proves challenging due to the weak signal in monolayers and the effect of temperature and disorder \cite{Vallejo_2021}. The experimental magnetization was found to be aligned with the predicted dependence for the Dirac spectrum, but the doping level remained elusive \cite{Sepioni2010}. In another study, isolating the residual contribution of paramagnetic spins proved unattainable \cite{Lianlian_2013,Li2015}.
 More recently, Vallejo Bustamante et al. managed to capture signatures of such susceptibility divergence by placing two giant magnetoresistance detectors below a sample of graphene sandwiched by layers of hexagonal boron nitride \cite{Vallejo_2021}.
 Note the strong diamagnetic character of graphene in Fig. \ref{fig:Fig2} c)} at low temperatures, as confirmed in an experiment with graphene nanocrystals obtained by sonic exfoliation \cite{Sepioni2010}.

\section{DHVA effect produced by a real magnetic field in curved graphene.} \label{Sec:DHVA-CURVED-GRAPHENE}

In this section, we will discuss the dHvA effect in strongly curved graphene with a fixed electron density $\rho$. To do so,  as mentioned earlier in Sec. \ref{sec4:Anew},  it is necessary to first compute the Helmholtz free energy by summing the energy for each filled level. For this purpose, it is possible to employ a generalization of the Euler-Maclaurin formula to perform the double sums. However, it is advisable to make certain considerations that allow us to simplify the calculation. In particular, \textcolor{black}{motivated by experimental findings in freestanding graphene, which reports values for the pseudomagnetic field ranging from a few to tens of Tesla} \cite{Xu2012, Xu20121, Downs2016}, we will consider the case of strongly curved graphene under a real magnetic field and the limit {\color{black}$|B_{T}^{(\tau_R)}|S <2 \pi \hbar/e$}. Therefore, {\color{black}  $m_{max}= \left\lfloor eB_{T}^{(\tau_R)}S/2 \pi \hbar \right \rfloor =0$ 
(see the corresponding effects on the LDOS in Fig. \ref{fig:Fig6}  when $B_s \gg B_N$). 
}{\color{black}{Note that in recent experiments employing trilayer graphene encapsulated with hBN it has been possible to obtain pseudomagnetic fields on the order of millitesla \cite{Zhou2023}, i.e., within the weak curvature regime. However, this type of system extends beyond the scope of what we have assumed in this work and will be explored in future studies due to its relevance. }}

\begin{figure}
    \centering
    \includegraphics[scale=0.47]{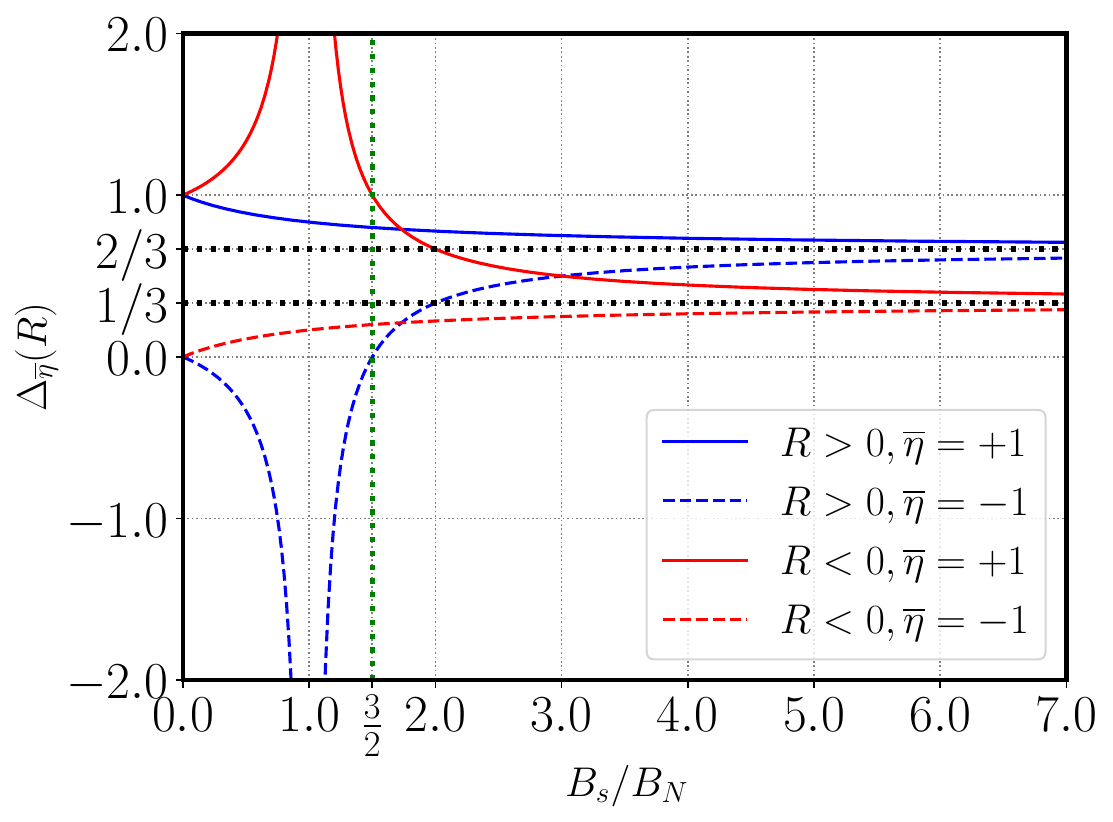}
    \caption{{\color{black}Plot of the gap-like term $\Delta_{\overline{\eta}}(R)$ (See Eq. \eqref{eq:gap-like term}). The blue and red solid (dashed) lines correspond to $R > 0, \overline{\eta}=1$ ($R > 0, \overline{\eta}=-1$) and $R < 0, \overline{\eta}=1$ ($R < 0, \overline{\eta}=-1$), respectively. Asymptotically, when $B_s/B_N \to \infty$ then $\Delta_{\overline{\eta}}(R)$ tends to $\Delta_{R}$. Meanwhile, in the limit $B_s/B_N \to 0$ then $\Delta_{\overline{\eta}}(R) \to \frac{1}{2}(1+\overline{\eta})$ analogous to a pseudo Zeeman term in flat graphene. } }
    \label{fig:delta-gap}
\end{figure}

\textcolor{black}{Thus, in the strongly curved graphene limit and} \textcolor{black}{from Eq. \eqref{eq:IV.16}, the eigenvalues can be rewritten only in terms of the principal quantum number, $n$, the pseudo-spin coupling index $\overline{\eta}$ and the curvature $R$, as }
{\color{black}
\begin{equation} \label{eq:V.2}
\begin{split}
    E_{n,\overline{\eta},\pm }(R)&=  \pm  \hbar \omega_{c, \tau_{R}}  \sqrt{  n+ \Delta_{\overline{\eta}}(R)}, ~~~n \geq n_{0}^{\overline{\eta}}, \\
\end{split}
\end{equation}}
where {\color{black} the cyclotron frequency is obtained from $\hbar \omega_{c, \tau_R }= \sqrt{2 e \hbar v_{F}^{2} |B_{T}^{(\tau_R)}|}$, and the gap-like term, $\Delta_{\overline{\eta}}(R)$, is given by
\begin{equation}\label{eq:gap-like term}
\begin{split}
   \Delta_{\overline{\eta}}(R) & \equiv \frac{1}{2}+ \frac{\text{sign(R)}}{6} \left( \frac{B_s+3\, \text{sign(R)}\overline{\eta} B_N}{|B_N + \text{sign(R)}\overline{\eta} B_s|} \right).\\
\end{split}
\end{equation}

From the energy eigenvalues (\ref{eq:V.2}), the lowest value of $n$ is $n_{0}^{\overline{\eta}} \equiv  
       \lceil- \Delta_{\overline{\eta}}(R)\rceil$, }{\color{black}
such that the LLs corresponding to $n<n_{0}^{\overline{\eta}}$ are not longer eigenstates as the pseudomagnetic field breaks the inversion symmetry and opens a gap.}
{\color{black}In other words, the primary effect of the curvature is to shift the Landau Level (LL) sequence and induce a gap at the zero level when $\Delta_{\overline{\eta}}(R) \geq 0$.  In Fig. \ref{fig:delta-gap}, we present the size of this gap term as a function of $B_s/B_{N}$ for different signs of the curvature and pseudo-spin coupling index $\overline{\eta}$.  Asymptotically, when $B_s/B_N \to \infty$,  $\Delta_{\overline{\eta}}(R)$ tends to $\Delta_{R}$, with
\begin{equation}\label{eq:deltar-1}
    \Delta_{R}\equiv \frac{1}{2}+ \frac{1}{6} \text{sign(R)},
\end{equation}
recovering the previous result \cite{Pavel2017}. Meanwhile, in the limit $B_s/B_N \to 0$, then $\Delta_{\overline{\eta}}(R) \to \frac{1}{2}(1+\overline{\eta})$ analogous to a pseudo-Zeeman term, recovering the result for the flat graphene. It is noteworthy that $\Delta_{\overline{\eta}}(R)$ is not well defined when $B_s/B_N=1$ and $\tau_{R}=-1$ because there is a resonance effect between the pseudo and external magnetic fields. Also, $\Delta_{\overline{\eta}}(R>0)$ changes sign when $B_s/B_N < 3/2$. 

}

{\color{black}
To calculate the free energy, we consider a system containing $N$ electrons within a sample of area $S$ in the low-temperature approximation, $T \approx 0$ K.
This approximation is valid because, from Eq. \eqref{eq:ciclotron frequency curvature}, the characteristic temperature $T_R$ associated with the cyclotron frequency  is
\begin{equation} \label{eq:V.1}
    \begin{split}
        T_R&= \hbar \omega_{c, \tau_{R}}/ k_B \sim 10^{3}\,\, {\rm{ K}}, 
    \end{split}
\end{equation}
for a field of $ B_{T}^{(\tau_R)} \sim 3\times 10^{2}~ \text{ T} $ \cite{Levi2010, Pavel2017}. Beyond this temperature, the approximation $T \to 0 ~{\rm K}$ is no longer applicable.
} The full occupation of LLs obeys, 
{\color{black}
\begin{eqnarray} \label{eq:V.4}
    \rho&=&\sum_{\overline{\eta}=\pm 1} \sum_{n=n_{0}^{\overline{\eta}}}^{n_{\overline{\eta},f}} g_{\overline{\eta}}(B_N,B_s)  f_{n},\nonumber\\
&=&    \sum_{\overline{\eta}=\pm 1} g_{\overline{\eta}}(B_N,B_s) (n_{\overline{\eta},f}+1-n_{0}^{\overline{\eta}}),
\end{eqnarray}
where  $f_{n}$ is the Fermi-Dirac distribution, which at low-temperature regime is $f_{n}\approx 1$, $n_{\overline{\eta}, f}$ is the highest LL occupied per pseudo-spin coupling index, and the degeneracy of the LLs is given by \cite{Pavel2017},
\begin{equation} \label{eq:V.5}
\begin{split}
g_{\overline{\eta}}(B_N,B_s)=   2 g_{s} |B_{T}^{(\tau_{R})}|/ \Phi_{0}= \frac{2e |B_{N}+\text{sign(R)} \overline{\eta}B_{s}|}{\hbar \pi},
\end{split}
\end{equation}
where $g_{s}$ is the spin degeneracy and $\Phi_{0}$ is the magnetic flux quantum. }

The Helmholtz free energy per unit area, $\mathcal{F}$, opposite to the flat case, can be expressed as the sum of two parts,
\begin{equation} \label{eq:free-energy-curved}
    \mathcal{F} = \mathcal{F}_{e}+ \mathcal{W}
\end{equation}
where $\mathcal{F}_{e}$ is the electronic free energy part obtained using the electronic spectra (\ref{eq:V.2}), whereas $\mathcal{W}$ is the elastic energy stored due to the geometrical deformation.  On the one hand, $\mathcal{W}$ per unit area is the mechanical work associated with the deformation part, and it can be expressed as \cite{landau1986theory, green1992theoretical}
 \begin{equation} \label{eq:work-energy}
     \mathcal{W}= \sigma_{ij} e_{ij}, 
 \end{equation}
i.e., it is the contraction between the deformation tensor, 
$e_{ij}$ 
and the stress tensor, 
$\sigma_{ij}$. {\color{black}On the other hand, the electronic part can be expressed as the sum of the correspondent electronic energy per pseudo-spin coupling index, $\mathcal{F}_{e}^{\overline{\eta}}$. Thus, from Eqs. \eqref{eq:V.2} and \eqref{eq:V.4}, the electronic free energy is   $\mathcal{F}_{e}\equiv \sum_{\overline{\eta}= \pm 1} \mathcal{F}_{e}^{\overline{\eta}}$, that is, 
\begin{widetext}
\begin{eqnarray} \label{eq:V.6}
\mathcal{F}_{e} 
&=& \sum_{\overline{\eta}= \pm 1} \sum_{n=n_{0}^{\overline{\eta}}}^{\infty} g_{\overline{\eta}}(B_N,B_s)  E_{n, \overline{\eta},-}(R)
 + \sum_{\overline{\eta}= \pm 1} \sum_{n=n_{0}^{\overline{\eta}}}^{n_{\overline{\eta},f}} g_{\overline{\eta}}(B_N,B_s)  E_{n, \overline{\eta},+}(R)\\
& =&\sum_{\overline{\eta}= \pm 1} \hbar \omega_{c, \tau_R} g_{\overline{\eta}}(B_{N}, B_{s}) 
\sum_{n=-\infty}^{n_{\overline{\eta},f}-n_{0}^{\overline{\eta}}} \text{sign(n)} \sqrt{|n|+n_{0}^{\overline{\eta}}+\Delta_{\overline{\eta}}(R)},
\end{eqnarray}
\end{widetext}
with the highest LL occupied $n_{\overline{\eta},f}$ calculated using equation (\ref{eq:V.4}) and the highest LL condition $E_{x, +, +}(R)=E_{y, -1, +}(R)$ using Eq. (\ref{eq:V.2}), where $x\equiv n_{+, f}$ and $y\equiv n_{-, f}$. Indeed, after a straightforward calculation using this procedure, the highest LL are obtained by
\begin{equation} \label{eq: Landau Level per pseudo-spin index}
    n_{\overline{\eta},f} = \left \lfloor \frac{1}{\tilde{\Lambda}_{\overline{\eta}}} - f_{\overline{\eta}}(R)\right\rfloor, 
\end{equation}
where $\tilde{\Lambda}_{\overline{\eta}}\equiv |B_{N}+ \text{sign(R)}\overline{\eta}B_s|/\tilde{B}_0$, being $\tilde{B}_{0}\equiv \hbar \pi \rho / 4e$, and 
\begin{eqnarray}
 f_{\overline{\eta}}(R) &\equiv& \frac{1}{2}+\frac{B_{>}+ \overline{\eta} B_N }{2(B_{>}+\text{sign(R)} \overline{\eta}B_{<})}\nonumber\\
 & -&\frac{\sum_{\overline{q}=\pm 1}n_{0}^{\overline{q}} (B_{>}+\overline{q}\text{ sign(R)} B_{<})}{2(B_{>}+\text{sign(R)} \overline{\eta}B_{<})}, 
\end{eqnarray}
where $B_{<}(B_{>})$ is the smaller (larger) of $B_{s}$ and $B_{N}$.  
Thus, we rewrite the electronic free energy in a very similar form to the flat case as 
{\color{black}
{\small\begin{eqnarray} \label{eq:V.7}
 \mathcal{F}_{e}= \frac{\rho}{2} \hbar \tilde{\omega}_{c0} \sum_{\overline{\eta}=\pm 1} \tilde{\Lambda}_{\overline{\eta}}^{3/2} \sum_{n=- \infty}^{n_{\overline{\eta},f}-n_{0}^{\overline{\eta}}} {\rm sign}(n) \sqrt{|n|+ \Delta_{\overline{\eta}}(R)+n_{0}^{\overline{\eta}}},\nonumber\\
\end{eqnarray}}
}
where $\hbar \tilde{\omega}_{c0}\equiv \sqrt{2 e \hbar v_{F}^{2} \tilde{B}_{0}}$.
Hence, after using the procedure detailed in Appendix \ref{firstAPP}, the electronic free energy is,
{\color{black}
\begin{equation} \label{eq:V.8}
   \frac{\mathcal{F}_{e}}{\hbar \tilde{\omega}_{c0} \rho}= \frac{1}{2} \sum_{\overline{\eta}= \pm 1} \mathfrak{F}_{0}\left(\tilde{\Lambda}_{\overline{\eta}},\Delta_{\overline{\eta}}(R), f_{\overline{\eta}}(R)\right), 
\end{equation}}
where the function $\mathfrak{F}_{0}(\lambda,\Delta, \overline{f})$ is given in Appendix \ref{firstAPP}. 

In Fig. \ref{fig:Fig7} a), we present the behavior of the free energy Eq.(\ref{eq:V.8}) as a function of the inverse real magnetic induction field with respect to the pseudomagnetic field. First, in  Fig. \ref{fig:Fig7} a), 
we observe oscillations that can be traced back to the interference between valleys. Secondly, there are critical values of $B_s/B_N$ at which the free energy has crossovers between negative and positive curvature cases. However, for $B_s \sim B_N$, the $R>0$ case always has lower free energy, while for $B_s \gg B_N$, the reverse case is seen, i.e., the case $R<0$ has lower free energy.
Thus, Eq. (\ref{eq:V.8}) is consistent with known examples of graphitic surfaces with positive curvature as fullerenes \cite{Terrones1995, Kroto1985} or those with negative curvature like Schwarzites, proposed many years ago by Mackay and Terrones \cite{Mackay1993,TERRONES199725,Terrones1992, Terrones1995} and other authors \cite{Zhang2017, Braun2018}.

To obtain the magnetization and magnetic susceptibility, we use the equations
\begin{equation} \label{eq:V.6.1}
\begin{split}
    M&= -\left( \frac{\partial \mathcal{F}}{\partial B_{N}} \right)_{\mu,T, e_{ij} }= -\left( \frac{\partial \mathcal{F}_{e}}{\partial B_{N}} \right)_{\mu, e_{ij} }\\
    \chi &= \left( \frac{\partial M}{\partial H} \right)_{\mu,  e_{ij}},
\end{split}    
\end{equation}
in particular, maintaining the deformation tensor, $e_{ij}$, constant.}  The resulting magnetization per unit area, $M$, {\color{black}depends whether $B_s$ is bigger or smaller than $B_N$; i.e.
\begin{itemize}
    \item[a)] in the case $B_{s} > B_{N}$,
    \begin{equation} \label{eq:V.9.1}
    \begin{split}
       \frac{\tilde{B_{0}}}{\hbar \tilde{\omega}_{c0} \rho} M&= \frac{\text{sign(R)} }{2} \sum_{\overline{\eta}= \pm 1} \overline{\eta}  \left[\mathfrak{M}_{0}\left(\tilde{\Lambda}_{\overline{\eta}},\Delta_{\overline{\eta}}(R), f_{\overline{\eta}}(R)\right) \right. \\
        &+ \left. \frac{\text{sign(R)} B_{s}}{\tilde{B_{0}}}\mathfrak{M}_{1}\left( \tilde{\Lambda}_{\overline{\eta}}, \Delta_{\overline{\eta}}(R), f_{\overline{\eta}}(R)\right) \right],
    \end{split}
\end{equation}
    \item[b)] in the case $B_N > B_s$,
    \begin{equation} \label{eq:V.9.2}
    \begin{split}
\frac{\tilde{B_{0}}}{\hbar \tilde{\omega}_{c0} \rho} M&= \frac{1}{2} \sum_{\overline{\eta}= \pm 1} \left[\mathfrak{M}_{0}\left(\tilde{\Lambda}_{\overline{\eta}},\Delta_{\overline{\eta}}(R),f_{\overline{\eta}}(R)\right) \right. \\
        &+ \left. \frac{\text{sign(R)} B_{s}}{\tilde{B_{0}}}\mathfrak{M}_{1}\left(\tilde{\Lambda}_{\overline{\eta}}, \Delta_{\overline{\eta}}(R), f_{\overline{\eta}}(R)\right) \right], 
    \end{split}
\end{equation}
\end{itemize}}
where the functions $\mathfrak{M}_{0}(\lambda,\Delta, \overline{n},\overline{f})$ and $\mathfrak{M}_{1}(\lambda,\Delta, \overline{n})$   are given in Appendix \ref{firstAPP}.   In Fig. \ref{fig:Fig7} b), we present the behavior of the magnetization as a function of the inverse magnetic induction field.

\begin{figure}[!ht]
    \centering
 \includegraphics[scale=0.65]{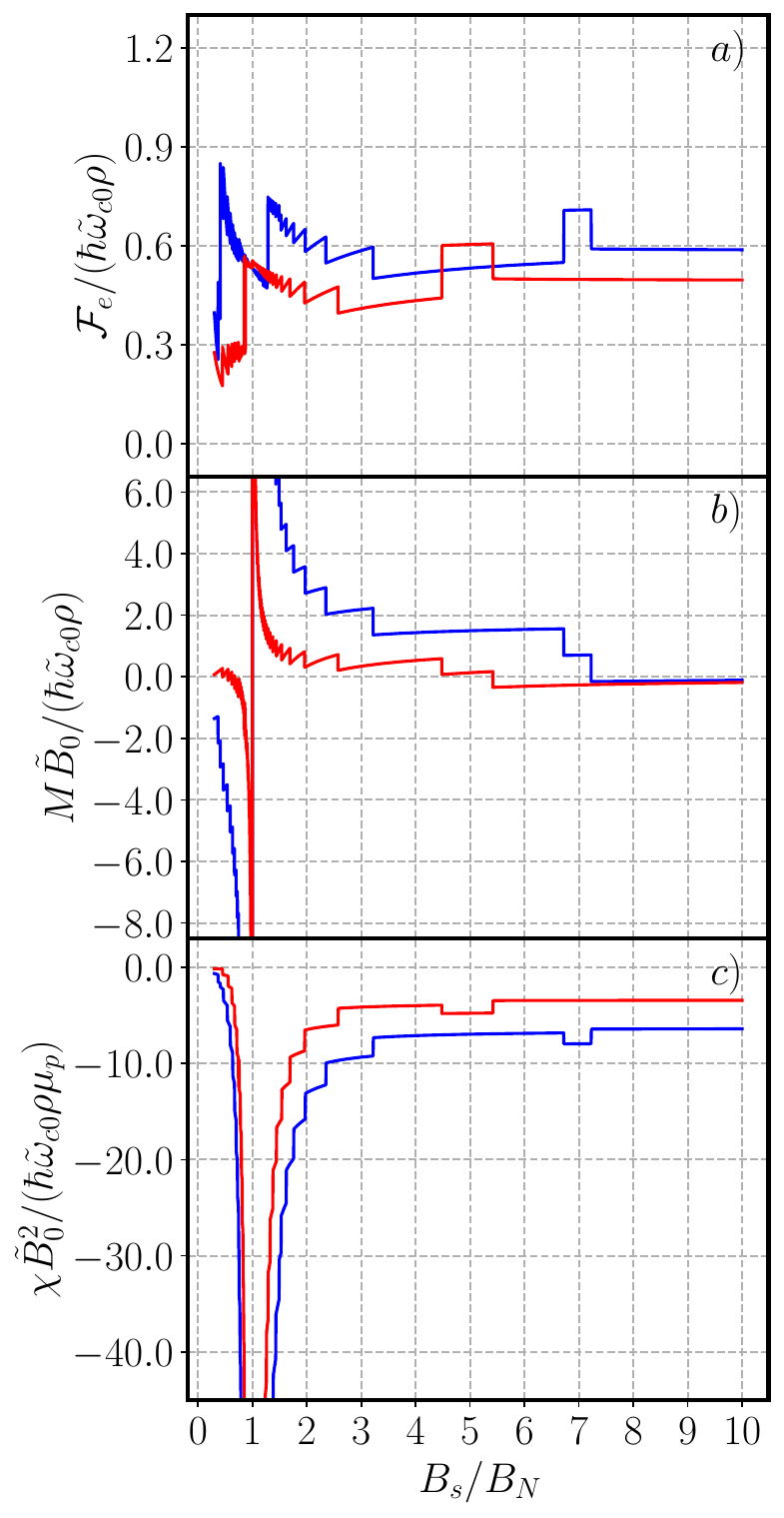}
    \caption{ Electronic part of the Helmholtz free energy (Eq. \eqref{eq:V.8}), magnetization (Eq. \eqref{eq:VI.1}), and magnetic susceptibility (Eq. \eqref{eq:VI.3}) as a function of $B_{s}/B_N$ for two different curvatures, $R>0$ (blue lines) and $R<0$ (red lines). Taking into account $\rho \sim 2 \times 10^{18} m^{-2}$ \cite{Manninen2022-1}, $B_{s} \sim 300$ T \cite{Levi2010}, and from Eq. $\tilde{B}_{0}= \hbar \pi \rho / 4e \sim 1033.47$ T, the magnetic susceptibility reaches a nearly constant value in weak magnetic fields $B_N<<B_s$, due to the inclusion of the pseudomagnetic field. This prevents the theoretical issue of the diamagnetic divergence in flat graphene at low temperatures. On the other hand, for strong magnetic fields {\color{black} such that $B_s/B_N \to 1$,  a resonance effect appears as the cyclotron frequency becomes $\omega_{\tau_{R}}=0$ in one valley and $\omega_{\tau_{R}}=2|B_N|^{1/2}$ in the other. This effect changes the magnetization from being negative to positive. However, the magnetic susceptibility is negative}. }
    \label{fig:Fig7}
\end{figure}

In Fig. \ref{fig:Fig7} c), we present the magnetic susceptibility, given by,
{\color{black}
\begin{equation} \label{eq:V.10}
    \begin{split}
        \frac{\tilde{B_{0}}^{2}}{\hbar \tilde{\omega}_{c0} \rho \mu_{p}} \chi&= \frac{1}{2} \sum_{\overline{\eta}=\pm 1}\left[\mathfrak{S}_{0}\left(\tilde{\Lambda}_{\overline{\eta}},\Delta_{\overline{\eta}}(R),  f_{\overline{\eta}}(R)\right) \right.\\
        & \left. + \mathfrak{S}_{1}\left(\frac{\text{sign(R)} B_{s}}{\tilde{B}_{0}},\tilde{\Lambda}_{\overline{\eta}},\Delta_{\overline{\eta}}(R),f_{\overline{\eta}}(R)\right)  \right], 
\end{split}
\end{equation}
}
where the functions $\mathfrak{S}_{0}(\lambda,\Delta, \overline{f})$ and $\mathfrak{S}_{1}(\mathfrak{b}_{1},\lambda,\Delta, \overline{f})$ are given in Appendix \ref{firstAPP}. 

As observed in Fig. \ref{fig:Fig7} (b) and (c), the dependence of the magnetization on $\sim \sqrt{B_{s}+ \text{sign(R)} \overline{\eta} B_{N}}$ implies that the susceptibility $\chi \sim (B_{s}+ \text{sign(R)} \overline{\eta} B_{N})^{-1/2}$ avoids the divergence problem that exists for a strictly flat sheet of graphene at zero value external magnetic field. It is noteworthy that diamagnetism in graphene decreases as the curvature increases. Therefore, corrugations appear to be essential for understanding experimentally measurable thermodynamic properties. In the literature, there are currently limited available results.

{\color{black}Interestingly, Fig. \ref{fig:Fig7} shows a resonance effect for very strong magnetic fields such that $B_s/B_N \to 1$. To understand this, we observe that the susceptibility $\chi \sim (B_{s}+ \text{sign(R)} \overline{\eta} B_{N})^{-1/2}$, displays a singularity if $\text{sign(R)} \overline{\eta}= -1$. Therein, the magnetization changes from negative to positive values.  The explanation of such phenomena is that the cyclotron frequency $\omega_{\tau_{R}} \sim |B_{N}+ \tau_{R} B_{s}|^{1/2}$ becomes zero for one valley implying the Landau Levels collapse to zero modes. In the other valley, $\omega_{\tau_{R}} \sim |2B_{N}|^{1/2}$, implying that only one valley contributes to the free energy.} 

\section{The emergence of a pseudo-dHvA effect produced by curvature without real magnetic fields} \label{sec:pseudo-dhva}

 In this section, we demonstrate how a pseudo-dHvA effect arises as a result of the mechanical properties of curved graphene. For this purpose, note that the treatment performed in Section \ref{Sec:DHVA-CURVED-GRAPHENE} can be applied with the pseudomagnetic field and define the quantities $M_{s} \equiv - \left(\partial \mathcal{F}/ \partial B_{s} \right)_{\mu}$ and $\chi_{s} \equiv \left(\partial M_{s} / \partial H_{s} \right)_{\mu}$ as the pseudomagnetization and pseudomagnetic susceptibility per unit area \cite{Dong2020, NingMa2019}, respectively. Here, $B_{s}= \mu_{p} H_{s}$ where $H_{s}$ is a pseudo-magnetic field intensity. 
 
 In the case of curved graphene under a magnetic field $B_{N}$, the pseudo magnetization per unit area, $M_{s}$ is given by the sum of an electronic pseudo magnetization part, $M_{se}$, and a pseudo magnetization part of the deformation, $M_{sd}$. The resulting pseudo magnetization per unit area, $M$, {\color{black}depends whether $B_s$ is bigger or smaller than $B_N$; i.e.}
{\color{black}
\begin{itemize}
    \item[a)] in the case $B_{s} > B_N$,
 \begin{equation} \label{eq:VI.1.1}
    \begin{split}
       \frac{\tilde{B_{0}}}{\hbar \tilde{\omega}_{c0} \rho}  M_{se}&= \frac{1}{2} \sum_{\overline{\eta}= \pm 1} \left[\mathfrak{M}_{0}\left(\tilde{\Lambda}_{\overline{\eta}},\Delta_{\overline{\eta}}(R),f_{\overline{\eta}}(R)\right) \right.\\
        &- \frac{\overline{\eta} B_{N}}{\tilde{B}_{0}} \left. \mathfrak{M}_{1}\left(\tilde{\Lambda}_{\overline{\eta}},\Delta_{\overline{\eta}}(R),f_{\overline{\eta}}(R)\right) \right], 
    \end{split}
\end{equation}
    \item[b)] in the case $B_{N} > B_{s}$,

    \begin{equation} \label{eq:VI.1}
    \begin{split}
  \frac{\tilde{B_{0}}}{\hbar \tilde{\omega}_{c0} \rho}   M_{se}&= \sum_{\overline{\eta}= \pm 1} \frac{\text{sign(R)} \overline{\eta}}{2}\left[\mathfrak{M}_{0}\left(\tilde{\Lambda}_{\overline{\eta}},\Delta_{\overline{\eta}}(R),f_{\overline{\eta}}(R)\right) \right.\\
        &- \frac{\overline{\eta}B_{N}}{\tilde{B}_{0}} \left. \mathfrak{M}_{1}\left(\tilde{\Lambda}_{\overline{\eta}},\Delta_{\overline{\eta}}(R),f_{\overline{\eta}}(R)\right) \right],
    \end{split}
\end{equation}
\end{itemize}
}
(\text{see  Eq. \eqref{eq:Ap.Function.General.Magnetization}}). The deformation pseudo magnetization part, $M_{sd}$, is not null since the stress tensor and deformation tensor depend on the pseudomagnetic field, thus it is expressed as
\begin{equation} \label{eq:VI.2}
    M_{sd}  =  \left( \frac{\partial \mathcal{W}}{\partial B_{s}} \right)= \sigma_{ij} \left(\frac{\partial  e_{ij}}{\partial B_{s}} \right)+ e_{ij} \left(\frac{\partial  \sigma_{ij}}{\partial B_{s}} \right).
\end{equation}

\begin{figure}[!ht]
    \centering
    \includegraphics[scale=0.6]{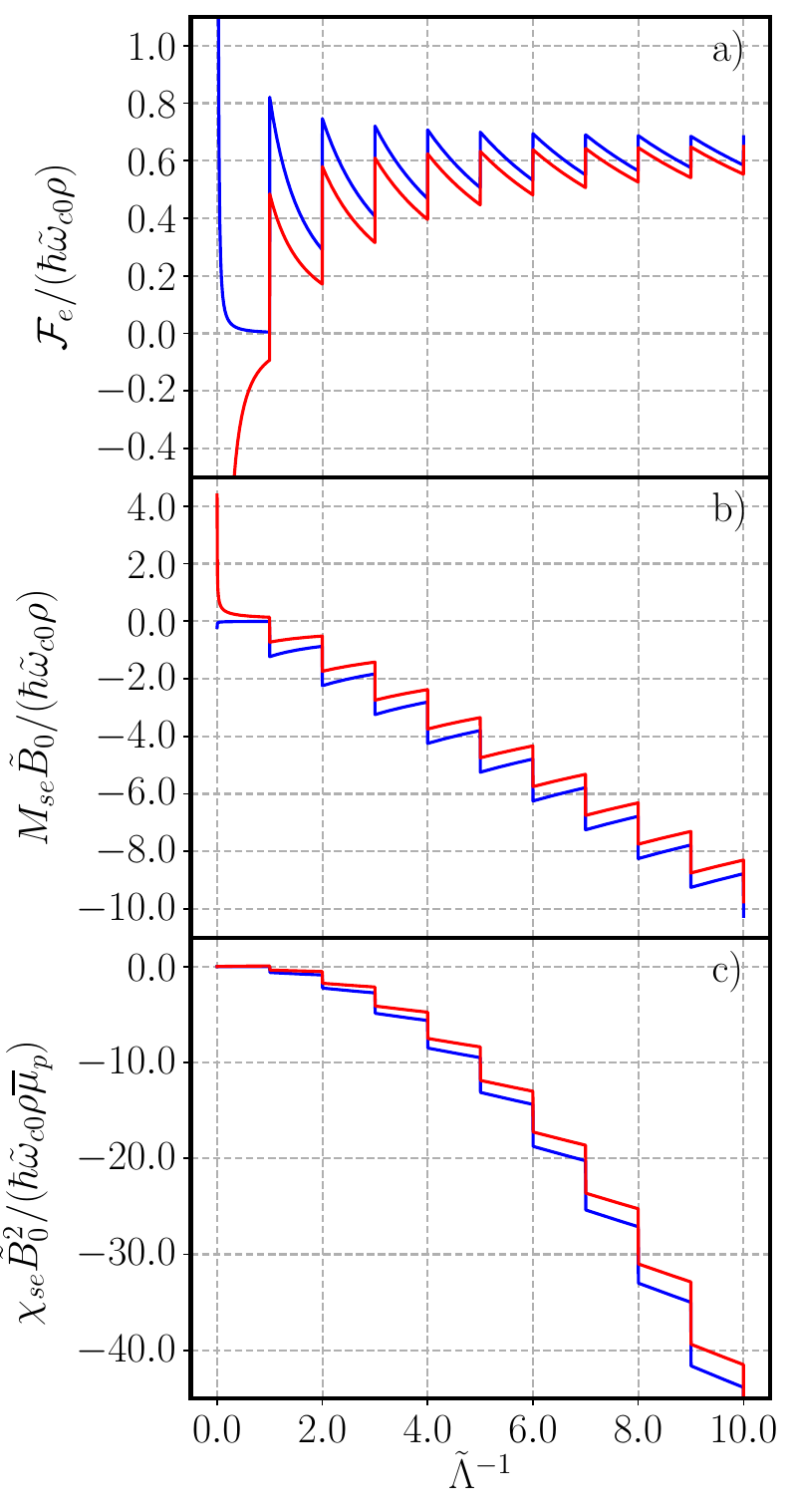}
    \caption{ Electronic part of a) the Helmholtz free energy (Eq. \eqref{eq:V.8}), b) the pseudo magnetization (Eq. \eqref{eq:VI.1}), and c) the pseudomagnetic susceptibility (Eq. \eqref{eq:VI.3}) without a magnetic field{\color{black}, i.e. $B_N=0$ T}, as a function of $\tilde{\Lambda}^{-1}= \tilde{B}_{0}/B_{s}$, for two different curvatures $R>0$ (blue lines) and $R<0$ (red lines). As mentioned in Sec. \ref{sec:pseudo-dhva}, the pseudo magnetization is associated with a mechanical stress tensor, and the pseudo susceptibility with internal reaction forces that oppose deformation, thus the discontinuities of these forces with a period of $1/B_{s}$ give rise to a pseudo-de Haas van Alphen effect.}
    
    \label{fig:Fig8}
\end{figure}
Similarly, the pseudo-susceptibility, $\chi_{s}$, will be the sum of a part due to electronics $\chi_{se}$ and a part due to the deformation $\chi_{sd}$.
{\color{black}
\begin{equation} \label{eq:VI.3}
    \begin{split}
     \frac{\tilde{B_{0}}^{2}}{\hbar \tilde{\omega}_{c0} \rho \mu_{p}}   \chi_{se}&= \frac{1}{2}\sum_{\overline{\eta}= \pm 1} \left[\mathfrak{S}_{0}\left(\tilde{\Lambda}_{\overline{\eta}},\Delta_{\overline{\eta}}(R),f_{\overline{\eta}}(R)\right)\right.\\
        & +\left. \mathfrak{S}_{1} \left(-\frac{ \overline{\eta}B_{N}}{\tilde{B}_0},\tilde{\Lambda}_{\overline{\eta}},\Delta_{\overline{\eta}}(R),f_{\overline{\eta}}(R)\right)\right], 
    \end{split}
\end{equation}
}
\text{(see  Eq. \eqref{eq:Ap.Function.General.Susceptibility})} and $\chi_{sd}$ is given by
{\small\begin{equation} \label{eq:VI.4}
\begin{split}
    \chi_{sd} &= 2\left( \frac{\partial \sigma_{ij}}{\partial B_{s}} \right) \left( \frac{\partial e_{ij}}{\partial B_{s}} \right)  + \sigma_{ij} \left( \frac{\partial^{2} e_{ij}}{\partial B_{s}^{2}} \right)+ e_{ij} \left( \frac{\partial^{2} \sigma_{ij}}{\partial B_{s}^{2}} \right)\\
    &= 2 F_{i} \left( \frac{\partial B_{s}}{\partial y_{j}} \right)^{-1} \left( \frac{\partial e_{ij}}{\partial B_{s}} \right)  + \sigma_{ij} \left( \frac{\partial^{2} e_{ij}}{\partial B_{s}^{2}} \right)+ e_{ij} \left( \frac{\partial^{2} \sigma_{ij}}{\partial B_{s}^{2}} \right),
\end{split}
\end{equation}}
where $F_{j}$ is the force density acting on the material in the direction $j$. In the previous expression, we used the fact that \cite{landau1986theory, green1992theoretical}
\begin{equation} \label{eq.VI.5}
    F_{i}= \partial \sigma_{ij}/ \partial y_{j}
\end{equation}

As seen in Fig. \ref{fig:Fig8}, we recover the divergence case when $B_{N}=0$ and $B_{s}=0$ (flat situation). However, it is necessary to note that this pseudo magnetization, $M_{se}$, and pseudo susceptibility, $\chi_{se}$, are related to mechanical observables and arise from the electronic part, and as shown in Eq. \eqref{eq:VI.2}, the pseudo magnetization of the deformation part is related to the stress tensor. From Eq. \eqref{eq:VI.3}, the pseudo susceptibility is described by oscillating internal forces acting directly on the graphene sheet, resulting in a mechanically induced pseudo de Haas van Alphen effect (pseudo-dHvA). Therefore, electronic forces act on the graphene sheet opposing flatness when $B_{N}=0$ and $B_{s} \to 0$. {\color{black} This means that spontaneous corrugations will appear to reduce the free energy.} This is related to discussions in previous works on how graphene (without substrate) achieves mechanical equilibrium by corrugation \cite{Geim2007Nature, Fasolino2007, Toby_Wiseman2018}.  

As a further result, in the strong curvature regime, the electronic forces opposing deformation are smaller in negative curvature surfaces. {\color{black}  This result is consistent with the numerical prediction by Terrones et al. concerning the stability of negative curved graphitic structures \cite{TERRONES199725}.}

{\color{black}
\section{Physical differences between magnetization and pseudomagnetization} 
\label{Sec: Physical differences between magnetization and pseudomagnetization}
In this section, we briefly discuss the physical difference between real magnetization and pseudo magnetization. To begin with, it is noteworthy to mention that magnetization represents the ratio of a change of free energy to a change in a real magnetic field, whereas pseudo-magnetization gives the ratio of the change of free energy to a change in a pseudomagnetic field. In other words, pseudo magnetization is related to how the free energy is modified when the curvature of the graphene sheet is changed. Indeed, from the quantitative viewpoint, let us consider the case $B_{s}>B_{N}$
. In this case, the magnetization per unit area can be obtained from Eq.  (\ref{eq:V.9.1}) that can be rewritten exactly as 
\begin{eqnarray}
M= \text{sign}(R) \left[ \tilde{\mu}_{+}(B_N,B_{s})-\tilde{\mu}_{-}(B_N,B_{s})\right],
\end{eqnarray}
where, we have defined $\tilde{\mu}_{\overline{\eta}}(B_N,B_s)$ as a magnetic-like moment given by 
\begin{eqnarray}
\tilde{\mu}_{\overline{\eta}}(B_N,B_s)=\tilde{\mu}^{(0)}_{\overline{\eta}}(B_N,B_s)+\frac{\text{sign(R)} B_{s}}{\tilde{B_{0}}}\tilde{\mu}^{(1)}_{\overline{\eta}}(B_N,B_s),\nonumber\\
\end{eqnarray}
where
\begin{eqnarray}
\tilde{\mu}^{(0)}_{\overline{\eta}}(B_N,B_{s})&=&\frac{\hbar \tilde{\omega}_{c0} \rho}{2\tilde{B_{0}}}\mathfrak{M}_{0}\left(\tilde{\Lambda}_{\overline{\eta}},\Delta_{\overline{\eta}}(R), f_{\overline{\eta}}(R)\right)\\
\tilde{\mu}^{(1)}_{\overline{\eta}}(B_N,B_{s})      &=& \frac{\hbar \tilde{\omega}_{c0} \rho}{2\tilde{B_{0}}}\mathfrak{M}_{1}\left( \tilde{\Lambda}_{\overline{\eta}}, \Delta_{\overline{\eta}}(R), f_{\overline{\eta}}(R)\right) .\nonumber\\
\end{eqnarray}
Therefore, the magnetization $M$ is given by an imbalance between the charge carriers with different $\overline{\eta}$. In contrast,  using Eq. \eqref{eq:VI.1.1}, the pseudo magnetization per area unit can be written exactly as  
\begin{eqnarray}
M_{se}=\tilde{\nu}_{+}(B_{N}, B_{s})+\tilde{\nu}_-(B_{N}, B_{s}),
\end{eqnarray}
where $\tilde{\nu}_{\overline{\eta}}(B_N,B_s)$ as another magnetic-like moment given by 
\begin{eqnarray}
\tilde{\nu}_{\overline{\eta}}(B_N,B_s)=\tilde{\mu}^{(0)}_{\overline{\eta}}(B_N,B_s)-\overline{\eta}\frac{ B_{N}}{\tilde{B_{0}}}\tilde{\mu}^{(1)}_{\overline{\eta}}(B_N,B_s).
\end{eqnarray}
Therefore, the pseudomagnetization $M_{se}$ is given by the sum of the magnetic like-moments $\tilde{\nu}_{\overline{\eta}}(B_{N}, B_{s})$ of charge carriers with different $\overline{\eta}$. 

Next, we look at the difference between $M$ and $M_{se}$ in the approximation $B_{s} \gg B_{N}$. At first order in $B_{N}/B_{s}$, the arguments of $\mathfrak{M}_{0}$ and $\mathfrak{M}_{1}$ are,
\begin{eqnarray}
\tilde{\Lambda}_{\overline{\eta}}&\approx& \frac{B_{s}}{\tilde{B}_{0}}\left(1+{\rm sign}(R)\overline{\eta}\frac{B_{N}}{B_{s}}\right),\\
\Delta_{\overline{\eta}}\left(R\right)&\approx&\Delta_{R}+\frac{1}{3}\overline{\eta}\frac{B_{N}}{B_{s}},\\
f_{\overline{\eta}}(R)&\approx&1+\frac{1}{2}(1-{\rm sign}(R))\overline{\eta}\frac{B_{N}}{B_{s}},
\end{eqnarray}
implying that in the limit when $B_{N}\to 0$ 
the time-reversal symmetry is restored, and the LLs are degenerated in pseudospin coupling index $\overline{\eta}$, since $\overline{\eta}$ appears as a coefficient in the term $B_{N}/B_{s}$. In addition, in this limit one has $\tilde{\mu}^{(k)}_{+}(B_N,B_{s})=\tilde{\mu}^{(k)}_{-}(B_N,B_{s})$, for $k=0,1$,  $\tilde{\mu}_{+}(B_N,B_{s})=\tilde{\mu}_{-}(B_N,B_{s})$, and  $\tilde{\nu}_{+}(B_N,B_{s})=\tilde{\nu}_{-}(B_N,B_{s})$, then the real magnetization, $M$, is zero as expected, whereas the pseudomagnetization is different than zero; indeed, $M^{0}_{se}:=\lim_{B_{N}\to 0}M_{se}=2\lim_{B_{N}\to 0}\tilde{\mu}^{(0)}_{+}(B_{N}, B_{s} )$, that is 
\begin{eqnarray}
M^{0}_{se}=\frac{\hbar \tilde{\omega}_{c0} \rho}{2\tilde{B_{0}}}\mathfrak{M}_{0}\left(\frac{B_{s}}{\tilde{B}_{0}},\Delta_{R}, 1\right)\label{Mse}
\end{eqnarray}
From this equation, we observe that the pseudo-magnetization emerges as a consequence of the corrugation of the sheet of graphene; in particular, 
it represents the variation of mechanical energy used by redistributing the charge carriers with pseudospin $\overline{\eta}=\pm 1$. Note that the behavior of the pseudo magnetization versus $\tilde{B}_{0}/B_{s}$, (\ref{Mse}), is shown in Fig. \ref{fig:Fig8} b). 

Furthermore, if one still considers  a non-zero small value of the real magnetic field $B_{N}$, one can easily show that 
\begin{eqnarray}
M_{se}\approx M^{0}_{se}+\left[\left.\left(\frac{\partial M}{\partial B_{s}}\right)\right|_{B_{N}=0}\right]B_{N},
\end{eqnarray}
where the second term indicates that real magnetic fields contribute to the resulting pseudo-magnetization, affecting the form in which the graphene can be deformed due to the charge redistribution by the magnetization.

}

\section{Concluding remarks }\label{sec5}

In this paper, we have investigated the magnetization and magnetic susceptibility of strongly curved graphene with and without magnetic fields, employing a continuous effective Dirac equation and a complementary tight-binding study. The results reveal that the magnetization and magnetic susceptibility exhibit discontinuities, following a period of $1/B$, indicative of the Haas van Alphen effect, with a dependence on the sign of the curvature. 
A nondivergent diamagnetic behavior is observable for low-intensity magnetic fields for both positive and negative curvature cases ($R<0$ and $R>0$).

Furthermore, a mechanical effect is also introduced due to the electronic contribution of graphene that gives rise to a pseudo-dHvA effect; this effect is related to oscillating (electronic) forces that oppose the deformations. {\color{black}These forces are divergent in flat graphene, indicating that graphene (without substrate) achieves mechanical equilibrium by corrugations as suggested in other works \cite{Geim2007Nature, Fasolino2007, Toby_Wiseman2018}. This implies that in free-standing graphene, the local susceptibility does not diverge. Nevertheless, when strain is applied, or encapsulation is employed, traces of such a divergent susceptibility become experimentally observable as corrugations are eliminated \cite{Vallejo_2021}}. In the strong curvature regime, the electronic forces opposing deformation are smaller for negatively curved surfaces.  

{\color{black}It is worth noting that our model captures the essential physics for low-energy states in relation to a tight-binding model. This is why, in very recent works \cite{Roberts2024}, efforts have been made to construct low-energy operators containing powers of the Dirac operator based on TB models. This approach aims to incorporate accurate information about high-energy states and, as established in our previous works \cite{Naumis_2024, Naumis2017}, to account for the shift of Dirac cones resulting from strain} {\color{black}and the effect of opposite curvatures.  Regarding this last point, if we assume the validity of our model as a first approximation, interference and resonance-type effects should be observed. This is because the sign of curvature appears coupled in the same way as valley indices.}

{\color{black} Finally, we conclude by suggesting that the proposed Haas van Alphen effect can be measured in experiments by combining an external magnetic field with induced curvature in a controlled manner. Several strategies exist to do this \cite{Naumis2017,Naumis_2024}. One possibility is to use graphene with a nanoscale ripple under an external magnetic field, similar to the device that experimentally measures the valley-dependent Landau levels \cite{ExperimentalValley}. Perhaps the most similar experiment to produce the pseudomagnetic field proposed here is the application of a local indentation by an atomic force microscope. Applying a controlled force at the tip of the microscope allows us to obtain a field akin to the one seen in Fig. \ref{fig:Fig4}. Another option is to use nanopillars on a suitable dielectric substrate. Therefore, we predict that the applied force needed to keep a specific curvature fixed will show oscillations if the magnetic field changes. 
All these results and proposals underscore the burgeoning field of \textit{curvatronics} as an area of abundant opportunities.}

\acknowledgments
PCV acknowledges the financial support of SNI-CONAHCyT. AJEC and GGN thanks the CONAHCyT fellowship (No. CVU 1007044) and the Universidad Nacional Autónoma de México (UNAM) for providing financial support (UNAM DGAPA PAPIIT IN101924 and CONAHCyT project 1564464). The authors acknowledge and express gratitude to Thomas Stegmann, and Ram\'on Carrillo-Bastos for enlightening discussions on the effects of curvature and Carlos Ernesto L\'opez Natar\'en from Secretaria T\'ecnica de C\'omputo y Telecomunicaciones for his valuable support to implement high-performance numerical calculations. 
\appendix

\section{Geometric properties of a curved surface with polar symmetry} \label{Sec: Ap.A.Geometric}
This Appendix considers a surface with a smooth deformation that preserves polar symmetry embedded in a three-dimensional space described in cylindrical coordinates. The surface is defined by a function $z(r)$.  The  differential line element for this surface is
\begin{equation} \label{ap.a.1}
    dl^{2}= dr^{2}+ r^{2} d\theta^{2} +dz^{2}= (1+ \alpha f(r)) dr^{2}+ r^{2} d \theta^{2},
\end{equation}
where
\begin{equation} \label{ap.a.2}
    dz^{2}= \left( \frac{\partial z(r)}{\partial r} \right)^{2} dr^{2} \equiv \alpha f(r)dr^{2}.
\end{equation}
Therefore, the spatial part of the metric tensor is 
\begin{equation} \label{ap.a.3}
    g_{ij}= \left( \begin{array}{cc}
        1+ \alpha f(r) & 0  \\
        0 &  r^{2}
    \end{array} \right),
\end{equation}
the symbol $\Gamma_{ij}^{k}$ gives the affine connection for the above metric, where the non-zero and non-equivalent terms are
\begin{equation} \label{ap.a.4}
    \begin{split}
        \Gamma_{rr}^{r}&= \frac{\alpha \partial_{r}f(r) }{2(1+\alpha f(r))},\\
        \Gamma_{\theta \theta}^{r}&= - \frac{r}{1+ \alpha f(r)},\\
        \Gamma_{r \theta}^{\theta}&= \frac{1}{r}.
    \end{split}
\end{equation}
On the other hand, since dreibeins satisfy $g_{\alpha \beta}= e^{\mathcal{A}}_{\alpha} e^{\mathcal{B}}_{\beta} \eta_{\mathcal{AB}}$, we make the following choice of $e_{\alpha}^{\mathcal{A}}$,

\begin{equation} \label{ap.a.5}
\begin{split}
    e^{1}_{r}&=(1+ \alpha f(r))^{1/2} \cos(\theta),\\
    e^{1}_{\theta}&= -r \sin(\theta),\\
    e^{2}_{r}&= (1+ \alpha f(r))^{1/2} \sin(\theta),\\
    e^{2}_{\theta}&= r \cos(\theta).
\end{split}
\end{equation}
Due to this, the non-zero spin connection coefficients $\omega_{\mu}^{\mathcal{AB}}$, given by $\omega_{\mu}^{\mathcal{AB}}= e^{\mathcal{A}}_{\nu}(\partial_{\nu}+ \Gamma_{\mu \lambda}^{\nu}) e^{\mathcal{B} \lambda}$, are
\begin{equation} \label{ap.a.6}
    \omega_{\theta}^{12}=1-(1+ \alpha f(r))^{-1/2},
\end{equation}
and the spin connection is 
\begin{equation} \label{ap.a.7}
    \Omega_{r}=0, \,\, \Omega_{\theta}= \frac{1-(1+\alpha f(r))^{-1/2}}{2} \gamma^{1} \gamma^{2}.
\end{equation}
Finally, from the definition of the covariant Riemann tensor, 
\begin{equation} \label{ap.a.8}
    R^{\mu}_{\alpha \nu \beta}= \partial_{\nu} \Gamma_{\alpha \beta}^{\mu}- \partial_{\beta} \Gamma_{\alpha \nu}^{\mu}+ \Gamma^{\sigma}_{\alpha \beta} \Gamma^{\mu}_{\sigma \nu}- \Gamma^{\sigma}_{\alpha \nu} \Gamma_{\sigma \beta}^{\mu},
\end{equation}
we obtain the Ricci's curvature tensor $R_{\sigma \mu}=R^{\lambda}_{\sigma \mu \lambda}$
\begin{equation} \label{ap.a.9}
    \begin{split}
        R_{rr}&= \frac{\alpha \partial_{r} f(r)}{2r (1+ \alpha f(r))},\\
        R_{\theta \theta}&= \frac{\alpha r \partial_{r} f(r)}{2 (1+ \alpha f(r))^{2}}.\\
    \end{split}
\end{equation}
Therefore, the scalar curvature $R= g^{\sigma \mu}R_{\sigma \mu}$ is 
\begin{equation} \label{ap.a.10}
    R= \frac{ \alpha \partial_{r} f(r)}{r(1+ \alpha f(r))^{2}}.
\end{equation}

In the case of a specific example given by a Gaussian deformation defined by the function
\begin{equation}  \label{ap.a.11}
z(r)= z_{0} e^{- r^{2}/(2 \varpi^{2})},
\end{equation}
where $z_{0}$ is the maximum height of the bump and $\varpi$ is the standard deviation around the origin of coordinates. Thus, we obtain that $\alpha f(r)$ and the scalar curvature $R$ are
\begin{equation} \label{ap.a.12}
\begin{split}
\alpha f(r)&= (z_{0}/ \varpi^{2})^{2} r^{2} \exp(-r^{2}/ \varpi^{2}),\\
    R&= \frac{2 \left( z_{0}/\varpi^{2} \right)^{2}  \left(1- (r^{2}/\varpi^{2} ) \right)}{ \left(1+ (z_{0}/ \varpi^{2})^{2} r^{2} \exp(-r^{2}/ \varpi^{2})\right)^{2}} e^{ - r^{2}/\varpi^{2} },
\end{split}
\end{equation}
in this case $\alpha \equiv \left( z_{0}/ \varpi^{2} \right)^{2}$ controls the type of regime we are in, so $\alpha \ll 1$ and $\alpha \gg 1$ indicate the weak and strong curvature regime, respectively.

\section{The Schrödinger-Lichnerowicz formula}
\label{appendix B}
This section will prove the Shr${\rm \ddot{o}}$dinger-Lichnerowicz formula. The starting point is the Euclidean Dirac operator ${\color{black}\slashed{D}_{\xi}= \underline{\gamma}_{\xi}^{j}(x)\boldsymbol{\nabla}_{j}^{\xi}} $ where ${\color{black}\boldsymbol{\nabla}_{j}^{\xi}:=(\nabla_{j}^{\xi}-iqA_{j}/ \hbar)}$ and ${\color{black}\nabla_{j}^{\xi}}$ is the covariant derivative acting on spinors. Now, let us square the operator ${\color{black}\slashed{D}_{\xi}}$, 
\begin{eqnarray}
{\color{black}\slashed{D}_{\xi}^2=\left[\frac{1}{2}\{\underline{\gamma}^{i}_{\xi}, \underline{\gamma}^{j}_{\xi}\}+\frac{1}{2}\left[\underline{\gamma}^{i}_{\xi}, \underline{\gamma}^{j}_{\xi}\right]\right]\boldsymbol{\nabla}_{i}^{\xi}\boldsymbol{\nabla}_{j}^{\xi}.}
\end{eqnarray}
Now, we use the Clifford algebra $\{\underline{\gamma}^{i}_{\xi}, \underline{\gamma}^{j}_{\xi}\}=2g^{ij}$ and the antisymmetric property of the commutator of $\gamma's$ to obtain 
\begin{eqnarray}
{\color{black}
\slashed{D}_{\xi}^2=g^{ij}\boldsymbol{\nabla}_{i}\boldsymbol{\nabla}_{j}+\frac{1}{2}\underline{\gamma}^{i}\underline{\gamma}^{j}\left[\boldsymbol{\nabla}_{i}, \boldsymbol{\nabla}_{j}\right].}
\end{eqnarray}
Next, we apply this operator on a spinor $\psi$ and use the explicit expression of the covariant derivative ${\color{black}\boldsymbol{\nabla}_{j}^{\xi}:=(\nabla_{j}^{\xi}-iqA_{j}/ \hbar)}$, just in the second term. Then, one has the second term

\begin{eqnarray}
{\color{black}\frac{1}{2}\underline{\gamma}^{i}_{\xi}\underline{\gamma}^{j}_{\xi}\left[\boldsymbol{\nabla}_{i}^{\xi}, \boldsymbol{\nabla}_{j}^{\xi}\right]}
&=&{\color{black}\frac{1}{2}\underline{\gamma}^{i}_{\xi}\underline{\gamma}^{j}_{\xi}\left\{\left[{\nabla}_{i}^{\xi}, {\nabla}_{j}^{\xi}\right]-i\frac{q}{\hbar}\left[{\nabla}_{i}^{\xi},A_{j}\right]\nonumber\right.}\\
&-&{\color{black}i\left. \frac{q}{\hbar}\left[A_{i},{\nabla}_{j}^{\xi}\right]
-\frac{q^2}{\hbar^{2}}\left[A_{i},A_{j}\right]\right\},}
\end{eqnarray}
the last term is zero since the gauge field $A_{i}$ is abelian. The terms in the middle can be simplified as follows  ${\color{black}\left[{\nabla}_{i}^{\xi},A_{j}\right]\psi=\left(\partial_{i}A_{j}\right)\psi}$. Thus, the last equation can be written as 
\begin{eqnarray}
{\color{black}\frac{1}{2}\underline{\gamma}^{i}_{\xi}\underline{\gamma}^{j}_{\xi}\left[\boldsymbol{\nabla}_{i}^{\xi}, \boldsymbol{\nabla}_{j}^{\xi}\right]
=\frac{1}{2}\underline{\gamma}^{i}_{\xi}\underline{\gamma}^{j}_{\xi}\left\{\left[{\nabla}_{i}^{\xi}, {\nabla}_{j}^{\xi}\right]-i\frac{q}{\hbar}(\partial_{i}A_{j}-\partial_{j}A_{i})\right\}},\nonumber\\
\end{eqnarray}
Let us consider $\underline{\gamma}^{i}_{\xi}=e^{i}_{a}\gamma^{a}_{\xi}$, $A_{a}=e_{a}^{i}A_{i}$,  and $\nabla_{a}^{\xi}=e_{a}^{i}\partial_{i}$ the covariant derivative acting on vector fields. Therefore, this last expression is, 
\begin{eqnarray}
{\color{black}
\frac{1}{2}\underline{\gamma}^{i}_{\xi}\underline{\gamma}^{j}_{\xi}\left[\boldsymbol{\nabla}_{i}^{\xi}, \boldsymbol{\nabla}_{j}^{\xi}\right]
=\frac{1}{2}\underline{\gamma}^{i}_{\xi}\underline{\gamma}^{j}_{\xi}\left[{\nabla}_{i}^{\xi}, {\nabla}_{j}^{\xi}\right]-\frac{iq}{2 \hbar}\gamma^{a}_{{\xi}}\gamma^{b}_{{\xi}}F_{ab},}
\end{eqnarray}
where $F_{ab}=\nabla_{a}^{\xi}A_{b}-\nabla_{b}^{\xi}A_{a}$ is the covariant magnetic strength tensor. The last term in the equation involves the Dirac matrices $\gamma's$ (without the underline). Now, we use the following identities
\begin{eqnarray}
{\color{black}\left[\nabla_{i}^{\xi}, \nabla_{j}^{\xi}\right]\psi}&=&{\color{black}\frac{1}{4}R_{ijkl}\underline{\gamma}^{k}_{{\xi}}\underline{\gamma}^{l}_{\xi}\psi,}\\
{\color{black} \underline{\gamma}^{i}_{\xi}\underline{\gamma}^{j}_{\xi}\underline{\gamma}^{k}_{\xi}\underline{\gamma}^{l}_{\xi}R_{ijkl}}&=&-2R,
\end{eqnarray}
where $R_{ijkl}$ is the Riemann tensor and $R$ is the Ricci scalar curvature. The above identities can be proven using the $SO(2)$ algebra 
\begin{eqnarray}
\frac{1}{4}\left[\phi_{ab}, \phi_{cd}\right]=\delta_{ac}\phi_{db}-g_{ad}\phi_{cb}-\delta_{bc}\phi_{da}+\delta_{bd}\phi_{ca},
\end{eqnarray}
where $\phi_{ab}$ is a 2nd order tensor in $SO(2)$. Also, it is important to use the following expression of the Riemann tensor in terms of the $2-$form $\omega\indices{_{j}^{a}_{b}}$
\begin{eqnarray}
R\indices{_{ij}^{ab}}=\partial_{i}\omega_{j}^{ab}-\partial_{j}\omega_{i}^{ab}+\omega\indices{_{i}^{a}_{e}}\omega\indices{_{j}^{eb}}-\omega\indices{_{j}^{a}_{e}}\omega\indices{_{i}^{eb}}.
\end{eqnarray}
Using these identities, it is not difficult to prove that
\begin{eqnarray}
{\color{black}\slashed{D}^2_{\xi}=g^{ij}\boldsymbol{\nabla}_{i}^{\xi}\boldsymbol{\nabla}_{j}^{\xi}-\frac{1}{4}R-\frac{iq}{4 \hbar}[\gamma^{a}_{\xi},\gamma^{b}_{\xi}]F_{ab}.}
\end{eqnarray}
Using the explicit representation of the $\gamma's$ matrices $\gamma^{1}_{\xi}\gamma^{0}_{\xi}= \sigma^{1}$ and $\gamma^{2}_{\xi}\gamma^{0}_{\xi}=\xi \sigma^{2}$, we can show that ${\color{black}[\gamma^{a}, \gamma^{b}]=2i\xi\epsilon^{ab}\sigma_{3}}$. Thus, the Schr$\ddot{o}$dinger-Lichnerowicz formula is given by 
\begin{eqnarray}
{\color{black}\slashed{D}_{\xi}^2=g^{ij}\boldsymbol{\nabla}_{i}^{\xi}\boldsymbol{\nabla}_{j}^{\xi}-\frac{1}{4}R+ \xi \frac{q}{2 \hbar}\sigma_{3}\epsilon_{ij}F^{ij}.}
\end{eqnarray}

\section{Commutation relations} \label{Ap.Commutation}
In this section, we will use commutation relations to determine the eigenvalues in two scenarios: a) Weak curvature and b) Strong curvature.

\subsection{Weak curvature regime.} \label{Ap.Weak}
For this, we introduce the operators $\hat{\upsilon}_{i}$ and $\pi_{i}$,  given by
\begin{equation} \label{Ap.C.1}
\begin{split}
    \hat{\upsilon}_{i}&=  y_{i} + \frac{1}{ e B_{N}} \epsilon_{ij} \pi^{j},\\
    \pi_{i}&= p_{i}+ \frac{e}{2} B_{N} \epsilon_{ij} y^{j},\\
\end{split}
\end{equation}
that satisfies the following commutation relations 
\begin{equation} \label{Ap.C.2}
\begin{split}
[\pi_{i}, \pi_{j}]&= i \hbar  e B_{N} \epsilon_{ij},\\     
[\hat{\upsilon}_{i}, \pi_{j}]&= 0,\\
[\hat{\upsilon}_{i}, \hat{\upsilon}_{j}]&= - \frac{i \hbar}{eB_{N}} \epsilon_{ij}.
\end{split}
\end{equation}
From \eqref{Ap.C.1} and the definition of angular momentum operator $\hat{L}$, we obtain that
\begin{equation} \label{Ap.C.3.1}
    - \hat{L}= \delta^{ij} \left(\frac{1}{2e B_{N}} \pi_{i} \pi_{j}- \frac{eB_{N}}{2} \hat{\upsilon}_{i} \hat{ \upsilon}_{j} \right),
\end{equation}
due to the algebraic structure, $\hat{L}$ satisfies that
\begin{equation} \label{Ap.C.3.2}
    [\hat{L},\delta^{ij} \pi_{i} \pi_{j}]=0.
\end{equation}
Therefore, it is possible to simultaneously diagonalize $\hat{L}$ and $\boldsymbol{\pi}^{2}$; for this purpose, we introduce creation operators $\hat{a}^{\dag}, \hat{b}^{\dag}$ and annihilation operators $\hat{a}, \hat{b}$ 
\begin{equation} \label{Ap.C.3}
    \begin{split}
     \hat{a}&= \sqrt{\frac{1}{2 \hbar e B_{N}}} \left( \pi_{1}+ i \pi_{2} \right),\\
     \hat{a}^{\dag}&= \sqrt{\frac{1}{2 \hbar e B_{N}}} \left( \pi_{1}- i \pi_{2} \right),\\
     \hat{b}&= \sqrt{\frac{eB_{N}}{2 \hbar}} \left( \hat{\upsilon}_{1}- i \hat{\upsilon}_{2} \right),\\
     \hat{b}^{\dag}&= \sqrt{\frac{eB_{N}}{2 \hbar}}  \left( \hat{\upsilon}_{1}+ i \hat{\upsilon}_{2} \right),
    \end{split}
\end{equation}
that satisfies the following relations
\begin{equation} \label{Ap.C.4}
\begin{split}
[\hat{a}, \hat{a}^{\dag}]&= 1,\\
[\hat{b}, \hat{b}^{\dag}]&= 1,\\
[\hat{a}, \hat{b}]&=0,\\
[\hat{a}^{\dag}, \hat{b}^{\dag}]&=0,\\
[\hat{b}^{\dag}, \hat{a}]&=0,\\
[\hat{a}^{\dag}, \hat{b}]&= 0.
\end{split}
\end{equation}
Then, from Eqs. \eqref{Ap.C.1} and \eqref{Ap.C.3.2}, we rewrite the operators $\pi_{i}$ and $\hat{\upsilon}_{i}$, given by
\begin{equation} \label{Ap.C.4.2}
\begin{split}
\pi_{1}&=  \sqrt{\frac{\hbar e B_{N}}{2}} (\hat{a}+ \hat{a}^{\dag}),\\
\pi_{2}&= i \sqrt{\frac{\hbar e B_{N}}{2}} (\hat{a}^{\dag}- \hat{a}),\\
\hat{\upsilon}_{1}&=  \sqrt{\frac{\hbar}{2eB_{N}}} (b+b^{\dag}),\\
\hat{\upsilon}_{2}&=  i \sqrt{\frac{\hbar}{2 eB_{N}}} (b-b^{\dag}),\\
\end{split}
\end{equation}
so that the angular momentum operator and the hamiltonian \eqref{eq:IV.10} are expressed as 
\begin{equation} \label{Ap.C.5}
\begin{split}
-\hat{L}&= \hbar \left( a^{\dag}a-b^{\dag}b\right),    \\
& \\
\tilde{\mathcal{H}}_{{\color{black}\xi}}^{2}&= \left( \begin{array}{cc}
          2 \hbar v_{F}^{2} e B_{N} (\hat{a}^{\dag}\hat{a}+1/2)  & 0 \\
            0 &  2\hbar v_{F}^{2} e B_{N} (\hat{a}^{\dag}\hat{a}+1/2)
        \end{array}\right)\\
 & + {{\color{black}\xi}}\left( e \hbar B_{N} v_{F}^{2}+ \hbar^{2} v_{F}^{2} \frac{R}{4}(\hat{b}^{\dag}\hat{b}-\hat{a}^{\dag}\hat{a}) \right) \sigma_{3}\\
    & + \hbar^{2} v_{F}^{2} \frac{R}{6}\left((\hat{b}^{\dag}\hat{b}-\hat{a}^{\dag}\hat{a})^{2} + \frac{1}{2} \right) \mathbb{1}.
\end{split}
\end{equation}

The resulting quantum states are quantum harmonic oscillator states like $\ket{n,l}$ with radial quantum number $n$ and angular momenta $l$ that satisfies,
\begin{equation} \label{eq:quantum-weak}
    \begin{split}
        \hat{a}\ket{n,l}&= \sqrt{n} \ket{n-1,l},\\
        \hat{a}^{\dag} \ket{n,l}&= \sqrt{n+1} \ket{n+1,l},\\
        \hat{b} \ket{n,l}&= \sqrt{l} \ket{n,l-1},\\
        \hat{b}^{\dag} \ket{n,l}&= \sqrt{l+1} \ket{n,l+1}.
    \end{split}
\end{equation}
Therefore, the square root of the eigenvalues of Hamiltonian \eqref{Ap.C.5} are
\begin{equation} \label{eq:Eig.Weak}
    E_{n,m,\tau, \xi, \pm }= \pm  \hbar \omega_{c}\sqrt{  n+\frac{1}{2}+{\color{black}\xi} \tau \frac{1+\lambda m}{2} + \frac{\lambda}{3} \left(m^{2}+ \frac{1}{2} \right)},
\end{equation}
where $\hbar \omega_{c }= \sqrt{2 e \hbar v_{F}^{2} B_{N}}$, $m=l-n$, $\tau= \pm 1$ is the eigenvalue of $\sigma_{3}$ and represent a pseudo-spin index, $\lambda= \text{sign}(R) B_{s}/B_{N}$ with $B_{s}= \hbar |R|/4e$ is the  pseudomagnetic field, $n \in \mathbb{N}$ and $m=- m_{max}, \ldots, m_{max}$. As detailed elsewhere \cite{Pavel2017}, $m_{max}= e (B_{N}+B_{s})S/2 \pi \hbar$.

\subsection{Strong curvature regime.} \label{Ap.Strong}
In a similar form as in Sec. \ref{Ap.Weak}, we introduce the operator $\hat{\Upsilon}_{i}^{({{\color{black}\tau_{R}}})}$ and $\Pi_{i}^{({{\color{black}\tau_{R}}})}$,  given by
\begin{equation} \label{Ap.D.1}
\begin{split}
\hat{\Upsilon}_{i}^{({{\color{black}\tau_{R}}})}&=  y_{i} + \frac{1}{ e B_{T}^{({{\color{black}\tau_{R}}})}} \epsilon_{ij} \Pi^{j}_{({{\color{black}\tau_{R}}})},\\
    \Pi_{i}^{({{\color{black}\tau_{R}}})}&= p_{i}+ \frac{e}{2} B_{T}^{({{\color{black}\tau_{R}}})} \epsilon_{ij} y^{j},\\
\end{split}
\end{equation}
where we defined the factor $\tau_{R}\equiv \tau \xi \rm{sign}(R)=\pm 1$ and  $B_{T}^{({{\color{black}\tau_{R}}})}=B_{N}+ \tau_{R} B_{s}$. These operators satisfy the following commutation relations, 
\begin{equation} \label{Ap.D.2}
\begin{split}
[\Pi_{i}^{({{\color{black}\tau_{R}}})}, \Pi_{j}^{({{\color{black}\tau_{R}}})}]&= i \hbar  e B_{T}^{({{\color{black}\tau_{R}}})} \epsilon_{ij},\\     
[\hat{\Upsilon}_{i}^{({{\color{black}\tau_{R}}})}, \Pi_{j}^{({{\color{black}\tau_{R}}})}]&= 0,\\
[\hat{\Upsilon}_{i}^{({{\color{black}\tau_{R}}})}, \hat{\Upsilon}_{j}^{({{\color{black}\tau_{R}}})}]&= - \frac{i \hbar}{eB_{T}^{({{\color{black}\tau_{R}}})}} \epsilon_{ij}.
\end{split}
\end{equation}
From \eqref{Ap.D.1} and the definition of angular momentum operator $\hat{L}$, we obtain that

\begin{equation} \label{Ap.D.3.1}
    - {{\color{black}\tau_{R}}} \hat{L}=  \delta^{ij} \left(\frac{1}{2e |B_{T}^{({{\color{black}\tau_{R}}})}|} \Pi_{i}^{({{\color{black}\tau_{R}}})} \Pi_{j}^{({{\color{black}\tau_{R}}})}- \frac{e|B_{T}^{({{\color{black}\tau_{R}}})}|}{2} \hat{\Upsilon}_{i}^{({{\color{black}\tau_{R}}})} \hat{ \Upsilon}_{j}^{({{\color{black}\tau_{R}}})} \right),
\end{equation}
due to the algebraic structure, $\hat{L}$ satisfies that
\begin{equation} \label{Ap.D.3.2}
    [\hat{L},\delta^{ij} \Pi_{i}^{({{\color{black}\tau_{R}}})} \Pi_{j}^{({{\color{black}\tau_{R}}})}]=0.
\end{equation}
Therefore, it is possible to simultaneously diagonalize $\hat{L}$ and $\boldsymbol{\Pi}^{2}$; for this purpose, we introduce creation operators $\hat{a}_{{{\color{black}\tau_{R}}}}^{\dag}, \hat{b}_{{{\color{black}\tau_{R}}}}^{\dag}$ and annihilation operators $\hat{a}_{{{\color{black}\tau_{R}}}}, \hat{b}_{{{\color{black}\tau_{R}}}}$ 
\begin{equation} \label{Ap.D.3}
    \begin{split}
     \hat{a}_{{{\color{black}\tau_{R}}}}&= \sqrt{\frac{1}{2 \hbar e |B_{T}^{({{\color{black}\tau_{R}}})}|}} \left( \Pi_{1}^{({{\color{black}\tau_{R}}})}+ i {{\color{black}\tau_{R}}} \Pi_{2}^{({{\color{black}\tau_{R}}})} \right),\\
     \hat{a}_{{{\color{black}\tau_{R}}}}^{\dag}&= \sqrt{\frac{1}{2 \hbar e |B_{T}^{({{\color{black}\tau_{R}}})}|}} \left( \Pi_{1}^{({{\color{black}\tau_{R}}})}- i{{\color{black}\tau_{R}}} \Pi_{2}^{({{\color{black}\tau_{R}}})} \right),\\
     \hat{b}_{{{\color{black}\tau_{R}}}}&= \sqrt{\frac{e|B_{T}^{({{\color{black}\tau_{R}}})}|}{2 \hbar}} \left( \hat{\Upsilon}_{1}^{({{\color{black}\tau_{R}}})}- i {{\color{black}\tau_{R}}}  \hat{\Upsilon}_{2}^{({{\color{black}\tau_{R}}})} \right),\\
     \hat{b}_{{{\color{black}\tau_{R}}}}^{\dag}&= \sqrt{\frac{e|B_{T}^{({{\color{black}\tau_{R}}})}|}{2 \hbar}}  \left( \hat{\Upsilon}_{1}+ i {{\color{black}\tau_{R}}}  \hat{\Upsilon}_{2}^{({{\color{black}\tau_{R}}})} \right),
    \end{split}
\end{equation}
that satisfies the following relations
\begin{equation} \label{Ap.D.4}
\begin{split}
[\hat{a}_{{{\color{black}\tau_{R}}}}, \hat{a}_{{{\color{black}\tau_{R}}}}^{\dag}]&= 1,\\
[\hat{b}_{{{\color{black}\tau_{R}}}}, \hat{b}^{\dag}_{{{\color{black}\tau_{R}}}}]&= 1,\\
[\hat{a}_{{{\color{black}\tau_{R}}}}, \hat{b}_{{{\color{black}\tau_{R}}}}]&=0,\\
[\hat{a}_{{{\color{black}\tau_{R}}}}^{\dag}, \hat{b}_{{{\color{black}\tau_{R}}}}^{\dag}]&=0,\\
[\hat{b}_{{{\color{black}\tau_{R}}}}^{\dag}, \hat{a}_{{{\color{black}\tau_{R}}}}]&=0,\\
[\hat{a}_{{{\color{black}\tau_{R}}}}^{\dag}, \hat{b}_{{{\color{black}\tau_{R}}}}]&= 0.
\end{split}
\end{equation}

Then, from Eqs. \eqref{Ap.D.1} and \eqref{Ap.D.3.2}, we rewrite the operators $\Pi_{i}^{({{\color{black}\tau_{R}}})}$ and $\hat{\Upsilon}_{i}^{({{\color{black}\tau_{R}}})}$, given by
\begin{equation} \label{Ap.D.4.2}
\begin{split}
\Pi_{1}^{({{\color{black}\tau_{R}}})}&=  \sqrt{\frac{\hbar e |B_{T}^{({{\color{black}\tau_{R}}})}|}{2}} (\hat{a}_{{{\color{black}\tau_{R}}}}+ \hat{a}_{{{\color{black}\tau_{R}}}}^{\dag}),\\
\Pi_{2}^{({{\color{black}\tau_{R}}})}&= i {{\color{black}\tau_{R}}} \sqrt{\frac{\hbar e |B_{T}^{({{\color{black}\tau_{R}}})}|}{2}} (\hat{a}_{{{\color{black}\tau_{R}}}}^{\dag}- \hat{a}_{{{\color{black}\tau_{R}}}}),\\
\hat{\Upsilon}_{1}^{({{\color{black}\tau_{R}}})}&=  \sqrt{\frac{\hbar}{2e|B_{T}^{({{\color{black}\tau_{R}}})}|}} (b_{{{\color{black}\tau_{R}}}}+b_{{{\color{black}\tau_{R}}}}^{\dag}),\\
\hat{\Upsilon}_{2}^{({{\color{black}\tau_{R}}})}&=  i {{\color{black}\tau_{R}}} \sqrt{\frac{\hbar}{2 e|B_{T}^{({{\color{black}\tau_{R}}})}|}} (b_{{{\color{black}\tau_{R}}}}-b_{{{\color{black}\tau_{R}}}}^{\dag}),\\
\end{split}
\end{equation}
so that the angular momentum operator and the hamiltonian \eqref{eq:IV.14} are expressed as 

\begin{equation} \label{Ap.D.5}
\begin{split}
- {{\color{black}\tau_{R}}} \hat{L}= \hbar \left( a_{{{\color{black}\tau_{R}}}}^{\dag}a_{{{\color{black}\tau_{R}}}}-b_{{{\color{black}\tau_{R}}}}^{\dag}b_{{{\color{black}\tau_{R}}}}\right), \hspace{4.6cm}   \\
& \\
\tilde{\mathcal{H}}_{{\color{black}\xi}}^{2}
=2 \hbar e v_{F}^{2}  \left\{\text{sign(R)} \frac{B_{s}}{3}\left(\hat{L}^{2} + \frac{1}{2} \right)  \mathbb{1}\right. \hspace{3cm}\\
     +\frac{1}{2} \left( \begin{array}{cc}
       {\color{black}\xi} B_{N}   & 0 \\
        0 &  -{\color{black}\xi}B_{N}
    \end{array} \right) \hspace{5cm}\\
     + \left.\left( \begin{array}{cc}
           |B_{T}^{({\color{black}\xi_{R}})} |(\hat{n}_{\xi_R}+\frac{1}{2})  & 0 \\
            0 &   |B_{T}^{(-{\color{black}\xi_{R}})} |(\hat{n}_{-\xi_R}+\frac{1}{2})
        \end{array}\right)\right\}, \hspace{1.2cm}
\end{split}
\end{equation}
with $\hat{n}_{\xi_R}=\hat{a}_{{\color{black}\xi_{R}}}^{\dag}\hat{a}_{{\color{black}\xi_{R}}}$, and we have introduce the term ${\color{black}\xi_{R}}$ defined by $ {\color{black}\xi_{R}\equiv \rm{sign(R)}\xi}$. 
The resulting quantum states are quantum harmonic oscillator states like $\ket{n,l}$ that satisfies
\begin{equation} \label{eq:quantum-strong}
    \begin{split}
        \hat{a}_{{{\color{black}\tau_{R}}}}\ket{n,l}&= \sqrt{n} \ket{n-1,l},\\
        \hat{a}_{{{\color{black}\tau_{R}}}}^{\dag} \ket{n,l}&= \sqrt{n+1} \ket{n+1,l},\\
        \hat{b}_{{{\color{black}\tau_{R}}}} \ket{n,l}&= \sqrt{l} \ket{n,l-1},\\
        \hat{b}_{{{\color{black}\tau_{R}}}}^{\dag} \ket{n,l}&= \sqrt{l+1} \ket{n,l+1}.
    \end{split}
\end{equation}
Therefore, the square root of the eigenvalues of Hamiltonian \eqref{Ap.D.5} are
\begin{equation} \label{eq:Eig.Strong.Magnetic}
    E_{n,m,\tau, \xi, \pm }= \pm  \hbar \omega_{c,{\color{black}\tau_{R}}}\sqrt{  n_{{\color{black}\tau_{R}}}+ \frac{1}{2}+ \frac{\lambda_{{\color{black}\tau_{R}}}}{3} \left(m_{{\color{black}\tau_{R}}}^{2}+ \frac{1}{2} \right)+ {\color{black}\xi \tau} \Theta_{{\color{black}\tau_{R}}} },
\end{equation}
where $\hbar \omega_{c ,{\color{black}\tau_{R}}}= \sqrt{2 e \hbar v_{F}^{2} |B_{T}^{({\color{black}\tau_{R}})}|}$, $m_{{\color{black}\tau_{R}}}=l_{{\color{black}\tau_{R}}}-n_{{\color{black}\tau_{R}}}$, $\tau= \pm 1$ is the eigenvalue of $\sigma_{3}$ and represent a pseudo-spin index, $\lambda_{{\color{black}\tau_{R}}}= \text{sign}(R) B_{s}/|B_{T}^{({\color{black}\tau_{R}})}|$, $\Theta_{{\color{black}\tau_{R}}}= B_{N}/2 |B_{T}^{({\color{black}\tau_{R}})}|$ with $B_{s}= \hbar |R|/4e$ is the  pseudomagnetic field, $n \in \mathbb{N}$ and $m_{\tau}=- m_{max, \tau}, \ldots, m_{max, \tau}$. As follows from Ref. \cite{Pavel2017}, $m_{max, \tau}= e |B_{T}^{({\color{black}\tau_{R}})}|S/2 \pi \hbar$.

We should note that the treatment for the case $B_{N}=0$ has been done previously \cite{Pavel2017}, such that the eigenvalues have a degeneration in the sublattice pseudo-spin and valley index, $\tau$ and $\xi$, i.e., 
\begin{equation} \label{eq:Eig.Stron.NonMagnetic}
    E_{n,m,\tau, \xi, \pm }= \pm  \hbar \omega_{c}\sqrt{  n+ \frac{1}{2}+ \frac{\text{sign(R)}}{3} \left(m^{2}+ \frac{1}{2} \right) },
\end{equation}
where $\hbar \omega_{c}= \sqrt{2 e \hbar v_{F}^{2} B_{s}}$

\section{General functions for free energy, magnetization, and magnetic susceptibility \label{firstAPP}}

Let us start with the generic adimensional free-energy
\begin{equation} \label{GenericF}
   \mathfrak{F}_{0}(\lambda,\Delta, \overline{f})= 
   \lambda^{3/2} \sum_{n=- \infty}^{\left\lfloor \frac{1}{\lambda}-\overline{f}\right\rfloor} {\rm sign}(n) \sqrt{|n|+ \Delta}.
\end{equation}
This series can be written in terms of the Riemann zeta function $\zeta_{R}(p)$, and Hurwitz zeta function, $\zeta_{H}(p,q)$, defined as
\begin{equation} \label{eq:III.14}
\begin{split}
\zeta_{R}(p)& \equiv \sum_{n=1}^{\infty} \frac{1}{n^{p}}\,\,; ~~~\zeta_{H}(p,q) \equiv \sum_{n=0}^{\infty} \frac{1}{(n+q)^{p}}.
\end{split}
\end{equation}
\textcolor{black}{Using the sum property of the Hurwitz function, one can simplify the free energy Eq. (\ref{GenericF}), }
\begin{equation} \label{eq:Ap.Function.General.Free.Energy}
\begin{split}
\mathfrak{F}_{0}(\lambda,\Delta, \overline{f})&=  - 
\lambda^{3/2} \zeta_{H}\left( - \frac{1}{2}, \left\lfloor \frac{1}{\lambda}-\overline{f}\right\rfloor+1+ \Delta \right). \\
\end{split}
\end{equation}
To calculate the magnetization, we need to differentiate the previous equation with respect to $B_s$ or $B_N$. By using the following identities \cite{Pratama2021},
\begin{eqnarray}
    \partial \zeta_{H}(p,q)/\partial q &=& - p \zeta_{H}(p+1,q),\nonumber\\
    \partial \lfloor x \rfloor/\partial x &=& \sum_{n \in \mathbb{Z}} \delta(x-n),\label{Identity}
    \end{eqnarray} 
and because $\Delta$ depends on $B_N$ and $B_s$,  we can separate the adimensional magnetization per unit area into three additive terms. The first two terms are,
\begin{equation} \label{eq:Ap.Function.General.Magnetization}
    \begin{split}
  \mathfrak{M}_{0}(\lambda,\Delta,  \overline{f})= 
  \left\{ \frac{3 {\lambda}^{1/2}}{2} \zeta_{H} \left( -\frac{1}{2}, \left\lfloor \frac{1}{\lambda}-\overline{f}\right\rfloor+1 + \Delta \right) \right.\\
        \left. - \frac{1}{2}{\lambda}^{-1/2} \zeta_{H} \left(\frac{1}{2}, \left\lfloor \frac{1}{\lambda}-\overline{f}\right\rfloor+1 + \Delta\right) \sum_{n \in \mathbb{Z}} \delta\left( n-\frac{1}{\lambda}+\overline{f} \right) \right\}, \\
{\color{black} \mathfrak{M}_{1}( \lambda, \Delta, \overline{f})= 
\lambda^{-1/2} \zeta_{H}\left( \frac{1}{2}, \left\lfloor \frac{1}{\lambda}-\overline{f}\right\rfloor+1+\Delta\right).}
    \end{split}
\end{equation}
The third term is proportional to the derivative $\partial \overline{f}/\partial B$ where $B=B_s$ or $B=B_N$ depending on the required case. However, this term is multiplied by a sum of Dirac delta functions. It only produces a marginal contribution at each magnetization jump produced when an LL is filled. Therefore, we will not consider this correction here, although we numerically confirm that it is marginal to the result.    Finally, the  magnetic susceptibility, obtained by differentiating the previous two magnetizations with respect to the fields, is,
\begin{equation} \label{eq:Ap.Function.General.Susceptibility}
    \begin{split}
     \mathfrak{S}_{0}(\lambda,\Delta,\overline{f})&= 
    \left\{ \frac{3}{4} {\lambda}^{-1/2} \zeta_{H}\left(- \frac{1}{2}, \left\lfloor \frac{1}{\lambda}-\overline{f}\right\rfloor+1+ \Delta \right)\right.\\
         &- \frac{1}{2} {\lambda}^{-3/2} \zeta_{H} \left( \frac{1}{2}, \left\lfloor \frac{1}{\lambda}-\overline{f}\right\rfloor+1 + \Delta \right)\\
         & \times \sum_{n \in \mathbb{Z}} \delta \left(n- \frac{1}{\lambda} +\overline{f}\right)\\
       & - \frac{1}{4} {\lambda}^{-5/2} \zeta_{H} \left( \frac{3}{2}, \left\lfloor \frac{1}{\lambda}-\overline{f}\right\rfloor+1+ \Delta \right) \\
       &  \times \sum_{n, m \in \mathbb{Z}} \delta \left( m- \frac{1}{\lambda} +\overline{f}\right) \delta\left( n- \frac{1}{\lambda}+\overline{f} \right)\\
       & - \frac{1}{2} \lambda^{-5/2} \zeta_{H} \left( \frac{1}{2}, \left\lfloor \frac{1}{\lambda}-\overline{f}\right\rfloor+1+ \Delta \right) \\
       & \times \left. \left.\sum_{m \in \mathbb{Z}} \delta'(x)\right|_{x=m-\frac{1}{\lambda}+\overline{f}} \right\}, \,\, {\overline{n}} \geq 1,\\
\end{split}
\end{equation}

\begin{equation} \label{eq:Ap.Function.General.Susceptibility.1}
    \begin{split}
     {\color{black} \mathfrak{S}_{1}(\mathfrak{b}_{1}, \lambda, \Delta,\overline{f})}&= {\color{black}
     \frac{\mathfrak{b}_1}{12}}\\
     & {\color{black}\times \left\{ 2 \lambda^{-3/2} \zeta_{H}\left(\frac{1}{2}, \left\lfloor \frac{1}{\lambda}-\overline{f}\right\rfloor+1+\Delta\right)\right.}\\
     & {\color{black}- \frac{\mathfrak{b_{1}}}{3} \lambda^{-5/2} \zeta_{H}\left(\frac{3}{2},\left\lfloor \frac{1}{\lambda}-\overline{f}\right\rfloor+1+\Delta\right)}\\
     &{\color{black}+  2 \lambda^{-5/2} \zeta_{H}\left(\frac{3}{2},\left\lfloor \frac{1}{\lambda}-\overline{f}\right\rfloor+1+\Delta\right)}\\
     &{\color{black} \left. \times \sum_{n \in \mathbb{Z}} \delta\left(n- \frac{1}{\lambda}+\overline{f}\right) \right\}}
\end{split}
\end{equation}

In the previous equations, we again neglected terms proportional to  $\partial \overline{f}/\partial B$.  Notice that all the previous equations are mathematically valid whenever  
\begin{equation*}
    \left\lfloor \frac{1}{\lambda}-\overline{f}\right\rfloor+\Delta+1 \geq 0.
\end{equation*}
In our problem, such a condition is always satisfied as the Fermi level is always bigger than zero.

\begin{figure*}[!ht]
    \centering
 \includegraphics[scale=0.65]{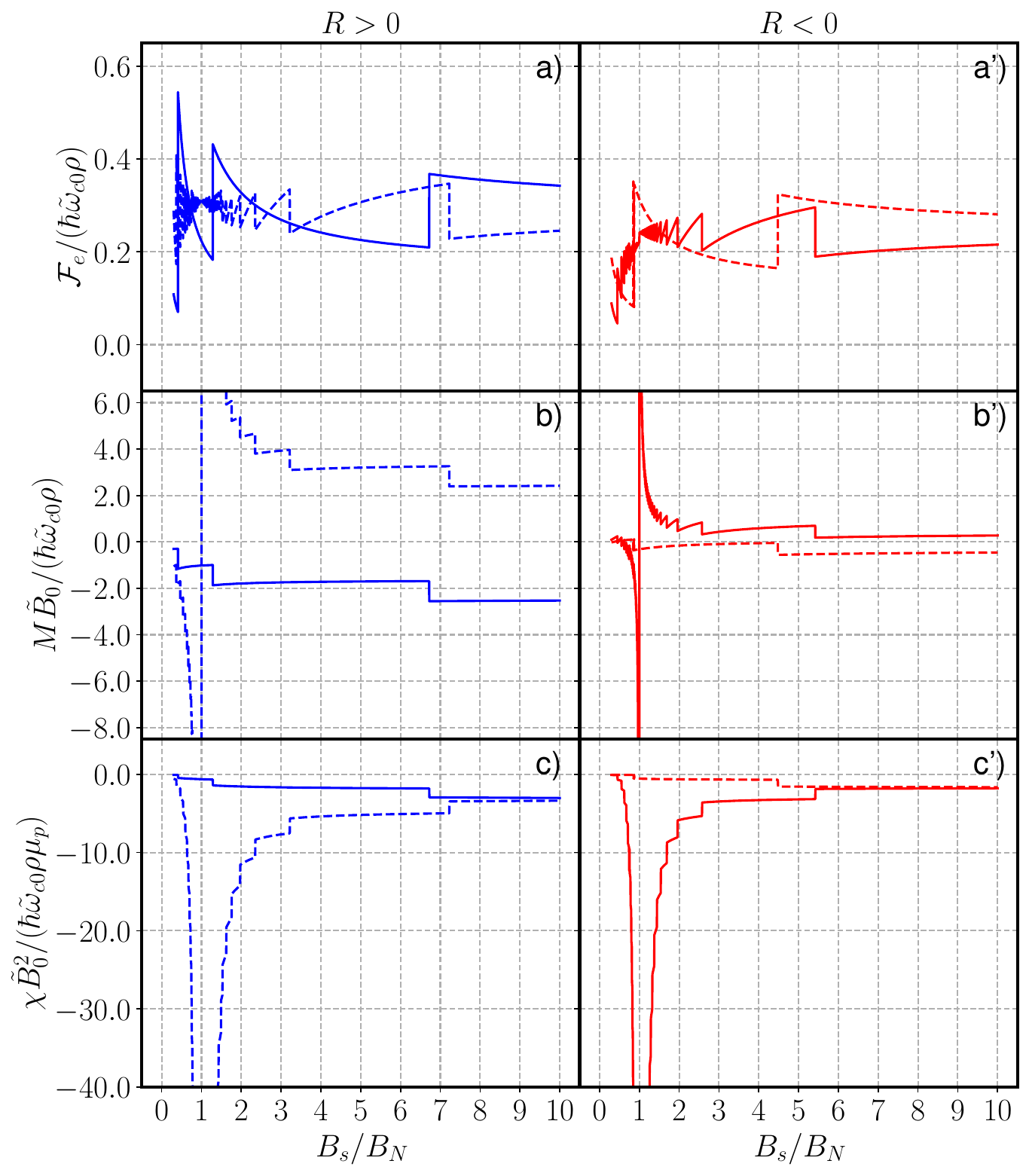}
    \caption{ {\color{black}Electronic contribution in the, a) Helmholtz free energy (Eq. \eqref{eq:V.8}), b) magnetization (Eq. \eqref{eq:VI.1}), and c), magnetic susceptibility (Eq. \eqref{eq:VI.3}) as a function of $B_{s}/B_N$ for two different curvatures $R>0$ (blue lines) and $R<0$ (red lines) separated into their respective two different pseudo-spin coupling index contributions, $\overline{\eta}=1$ (solid lines) and $\overline{\eta}=-1$ (dashed lines), respectively. The interference effect seen in Fig. \ref{fig:Fig7} is due to the different pseudo-spin coupling contributions presented here. Taken into account that $\rho \sim 2 \times 10^{18} m^{-2}$ \cite{Manninen2022-1}, $B_{s} \sim 300$ T \cite{Levi2010}, the total field used for making these plots is $\tilde{B}_{0}= \hbar \pi \rho / 4e \sim 1033.47$ T.}} 
    \label{fig:Fig9}
\end{figure*}

{\color{black}
In Fig. \ref{fig:Fig9}, we show the thermodynamical properties oscillating behavior for different curvatures and for the possible values of pseudo-spin coupling index $\overline{\eta}=\pm 1$.  If we compare with the total part seen in Fig. \ref{fig:Fig7}, we can understand how the interference effect appears due to the different contributions. 
}

\newpage
{.}
\newpage

\bibliographystyle{unsrt}
\bibliography{biblio}

\end{document}